\documentclass[amsmath,amssymb,amsfonts,aps,prl,superscriptaddress,bibnotes,showpacs,showkeys,longbibliography,notitlepage,10pt,twocolumn]{revtex4-2}
\pdfoutput=1

\usepackage{mathtools}
\usepackage{epsfig}
\usepackage[english]{babel}
\usepackage{color}
\usepackage{subcaption}
\usepackage{hyperref}
\usepackage{lipsum}
\usepackage{xcolor}
\usepackage{amsthm}
\usepackage{multirow}
\usepackage[utf8]{inputenc}

\AtBeginDocument{%
    \newwrite\bibnotes
    \def\bibnotesext{Notes.bib}
    \immediate\openout\bibnotes=\jobname\bibnotesext
    \immediate\write\bibnotes{@CONTROL{REVTEX41Control}}
    \immediate\write\bibnotes{@CONTROL{%
    apsrev41Control,author="08",editor="1",pages="1",title="0",year="1"}}
     \if@filesw
     \immediate\write\@auxout{\string\citation{apsrev41Control}}%
    \fi
}%

\begin{document}
\title{Energy-Momentum Tensor and Related Experimental Analysis of Electromagnetic Waves in Media}

\author{Gui-Xiong Liang}
\affiliation{Cangwu Wangfu Education committee office, Wuzhou city, China, District 543004}

\date{\today}

\begin{abstract}
The definition of the energy-momentum tensor for electromagnetic waves in free space is not in dispute. Thus, relating the energy-momentum tensor of an electromagnetic wave in a medium to its energy-momentum tensor in free space will help us to understand the energy-momentum tensor of an electromagnetic wave in a medium. Starting from the relationship between Noether's theorem and the Einstein field equations and the definition of energy-momentum tensor, this article uses boundary conditions to transform the energy-momentum tensor of electromagnetic waves from the free space outside the medium to the interior of the medium, and further explores the energy-momentum tensor of electromagnetic waves in the medium. We find that the energy-momentum tensor of electromagnetic waves in media is very similar to that of ordinary fluids, and concepts such as density, pressure, and energy transfer rate can be similarly defined. On this basis, we conducted a detailed theoretical analysis on the mean momentum and equivalent mass of photons in the medium, the energy transmission rate and pressure of beams in the medium, the relationship between pressure and polarization of beams, the influence of polarization energy and magnetization energy of the medium, the Bernoulli effect of beams, and the energy-momentum tensor of beams in moving media. We also obtain a conservation new energy-momentum tensor based on the interaction term between the electromagnetic field and the medium. From this energy-momentum tensor, we can derive both the Minkowski momentum and the Abraham momentum simultaneously. We find that Minkowski momentum is actually a canonical momentum that considers the influence of the interaction between electromagnetic waves and media, while Abraham momentum is actually a mechanical momentum that does not consider the influence of the interaction between electromagnetic waves and media. We also find that when the "pressure" given in this article significantly affects the experimental results, the experiment will support Minkowski momentum, namely canonical momentum. When the influence of the "pressure" given in this article on the experimental results is negligible, the experiment will support the Abraham momentum, namely mechanical momentum. Based on the theory obtained in this paper, we have provided theoretical explanations for Jones'experiment of light pressure in a medium, Ashkin's free liquid surface deformation experiment, Weilong's optical fiber deformation experiment, and frequency shift measurement experiment. The theory obtained in this paper can self-consistently explain the above experiments simultaneously. Unlike the Minkowski and Abraham tensors, according to the energy-momentum tensor proposed in this paper, a beam in a medium also generates a pressure on its side, and the direction of this pressure is related to the polarization of the beam. We hope that experimental workers will be able to design experiments to verify the presence of beam pressure on its side. The findings of this paper may shed new light on the application of light.
\end{abstract}

\maketitle

\section{Introduction}\label{sec:s1}
The correct form of the energy-momentum tensor of an electromagnetic wave, and hence the momentum of an electromagnetic wave in a dielectric medium, has been debated for more than a century. Brevik \textcolor[rgb]{0.184313725,0.188235294117647,0.564705882}{\cite{Brevik1979}}and Pfeifer\textcolor[rgb]{0.184313725,0.188235294117647,0.564705882}{\cite{Pfeifer}} reviewed this issue in 1979 and 2007, respectively. Pfeifer \textcolor[rgb]{0.184313725,0.188235294117647,0.564705882}{\cite{Pfeifer}} argues that the debate has been resolved, but some scholars argue that the debate is still open, and that the issue remains the subject of theoretical and experimental research \textcolor[rgb]{0.184313725,0.188235294117647,0.564705882}{\cite{Leonhardt}}. Indeed, research papers related to this issue continue to appear to this day\textcolor[rgb]{0.184313725,0.188235294117647,0.564705882}{\cite{Obukhov,Michael,Arthur,Angel,Matias}}.\par

Two different forms of the energy-momentum tensor were originally proposed by Minkowski\textcolor[rgb]{0.184313725,0.188235294117647,0.564705882}{\cite{Minkowski1,Minkowski2}} and Abraham\textcolor[rgb]{0.184313725,0.188235294117647,0.564705882}{\cite{Abraham1,Abraham2}}, though more have been added in later years
\textcolor[rgb]{0.184313725,0.188235294117647,0.564705882}{\cite{LinZonghan,Marx,Grot,Groot,Groot2,Groot3,Penfield,Peierls,Stephen}}.
The Minkowski and Abraham tensors give diametrically opposite results for the momentum of electromagnetic waves in a medium.According to the Minkowski tensor, the momentum of an electromagnetic wave in a medium with refractive index $n$ will be $n$ times as much as its momentum in free space. However, according to the Abraham tensor, the momentum of an electromagnetic wave in a medium with refractive index $n$ is its $1/n$ in free space.The two very different results have sparked controversy
\textcolor[rgb]{0.184313725,0.188235294117647,0.564705882}{\cite{Pfeifer}}.
Theoretical physicists on both sides of the dispute have provided theoretical arguments in favour of their side, and there have been more in-depth theoretical discussions
\textcolor[rgb]{0.184313725,0.188235294117647,0.564705882}{\cite{Penfield,Groot3,Gordon1973Radiation,1976Variational,1979About,Maugin1980Further,2007New,Stephen,2020Dielectric}}.
Experimental physicists try to experimentally verify which of these tensors is correct
\textcolor[rgb]{0.184313725,0.188235294117647,0.564705882}{\cite{JONES,JONES1954,1973Radiation,G1975Measurement,A1980Walker,Campbell2005,Weilong2008}}.\par
In the face of the century-long debate mentioned above, we believe that it is of interest to find a unified and universal definition of the energy-momentum tensor of electromagnetic waves in a medium. Therefore, we investigated this question based on Noether's theorem and Einstein field equations. The definition of the energy-momentum tensor for electromagnetic waves in free space is not controversial. Therefore, relating the energy-momentum tensor of an electromagnetic wave in the medium to its energy-momentum tensor in free space will help us to understand the problem. In this paper, we use boundary conditions to transfer the energy-momentum tensor of electromagnetic waves from free space outside the medium to the interior of the medium, and make a more in-depth exploration of the energy-momentum of electromagnetic waves in a linear non-ferromagnetic medium. After intensive investigation, we find that the energy-momentum tensor of electromagnetic waves in media is very similar to that of ordinary fluids, and concepts such as density, pressure and energy transfer rate can be similarly defined. Motivated by this, we have performed a detailed theoretical analysis of the mean momentum and pressure of light in the medium and presented our own unique insights. In this paper, we combine the interaction term between macroscopic electromagnetic waves and the medium to obtain a new energy-momentum tensor for macroscopic electromagnetic waves in the medium that can derive both Minkowski momentum and Abraham momentum. We find that Minkowski momentum is actually a canonical momentum that takes into account the contribution of the interaction term, while Abraham momentum is actually a mechanical momentum that does not take into account the contribution of the interaction term. In this respect, we are consistent with the ideas by Stephen of the University of Strathclyde in 2010\textcolor[rgb]{0.184313725,0.188235294117647,0.564705882}{\cite{Stephen}}, but the argumentation method we adopt is completely different. Our reasoning process is more helpful in revealing the connection and difference between these two types of momenta. In this paper, we discuss not only the energy-momentum tensor of the macroscopic pure electromagnetic field, but also the energy-momentum tensor including the polarization energy and magnetization energy of the medium, and compare it with the Minkowski tensor and the Abraham tensor. We also discuss the relation between the beam pressure and the polarization of the light, and the Bernoulli effect of the beam pressure. I also discuss the energy-momentum tensor of electromagnetic waves in moving media. Based on the theory obtained in this paper, we have provided theoretical explanations for Jones'experiment of light pressure in a medium\textcolor[rgb]{0.184313725,0.188235294117647,0.564705882}{\cite{JONES,JONES1954}},
Ashkin's free liquid surface deformation experiment\textcolor[rgb]{0.184313725,0.188235294117647,0.564705882}{\cite{1973Radiation}},
Weilong's optical fiber deformation experiment\textcolor[rgb]{0.184313725,0.188235294117647,0.564705882}{\cite{Weilong2008}}, and frequency shift measurement experiments\textcolor[rgb]{0.184313725,0.188235294117647,0.564705882}{\cite{Brevik1979}}. The theory obtained in this paper can self-consistently explain these experimental phenomena. This paper also presents some new predictions that need to be tested experimentally.\par
In this paper, like $\, _{\text{in}}F^{\mu  \nu }$, this article stipulates that the subscript written before the physical quantity is used to indicate the physical meaning of the physical quantity, such as "in" here indicating it as "incident beam". The Greek letters $\mu,\nu$, written after the physical quantities, range from 0 to 3 and denote the specific indices of the tensor, with 0 indicating the index of the time component. Latin letters such as $i,j$ from 1 to 3 denote specific indices of spatial components. Latin letters such as $a$ and $b$ denote abstract index notations of the tensors. In addition to the special cases already explained, for example, the magnetic permeability in free space is still recorded as $\mu_0$ instead of $\, _{0}\mu$, and the polarizability is still recorded as $\chi_e$ instead of $\, _{e}\chi$. In this paper, the upper and lower pairs of indicators indicate the execution of the Einstein summation convention. In this paper,  physical quantities are represented as vectors or tensors in bold type when indexes are not used.

\section{Noether's theorem and  Einstein field equations and the definition of the energy-momentum tensor}\label{sec:s2}
The reason for this debate is that the energy-momentum tensor of the electromagnetic field has been defined from different perspectives in different forms of media.Therefore, it is necessary to further explore how to define the energy-momentum tensor of the field, especially in the presence of interactions.The energy-momentum tensor involves the conservation of energy and momentum, so its definition is closely related to Noether's theorem. Let us briefly recall Noether's theorem\textcolor[rgb]{0.184313725,0.188235294117647,0.564705882}{\cite{2010Noether}}. Let us assume that $\mathcal{L}=\mathcal{L}\left(\boldsymbol\Phi ,\partial _a\boldsymbol\Phi ,\boldsymbol\Psi ,\partial _a\boldsymbol\Psi\right)$ is the Lagrangian density determined by the interacting matter fields $\boldsymbol\Phi$ and $\boldsymbol\Psi$, including the interaction terms $\boldsymbol\Phi$ and $\boldsymbol\Psi$.Now we move the matter field as a whole from the field point $x^\mu$ to the field point $x'^{\mu }=x^{\mu }+\epsilon  \xi ^{\mu }$, where $\epsilon$ is an arbitrary infinitesimal constant parameter, $\xi ^{\mu }$ is an arbitrary vector field called the infinitesimal generator, and $ x^{\mu } \rightarrow x'^{\mu }$ is an infinitesimal map.After translation, we obtain a new Lagrangian density $\mathcal{L}'$.Let $L=\sqrt{-g}\mathcal{L}$ and $L'=\sqrt{-g'}\mathcal{L}'$ and their corresponding actions be $S=\displaystyle\int Ld^4x$ and $S'=\displaystyle\int L' d^4x'$, respectively. Since $ x^{\mu } \rightarrow x'^{\mu }$ is an infinitesimal mapping, and since $\xi ^{\mu }$ is a vector field, the derivative with respect to it should be a covariant derivative, so we will get $d^4 x'= \left(1+\epsilon\nabla _{\mu }\xi ^{\mu }\right)d^4 x$ after omitting higher order infinitesimal quantities.Thus, we have the variation of the action:
\begin{equation}
    \label{eq:1}
 \delta  S=S'-S=\displaystyle\int \left(\delta  L+L\epsilon \nabla _{\mu }\xi ^{\mu }\right)d^4 x
\end{equation}
where:
\begin{equation}
    \label{eq:2}
 \begin{array}{ll}\delta  L=\left[L'\left(\boldsymbol\Phi' ,\partial '_a\boldsymbol\Phi' ,\boldsymbol\Psi' ,\partial' _a\boldsymbol\Psi '\right)-L'\left(\boldsymbol\Phi ,\partial _a\boldsymbol\Phi ,\boldsymbol\Psi ,\partial _a\boldsymbol\Psi\right)\right]\\ \text{ }\text{ }\text{ }\text{ }+\left[L'\left(\boldsymbol\Phi ,\partial _a\boldsymbol\Phi ,\boldsymbol\Psi ,\partial _a\boldsymbol\Psi \right)-L\left(\boldsymbol\Phi ,\partial _a\boldsymbol\Phi ,\boldsymbol\Psi ,\partial _a\boldsymbol\Psi\right)\right] \end{array}
\end{equation}
and have:
\begin{equation}
    \label{eq:3}
 \begin{array}{ll}
 L'\left(\boldsymbol\Phi' ,\partial '_a\boldsymbol\Phi' ,\boldsymbol\Psi' ,\partial' _a\boldsymbol\Psi '\right)-L'\left(\boldsymbol\Phi ,\partial _a\boldsymbol\Phi ,\boldsymbol\Psi ,\partial _a\boldsymbol\Psi\right)\\=L'\left(x'^{\mu }\right)-L'\left(x^{\mu }\right) \\= \frac{\partial  L}{\partial  x^{\mu }}\epsilon  \xi ^{\mu }
 \end{array}
\end{equation}
Make $\bar{\delta }L=L'\left(\boldsymbol\Phi ,\partial _a\boldsymbol\Phi ,\boldsymbol\Psi ,\partial _a\boldsymbol\Psi \right)-L\left(\boldsymbol\Phi ,\partial _a\boldsymbol\Phi ,\boldsymbol\Psi ,\partial _a\boldsymbol\Psi\right)$, so from Eq.\textbf{(}\ref{eq:1}\textbf{)}-Eq.\textbf{(}\ref{eq:3}\textbf{)}, we obtain:
\begin{equation}
    \label{eq:4}
 \delta  L=\frac{\partial  L}{\partial  x^{\mu }}\epsilon  \xi ^{\mu }+\bar{\delta }L
\end{equation}
$\bar{\delta }L$ is also a variation, Its algorithm is as follows\textcolor[rgb]{0.184313725,0.188235294117647,0.564705882}{\cite{Noether}}:
\begin{equation}
    \label{eq:5}
 \bar{\delta }L=\frac{\partial  L}{\partial  \boldsymbol\Phi }\bar{\delta }\boldsymbol\Phi +\frac{\partial  L}{\partial  \left(\partial _a\boldsymbol\Phi \right)}\bar{\delta }\left(\partial _a\boldsymbol\Phi \right)+\frac{\partial  L}{\partial  \boldsymbol\Psi}\bar{\delta }\boldsymbol\Psi +\frac{\partial  L}{\partial  \left(\partial _a\boldsymbol\Psi \right)}\bar{\delta }\left(\partial _a\boldsymbol\Psi \right)
\end{equation}
Substituting Eq.\textbf{(}\ref{eq:5}\textbf{)} into Eq.\textbf{(}\ref{eq:4}\textbf{)}, then substituting the resulting result into Eq.\textbf{(}\ref{eq:1}\textbf{)}. After simplification , we obtain:
\begin{equation}
    \label{eq:6}
 \delta  S=\displaystyle\int  \nabla _{\mu }\left(L \epsilon  \xi ^{\mu } +\frac{\partial  L}{\partial  \left(\partial _{\mu }\boldsymbol\Phi \right)}\bar{\delta }\boldsymbol\Phi  +\frac{\partial  L}{\partial  \left(\partial_{\mu }\boldsymbol\Psi \right)}\bar{\delta }\boldsymbol\Psi \right) d^4 x
\end{equation}
In the above derivation, we applied $\frac{\partial  L}{\partial  \boldsymbol\Phi }-\partial _{\mu }\frac{\partial  L}{\partial  \left(\partial _{\mu }\boldsymbol\Phi \right)} =0$ and $ \frac{\partial  L}{\partial  \boldsymbol\Psi}-\partial _{\mu}\frac{\partial  L}{\partial  \left(\partial _{\mu}\boldsymbol\Psi \right)} =0$. They can be derived from the least action principle.Due to $\bar{\delta }\boldsymbol\Phi =(\partial_\nu\boldsymbol\Phi) \epsilon\xi^\nu$, $\bar{\delta }\boldsymbol\Psi =(\partial_\nu\boldsymbol\Psi) \epsilon\xi^\nu$, $\xi ^{\mu }=\delta^\mu_\nu\xi^\nu$, and because $L=\sqrt{-g}\mathcal{L}$ and $\nabla _{\mu }\sqrt{-g}=\frac{\partial  \sqrt{-g}}{\partial  \left(g_{\alpha  \beta }\right)}\nabla _{\mu }g_{\alpha  \beta }=0$, from Eq.\textbf{(}\ref{eq:6}\textbf{)}, we obtain:
\begin{equation}
    \label{eq:7}
\delta  S=\displaystyle\int  \nabla _{\mu }\left(\mathcal{H}^{\mu \nu }\xi_\nu \right)\epsilon\sqrt{-g} d^4 x
\end{equation}
where:
\begin{equation}
    \label{eq:8}
\mathcal{H}^\mu _{\text{ }\nu} =\mathcal{L}\delta ^{\mu }_{\text{ }\nu} +\frac{\partial  \mathcal{L}}{\partial  \left(\partial _{\mu }\boldsymbol\Phi \right)}\partial_\nu\boldsymbol\Phi  +\frac{\partial  \mathcal{L}}{\partial  \left(\partial_{\mu }\boldsymbol\Psi \right)}\partial_\nu\boldsymbol\Psi
\end{equation}
If the system is invariant under the Infinitesimal transformation $x'^{\mu }=x^{\mu }+\epsilon  \xi ^{\mu }$, then the variation of the action is $\delta S=0$. Thus from $\epsilon$ being an arbitrary infinitesimal constant, we obtain:
\begin{equation}
    \label{eq:9}
 \nabla _{\mu }\left(\mathcal{H}^{\mu \nu} \xi_\nu \right)=0
\end{equation}
When using the rectangular coordinate system in flat spacetime, the Christoffel symbol is $\Gamma^\mu_{\nu\beta}=0$\textcolor[rgb]{0.184313725,0.188235294117647,0.564705882}{\cite{LiangCanbin2006}}, so we have:
\begin{equation}
    \label{eq:10}
 \partial _{\mu }\left(\mathcal{H}^{\mu  \nu } \xi _{\nu }\right)=0
\end{equation}
In this case, $J^\mu=\mathcal{H}^{\mu  \nu } \xi _{\nu }$ is a conserved flow. By Gauss's law, the corresponding conserved quantity is $Q=\displaystyle\int J^0d^3x$. Eq.\textbf{(}\ref{eq:7}\textbf{)} shows that when there is $\delta S=0$ under a certain Infinitesimal transformation, a conserved quantity can be given, which is the famous Noether's theorem.Where the infinitesimal transformation is a translation or rotation of the spacetime, $\mathcal{H}^{\mu  \nu } $ is here the energy-momentum tensor.When $\mathcal{C}^{\mu  \nu } $ makes $\displaystyle\int  \nabla _{\mu }\left(\mathcal{C}^{\mu \nu }\xi_\nu \right)\epsilon\sqrt{-g} d^4 x=0$ valid, $\mathcal{H}^{\mu  \nu } +\mathcal{C}^{\mu  \nu } $ makes Eq.\textbf{(}\ref{eq:7}\textbf{)} still valid and thus also gives a conserved flow.Therefore, the form of the energy-momentum tensor is not the only one, anything that can be written in the form of Eq.\textbf{(}\ref{eq:7}\textbf{)} can be called the energy-momentum tensor, which is the fundamental reason why people can define various forms of energy-momentum tensors for electromagnetic waves in the medium. However, if we combine General relativity, the situation is different.
Since the Lagrangian density of the Tensor field of matter is also related to the metric field $g_{\mu  \nu }$ of spacetime, there is $\mathcal{L}=\mathcal{L}\left(g_{\mu  \nu },\boldsymbol\Phi ,\partial _a\boldsymbol\Phi ,\boldsymbol\Psi ,\partial _a\boldsymbol\Psi\right)$. Therefore, the action of the infinitesimal transformation on the metric field should also be taken into account when performing the variation of $L=\sqrt{-g}\mathcal{L}$, so that we have:
\begin{equation}
    \label{eq:11}
 \delta  S=\displaystyle\int \left(\frac{\delta  L}{\delta  g_{\mu  \nu }}\text{$\delta $g}_{\mu  \nu }+\nabla _{\mu }\left(\mathcal{H}^{\mu  \nu } \xi _{\nu }\right)\epsilon \sqrt{-g}\right) d^4 x
\end{equation}
The Einstein field equations of the form $G^{\mu\nu}=\frac{8\pi G}{c^4} \frac{-2}{\sqrt{-g}}\frac{\delta L }{\delta  g_{\mu  \nu }} $ can be obtained from the least action principle\textcolor[rgb]{0.184313725,0.188235294117647,0.564705882}{\cite{LiangCanbin2006}}. Since the Einstein tensor $G^{\mu\nu}$ is a symmetric tensor with 0 covariant divergence, $\frac{\delta  L}{\delta  g_{\mu  \nu }}$ is also a symmetric tensor with 0 covariant divergence, namely $\nabla _{\mu }\frac{\delta  L}{\delta  g_{\mu  \nu }}=0$ and $\frac{\delta  L}{\delta  g_{\mu  \nu }}=\frac{\delta  L}{\delta  g_{\nu  \mu }}$ are true. For the infinitesimal transformation $x'^{\mu }=x^{\mu }+\epsilon  \xi ^{\mu }$, $\delta g_{\mu  \nu }$ is given by\textcolor[rgb]{0.184313725,0.188235294117647,0.564705882}{\cite{Dirac}}:
\begin{equation}
    \label{eq:12}
 \begin{array}{ll}\delta g_{\mu  \nu }=\epsilon \left(\nabla _{\mu }\xi _{\nu }+\nabla _{\nu }\xi _{\mu } \right)+\nabla_{\beta} (g_{\mu\nu }) \epsilon \xi ^{\beta }\\ \text{ }\text{ }\text{ }\text{ }\text{ }\text{ }= \epsilon \left(\nabla _{\mu }\xi _{\nu }+\nabla _{\nu }\xi _{\mu } \right)\end{array}
\end{equation}
Thus, from $\nabla _{\mu }\frac{\delta  L}{\delta  g_{\mu  \nu }}=0$ and $\frac{\delta  L}{\delta  g_{\mu  \nu }}=\frac{\delta  L}{\delta  g_{\nu  \mu }}$, we obtain:
\begin{equation}
    \label{eq:13}
 \frac{\delta  L}{\delta  g_{\mu  \nu }}\delta  g_{\mu  \nu }=\epsilon \nabla _{\mu }\left(\frac{2\delta  L}{\delta  g_{\mu  \nu }}\xi _{\nu }\right)
\end{equation}
Substituting Eq.\textbf{(}\ref{eq:13}\textbf{)} into Eq.\textbf{(}\ref{eq:11}\textbf{)}, we obtain:
\begin{equation}
    \label{eq:14}
 \delta  S=\displaystyle\int \nabla _{\mu }\left[\left(\frac{2}{\sqrt{-g}}\frac{\delta  L}{\delta  g_{\mu  \nu }}+\mathcal{H}^{\mu  \nu } \right)\xi _{\nu }\right]\epsilon \sqrt{-g} d^4 x
\end{equation}
Let $\frac{2}{\sqrt{-g}}\frac{\delta  L}{\delta  g_{\mu  \nu }}+\mathcal{H}^{\mu  \nu }=C^{\mu  \nu }$, then when $\delta  S=0$ ,there is $\displaystyle\int \nabla _{\mu }\left(-C^{\mu  \nu }\xi _{\nu }\right)\epsilon \sqrt{-g} d^4 x=0$. Adding it to Eq.\textbf{(}\ref{eq:7}\textbf{)} yields:
\begin{equation}
    \label{eq:15}
 \delta  S=\displaystyle\int \nabla _{\mu }\left(T^{\mu  \nu } \xi _{\nu }\right)\epsilon \sqrt{-g} d^4 x
\end{equation}
where $T^{\mu  \nu }=\mathcal{H}^{\mu  \nu }-C^{\mu  \nu }$. From $\frac{2}{\sqrt{-g}}\frac{\delta  L}{\delta  g_{\mu  \nu }}+\mathcal{H}^{\mu  \nu }=C^{\mu  \nu }$, we can obtain:
\begin{equation}
    \label{eq:16}
 T^{\mu  \nu }=\frac{-2}{\sqrt{-g}}\frac{\delta  L}{\delta  g_{\mu  \nu }}=\frac{-2}{\sqrt{-g}}\frac{\delta (\mathcal{L} \sqrt{-g})}{\delta  g_{\mu  \nu }}
\end{equation}
Comparing Eq.\textbf{(}\ref{eq:15}\textbf{)} and Eq.\textbf{(}\ref{eq:7}\textbf{)}, it can be seen that the physical meaning of $T^{\mu  \nu }$ is also an energy-momentum tensor. Compared with $\mathcal{H}^{\mu \nu} $ defined by Eq.\textbf{(}\ref{eq:8}\textbf{)}, it has the advantage of symmetry. According to Einstein's derivation, the Einstein field equations can be written in terms of $G^{\mu\nu}=\frac{8\pi G}{c^4} S^{\mu\nu}$, and its physical meaning implies that $S^{\mu\nu}$ on the right hand side of the field equations should be understood as the energy-momentum tensor of the matter field\textcolor[rgb]{0.184313725,0.188235294117647,0.564705882}{\cite{LiangCanbin2006}}
\textcolor[rgb]{0.184313725,0.188235294117647,0.564705882}{\cite{Wheeler}}. The Einstein field equations hold only if $S^{\mu\nu}=T^{\mu\nu}$. In this sense, the form of the energy-momentum tensor that enables the Einstein field equations to hold is unique. Only the energy-momentum tensor defined by Eq.\textbf{(}\ref{eq:16}\textbf{)} can make the Einstein field equations hold. Therefore, we believe that a unified definition of the energy-momentum tensor of a pure field based on Eq.\textbf{(}\ref{eq:16}\textbf{)} is a wise choice, as other forms of definition cannot make the Einstein field equations hold. We do not reject writing it in different forms, because according to Eq.\textbf{(}\ref{eq:7}\textbf{)}, we add a tensor $\mathcal{C}^\mu _{\text{ }\nu}$ that satisfies $\displaystyle\int  \nabla _{\mu }\left(\mathcal{C}^\mu _{\text{ }\nu} \xi^\nu \right)\epsilon\sqrt{-g} d^4 x=0$ on the basis of $\mathcal{H}^\mu _{\text{ }\nu}$, and all the results can be considered as the energy-momentum tensor of the system. But what can make the Einstein field equations hold true is only the energy-momentum tensor defined by Eq.\textbf{(}\ref{eq:16}\textbf{)}. In this paper, we mainly investigate the characteristics of the energy-momentum tensor of electromagnetic waves in a medium when the energy-momentum tensor is uniformly defined by Eq.\textbf{(}\ref{eq:16}\textbf{)}, and whether the obtained results can explain the relevant experiments.

\section{The energy-momentum tensor of a macroscopic electromagnetic wave propagating in an isotropic linear medium}\label{sec:s3}
In order to give the energy-momentum tensor of an electromagnetic wave propagating in an isotropic linear medium by Eq.\textbf{(}\ref{eq:16}\textbf{)}, we must first figure out the Lagrange density and its physical meaning when the electromagnetic wave propagates in an isotropic linear medium. Now, let us review the behavior of electromagnetic waves as they propagate through a medium.When an electromagnetic wave propagates through a medium, the medium in which it travels will be electric polarization, creating an electric dipole moment. The degree of electric polarization of the medium can be represented by the intensity of polarization $(P^i)$, which is defined as\textcolor[rgb]{0.184313725,0.188235294117647,0.564705882}{\cite{Shuohong}}:
\begin{equation}
    \label{eq:17}
 P^i=\lim_{\Delta  V\to 0} \frac{\sum p^i}{\text{$\Delta $V}}
\end{equation}
Here $\sum p^i$ is the vector sum of all electric dipole moments $p^i=qr^i$ in volume $\Delta  V$, where $r^i$ is the position vector from negative charge $-q$ to positive charge $q$. Similarly, when an electromagnetic wave passes through a medium, the medium it travels through is magnetized, and the degree of magnetization of the medium can be expressed in terms of the magnetization vector $(M^i)$, which is defined as:
\begin{equation}
    \label{eq:18}
 M^i=\lim_{\Delta  V\to 0} \frac{\sum m^i}{\Delta  V}
\end{equation}
Here $\sum m^i$ is the vector sum of all magnetic moments $m^i=Is^i$ in volume $\Delta  V$, where $I$ is the current strength in the closed loop, and $s^i$ is the area vector of the closed loop.
For the electromagnetic field, we may express it in terms of the electromagnetic field tensor. The electromagnetic field tensor is a second-order antisymmetric tensor consisting of the electric field intensity vector $(E^i)$ and the magnetic induction intensity vector $(B^i)$. When the sign difference of the metric is $(-,+,+,+)$, the electromagnetic field tensor can be expressed as:
\begin{equation}
    \label{eq:19}
 \begin{array}{ll}\left(F^{\mu  \nu }\right)=\left(\nabla ^{\mu }A^{\nu }-\nabla ^{\nu }A^{\mu }\right)\\\text{ }\text{ }\text{ }\text{ }\text{ }\text{ }\text{ }\text{ }=\left( \begin{array}{cccc}  0 & \frac{E^1}{c} & \frac{E^2}{c} & \frac{E^3}{c} \\  -\frac{E^1}{c} & 0 & B^3 & -B^2 \\  -\frac{E^2}{c} & -B^3 & 0 & B^1 \\  -\frac{E^3}{c} & B^2 & -B^1 & 0 \end{array} \right)\end{array}
\end{equation}
Here $(A^{\mu })$ is electromagnetic four-potential. Just as the electric field strength and magnetic induction strength can be unified to form the electromagnetic field tensor, the polarization and magnetization of a medium can be unified to form a second-order antisymmetric tensor in four dimensions, called the moment tensor $\left(M^{\mu  \nu }\right)$. For the moment tensor, we have:
\begin{equation}
    \label{eq:20}
\left(M^{\mu  \nu }\right)=\left( \begin{array}{cccc}  0 & -c P^1 & -c P^2 & -c P^3 \\  c P^1 & 0 & M^3 & -M^2 \\  c P^2 & -M^3 & 0 & M^1 \\  c P^3 & M^2 & -M^1 & 0 \end{array} \right)
\end{equation}
When the medium is polarized and magnetized, a four-dimensional Induced current density $\left(\, _Mj^{\mu }\right)$ is generated, which is related to the moment tensor as follows:
\begin{equation}
    \label{eq:21}
\, _Mj^{\mu }=\nabla _{\nu }M^{\mu  \nu }
\end{equation}
Thus,  the total four-current density $j^{\mu }$ in the spacetime is the vector sum of the usual conducting four-current density $\, _{\rho }j^{\mu }$ and the induced four-current density $\, _Mj^{\mu }$, so we have:
\begin{equation}
    \label{eq:22}
j^{\mu }=\, _{\rho }j^{\mu }+\, _Mj^{\mu }
\end{equation}
When an electromagnetic field interacts with a four-dimensional current density, its Lagrangian density is:
\begin{equation}
    \label{eq:23}
\mathcal{L}=\, _F\mathcal{L}+\, _I\mathcal{L}+\, _m\mathcal{L}
\end{equation}
where:
\begin{equation}
    \label{eq:24}
\, \, _F\mathcal{L}=-\frac{1}{4\mu _0}F^{\mu  \nu }F^{\alpha  \beta }g_{\mu  \alpha }g_{\nu  \beta }
\end{equation}
is the Lagrangian density of the free electromagnetic field and $\mu_0$ is the free space susceptibility.
\begin{equation}
    \label{eq:25}
\, _I\mathcal{L}=A^{\mu }j^{\nu }g_{\mu  \nu }
\end{equation}
is the interaction term between the electromagnetic field and the four-dimensional current density. $\, _m\mathcal{L}$ is the Lagrangian density of the medium itself, its specific form will not be investigated in this paper for the time being. It can be seen from Eq.\textbf{(}\ref{eq:22}\textbf{)}, Eq.\textbf{(}\ref{eq:23}\textbf{)} and Eq.\textbf{(}\ref{eq:25}\textbf{)} that the induced four-current density $\, _Mj^{\mu }$, as a component of the total four-current, appears in the Lagrangian density as an interaction term. It is important to be aware of this.\par

Now, we wish to obtain specific expressions for the  moment tensor $\left(M^{\mu  \nu }\right)$ and the  induced four-current density $\, _Mj^{\mu }$. For an observer at rest with respect to the medium, the intensity of polarization $(P^i)$ of the medium depends on the electric field strength and is a function of the electric field strength. For media with linear polarization , $(P^i)$ can be written as:
\begin{equation}
    \label{eq:26}
P^i=\chi_e\varepsilon _0E^i=\frac{\chi_e }{c\mu_0} F^{0i}
\end{equation}
Where, $\chi_e$ is the polarizability of the medium, which is related to its properties;  $\varepsilon _0$ is the permittivity in free space. \par
Similarly, for an observer at rest with respect to the medium, the magnetization $M^i$ of the medium is a function of the magnetic field induced strength $B^i$. For weak magnetic matter, there can also be a linear relationship between $M^i$ and $B^i$, $(M^i)$ can be written as:
\begin{equation}
    \label{eq:27}
\left(M^i\right)=\frac{\chi _m}{\left(1+\chi _m\right)\mu _0}\left(F^{23},F^{31},F^{12}\right)
\end{equation}
Substituting Eq.\textbf{(}\ref{eq:26}\textbf{)} and Eq.\textbf{(}\ref{eq:27}\textbf{)} into Eq.\textbf{(}\ref{eq:20}\textbf{)}, we obtain that for a medium that exhibits linear polarization and magnetization, its moment tensor can be expressed as:
\begin{equation}
    \label{eq:28}
\left(M^{\mu  \nu }\right)=\left( \begin{array}{cccc}  0 & -\frac{\chi _e}{\mu _0}F^{01} & -\frac{\chi _e}{\mu _0}F^{02} & -\frac{\chi _e}{\mu _0}F^{03} \\  -\frac{\chi _e}{\mu _0}F^{10} & 0 & \frac{\chi _m}{1+\chi _m}\frac{F^{12}}{\mu _0} & \frac{\chi _m}{1+\chi _m}\frac{F^{13}}{\mu _0} \\  -\frac{\chi _e}{\mu _0} F^{20} & \frac{\chi _m}{1+\chi _m}\frac{F^{21}}{\mu _0} & 0 & \frac{\chi _m}{1+\chi _m}\frac{F^{23}}{\mu _0} \\  -\frac{\chi _e}{\mu _0}F^{30} & \frac{\chi _m}{1+\chi _m}\frac{F^{31}}{\mu _0} & \frac{\chi _m}{1+\chi _m}\frac{F^{32}}{\mu _0} & 0 \end{array} \right)
\end{equation}
When we adopt the Lorentz gauge $\nabla _{\nu }A^{\nu }=0$, we can obtain from Eq.\textbf{(}\ref{eq:28}\textbf{)}, Eq.\textbf{(}\ref{eq:21}\textbf{)} and Eq.\textbf{(}\ref{eq:19}\textbf{)}:
\begin{equation}
    \label{eq:29}
\, _Mj^0=\frac{\chi _e}{\mu _0}\nabla _{\nu }\nabla ^{\nu }A^0
\end{equation}
Similarly, we can obtain:
\begin{equation}
    \label{eq:30}
    \begin{array}{ll}
\, _Mj^i=-\frac{\chi _m}{\left(1+\chi _m\right)\mu _0}\nabla _{\nu }\nabla ^{\nu }A^i\\\text{ }\text{ }\text{ }\text{ }\text{ }\text{ }\text{ }\text{ }\text{ }-\left(\frac{\chi _e}{\mu _0}+\frac{\chi _m}{\left(1+\chi _m\right)\mu _0}\right)\nabla _0\left(\nabla ^iA^0-\nabla ^0A^i\right)
    \end{array}
\end{equation}
According to Eq.\textbf{(}\ref{eq:23}\textbf{)}, based on the least action principle, we can obtain:
\begin{equation}
    \label{eq:31}
   \nabla _{\nu }F^{\mu  \nu }=\mu _0 j^{\mu }
\end{equation}
Since we have adopted the Lorentz gauge $\nabla _{\nu }A^{\nu }=0$ and assume the conducting four-current
density $\, _{\rho }j^{\mu }=0$ in the medium, we can obtain:
\begin{equation}
    \label{eq:32}
   \nabla _{\nu }\nabla ^{\nu }A^{\mu }=-\mu _0 (\, _Mj^{\mu })
\end{equation}
Since electromagnetic waves are shear waves, from the properties of shear waves and $\nabla _{\nu }A^{\nu }=0$, we can obtain $\nabla _{0 }A^{0 }=0$ when the Lorentz gauge is used. We assume that there is no electrostatic field in the medium so that $A^0=0$ can be obtained. Therefore, Eq.\textbf{(}\ref{eq:30}\textbf{)} can be further simplified as:
\begin{equation}
    \label{eq:33}
   \, _Mj^i=\frac{n^2-1}{\mu_0}\nabla _0\nabla ^0A^i
\end{equation}
where:
\begin{equation}
    \label{eq:34}
   n=\sqrt{\left(1+\chi _e\right) \left(1+\chi _m\right)}
\end{equation}
is the refractive index of the medium. Substituting Eq.\textbf{(}\ref{eq:33}\textbf{)} into Eq.\textbf{(}\ref{eq:32}\textbf{)}, we can get:
\begin{equation}
    \label{eq:35}
   n^2\nabla _0\nabla ^0A^i+\nabla _j\nabla ^jA^i=0
\end{equation}
namely:
\begin{equation}
    \label{eq:36}
   -n^2 \frac{\partial ^2A^i}{(\partial x^0)^2}+\frac{\partial ^2A^i}{\left(\partial x^1\right)^2}+\frac{\partial ^2A^i}{\left(\partial x^2\right)^2}+\frac{\partial ^2A^i}{\left(\partial x^3\right)^2}=0
\end{equation}
This is the well-known wave equation. Its general solution can be written as an arbitrary function $F(X)$ and its linear superposition, with $(X)=\left( -\frac{\omega  x^0}{c}\pm\frac{ \omega  n\text{sin$\theta $} \text{cos$\varphi $} \text { } x^1 +\omega  n \text{sin$\theta $} \text{sin$\varphi $}  \text { }x^2+\omega  n \text{cos$\theta $} \text { } x^3}{c} \right)$ as its independent variable. For simplicity, let the electromagnetic wave in the medium be a monochromatic wave propagating along the $x^1$ axis, and if only the $A^2$ component is nonzero, then we have:
\begin{equation}
    \label{eq:37}
   \left(A^{\mu }\right)=\left(0,0,F\left(-\frac{\omega  x^0}{c}+\frac{\omega n  x^1}{c}\right),0\right)
\end{equation}
Substituting Eq.\textbf{(}\ref{eq:37}\textbf{)} into $\left(F^{\mu  \nu }\right)=\left(\nabla ^{\mu }A^{\nu }-\nabla ^{\nu }A^{\mu }\right)$, we can get that the electromagnetic field tensor in the medium is:
\begin{equation}
    \label{eq:38}
   \left(F^{\mu  \nu }\right)=\left( \begin{array}{cccc}  0 & 0 & 1 & 0 \\  0 & 0 & n & 0 \\  -1 & -n & 0 & 0 \\  0 & 0 & 0 & 0 \end{array} \right)\frac{E}{c}
\end{equation}
where:
\begin{equation}
    \label{eq:39}
   E= \omega F' \left(-\frac{\omega  x^0}{c}+\frac{\omega  x^1}{c/n}\right)= E_0 f\left(-\frac{\omega  x^0}{c}+\frac{\omega  x^1}{c/n}\right)
\end{equation}
is the magnitude of the electric field strength in the medium, $ f =\omega F' /E_0$, $E_0$ is the module of the electric field intensity vector. Substituting Eq.\textbf{(}\ref{eq:23}\textbf{)} into Eq.\textbf{(}\ref{eq:16}\textbf{)}, we get:
\begin{equation}
    \label{eq:40}
   T^{\mu\nu}=\, _FT^{\mu  \nu }+\, _IT^{\mu  \nu }+\, _mT^{\mu  \nu }
\end{equation}
Where $\, _FT^{\mu  \nu }=\frac{-2}{\sqrt{-g}}\frac{\delta ( \, _F\mathcal{L}\sqrt{-g})}{\delta  g_{\mu  \nu }}$ is the energy-momentum tensor corresponding to $\, _F\mathcal{L}$, which belongs to  the electromagnetic field; $\, _IT^{\mu  \nu }=\frac{-2}{\sqrt{-g}}\frac{\delta ( \, _I\mathcal{L}\sqrt{-g})}{\delta  g_{\mu  \nu }}$ is the energy-momentum tensor corresponding to the interaction term $\, _I\mathcal{L}$, which belongs to the whole system rather than only one of the fields involved in the interaction; $\, _mT^{\mu  \nu }=\frac{-2}{\sqrt{-g}}\frac{\delta ( \, _m\mathcal{L}\sqrt{-g})}{\delta  g_{\mu  \nu }}$ is the energy-momentum tensor corresponding to $\, _m\mathcal{L}$, which belongs to the medium.\par

We will focus on the energy-momentum tensor $\, _FT^{\mu  \nu }$ corresponding to the electromagnetic field in the medium. The electromagnetic field $\left(F^{\mu  \nu }\right)$ in the medium here refers to the total macroscopic electromagnetic field, including the electromagnetic field transmitted from the outside, as well as the electromagnetic field generated by the induced current density, but does not include the contribution of polarization energy and magnetization energy belonging to the medium but not belonging to the macroscopic electromagnetic field. This is the main difference between the energy-momentum tensor $\, _FT^{\mu  \nu }$ presented in this paper and the Minkowski and Abraham tensors. The Minkowski and Abraham tensors include contributions from the polarization and magnetization energies, but $\, _FT^{\mu  \nu }$ is only for the macroscopic electromagnetic field. The macroscopic electromagnetic field in the medium is generated by the four-dimensional electromagnetic potential accord with Eq.\textbf{(}\ref{eq:36}\textbf{)}, so it is a unified and indivisible whole in the macro sense.  However, the polarization energy and magnetization energy of a medium are different. A homogeneous medium is assumed to be polarized and magnetized by an applied static electromagnetic field, and there is no free charge inside the medium.Since the interior of the medium is homogeneous, the polarization $(P^i)$ and magnetization $(M^i)$ are homogeneous. Since the applied electromagnetic field is an static electromagnetic field, they do not change in time. In this case, although the interior of the medium is polarized and magnetized, each component of the corresponding moment tensor $\left(M^{\mu  \nu }\right)$ is a constant. In this case, the total induced current density $\nabla _{\nu }M^{\mu  \nu }=0$, that is, there is neither bound charge nor magnetization current at the location of the homogeneous medium, and the bound charge and magnetization current of the entire medium are only at the surface of the medium. That is, the interior of the homogeneous medium actually does not contribute to the macroscopic electromagnetic field, and only the bound charges and magnetization currents at the surface of the medium contribute.That is, in this case, the macroscopic electromagnetic field in the interior of the homogeneous medium arises only from contributions from the exterior and boundary of the medium, and the interior of the homogeneous medium does not generate the macroscopic electromagnetic field itself. However, due to the polarization and magnetization in the medium, electric and magnetic dipole moments are generated and thus cannot be homogeneous at the microscopic level. Indeed, even if the medium is not polarized or magnetized, the interior of the medium cannot be perfectly homogeneous at the microscopic level. The absence of macroscopic electromagnetic fields in the interior of the medium is due to the fact that these irregularities are smoothed out in the macroscopic view. When the medium is polarized, the microscopic inhomogeneities in the medium are intensified and electromagnetic fields are generated around each electric or magnetic dipole moment. These electromagnetic fields are not contained in macroscopic electromagnetic fields, but exist and have energy at the microscopic level. These energies are the polarization and magnetization energies.It can be seen that the Lagrange density $\, _F\mathcal{L}$ here describes only the macro electromagnetic field, and the energy-momentum tensor $\, _FT^{\mu  \nu }$ is only the energy-momentum tensor of the macro electromagnetic field, which does not include the contribution of polarization energy and magnetization energy caused by the micro electromagnetic field in the medium. The energy-momentum tensor of the microscopic electromagnetic field in the medium, which is not included in the macroscopic electromagnetic field and therefore does not belong to $\, _F\mathcal{L}$ and hence to $\, _m\mathcal{L}$, is part of the energy-momentum tensor of the medium.Therefore, when we analyze the momentum of the electromagnetic field in the medium, we should only analyze the macroscopic electromagnetic field and should not mix the microscopic electromagnetic field belonging to the medium. However, since the polarization and magnetization energies in the medium are free energies, they also change when the macroscopic electromagnetic field is changed, which also gives rise to mechanical effects. In addition, some experiments will be accompanied by electrostrictive and other mechanical effects that also belong to the behavior of the medium. Therefore, when analyzing experimental phenomena, it is sometimes necessary to consider the full range of possible media factors in addition to the analysis of the effects induced by macroscopic electromagnetic fields.In this paper, we mainly analyze some basic properties of the energy-momentum tensor $\, _FT^{\mu  \nu }$, which does not include medium contributions such as polarization energy and magnetization energy. For the energy-momentum tensors including the contribution of polarization energy and magnetization energy, such as Minkowski tensor, Abraham tensor and Einstein-Laub tensor, many literature have made a detailed discussion\textcolor[rgb]{0.184313725,0.188235294117647,0.564705882}{\cite{Brevik1979,Pfeifer}}, which will not be discussed in depth in this paper. For the energy-momentum tensor of the macroscopic electromagnetic field determined by $\, _F\mathcal{L}$, by substituting  Eq.\textbf{(}\ref{eq:24}\textbf{)}  into  Eq.\textbf{(}\ref{eq:16}\textbf{)} , we get:
\begin{equation}
    \label{eq:41}
   \, _FT^{\mu  \nu }=\frac{1}{\mu _0}\left(F^{\mu  \beta }F_{\text{   }\beta }^{\nu }-\frac{1}{4}F_{\alpha  \beta }F^{\alpha  \beta }g^{\mu  \nu }\right)
\end{equation}
We will only discuss the problem in Minkowski space, where the metric field is $\left(g^{\mu  \nu }\right)=\left(
\begin{array}{cccc}
 -1 & 0 & 0 & 0 \\
 0 & 1 & 0 & 0 \\
 0 & 0 & 1 & 0 \\
 0 & 0 & 0 & 1
\end{array}
\right)$.\par
Substituting Eq.\textbf{(}\ref{eq:38}\textbf{)} into Eq.\textbf{(}\ref{eq:41}\textbf{)}, we get:
\begin{equation}
    \label{eq:42}
  (\, _FT^{\mu  \nu })=\left( \begin{array}{cccc}  \frac{n^2+1 }{2} & n & 0 & 0 \\  n & \frac{n^2+1 }{2} & 0 & 0 \\  0 & 0 & \frac{n^2-1 }{2} & 0 \\  0 & 0 & 0 & -\frac{n^2-1 }{2} \end{array} \right)\frac{E^2}{c^2\mu _0}
\end{equation}
Where $E$ is the electric field intensity in the medium, which is given by Eq.\textbf{(}\ref{eq:39}\textbf{)}.

\section{The relation between the energy-momentum tensor of an electromagnetic wave in a medium and its energy-momentum tensor in free space}\label{sec:s4}
An electromagnetic wave propagating in a medium is caused by an incoming electromagnetic wave from outside. The definition of the energy-momentum tensor for electromagnetic waves in free space is not controversial. Therefore, relating the energy-momentum tensor of an electromagnetic wave in a medium to its energy-momentum tensor in free space will help us to understand the energy-momentum tensor of an electromagnetic wave in a medium. The electromagnetic wave is assumed to be incident from medium 1 with refractive index $n_1$ to medium 2 with refractive index $n_2$ with incident angle $\theta$. Then, according to the boundary condition, the reflection angle is $\theta'=\theta$ and the refraction angle $\theta''$ satisfies $\frac{sin\theta}{sin\theta''}=n_2/n_1$. Let $\,_r\boldsymbol{e}$ be the normal vector to the interface. For simplicity, we only study non-ferromagnetic linear media, so the boundary conditions are\textcolor[rgb]{0.184313725,0.188235294117647,0.564705882}{\cite{Shuohong}}:
\begin{equation}
    \label{eq:43}
 \,_r\boldsymbol{e}\times \left(\, _{\text{tran}}\boldsymbol{\, E}-\, _{\text{inc}}\boldsymbol{\, E}\right)=0
\end{equation}
\begin{equation}
    \label{eq:44}
 \,_r\boldsymbol{e}\times \left(\, _{\text{tran}}\boldsymbol{\, H}-\, _{\text{inc}}\boldsymbol{\, H}\right)=0
\end{equation}
where $\, _{\text{tran}}\boldsymbol{\, E}$ and $\, _{\text{tran}}\boldsymbol{\, H}$ are the electric field intensity vectors and magnetic field intensity vectors of the transmitted wave, $\, _{\text{inc}}\boldsymbol{\, E}$ is the electric field intensity vector after the superposition of the incident wave electric field intensity vector $\, _{\text{in}}\boldsymbol{\, E}$ and the reflected wave electric field intensity vector $\, _{\text{re}}\boldsymbol{\, E}$, $\, _{\text{inc}}\boldsymbol{\, H}$ is the magnetic field intensity vector after the superposition of the incident wave magnetic field intensity vector $\, _{\text{in}}\boldsymbol{\, H}$ and the reflected wave magnetic field intensity vector $\, _{\text{re}}\boldsymbol{\, H}$. Thus, when the electric field intensity vector is perpendicular to the normal vector $\,_r\boldsymbol{e}$, the magnitudes of the above vectors satisfy the following relation:
\begin{equation}
    \label{eq:45}
 \, _{\text{tran}} E_0=\, _{\text{inc}}E_0=\, _{\text{in}}E_0+\, _{\text{re}}E_0
\end{equation}
\begin{equation}
    \label{eq:46}
    \begin{array}{ll}
 \, _{\text{tran}}H_0 \cos  \theta''=\, _{\text{inc}}H_0 \cos  \theta =\, \left.(_{\text{in}}H_0-\, _{\text{re}}H_0\right)\cos  \theta
 \\
 \\
 \end{array}
\end{equation}
Here we use the subscript 0 to denote the module of the vector. By Eq.\textbf{(}\ref{eq:38}\textbf{)} we have $\text{}_{\text{in}}H_0=\frac{n_1(\text{}_{\text{in}}E_0)}{c \mu _1}$, $\text{}_{\text{re}}H_0=\frac{n_1(\text{}_{\text{re}}E_0)}{c \mu _1}$ and $\, _{\text{tran}}H_0=\frac{n_2 (\, _{\text{tran}} E_0)}{c \mu_2 }$, so Eq.\textbf{(}\ref{eq:46}\textbf{)} can be reduced to:
\begin{equation}
    \label{eq:47}
\frac{n_2 (\, _{\text{tran}} E_0)}{\mu_2 } \cos  \theta ''=\frac{n_1 }{\mu _1}\, \left(\text{}_{\text{in}}E_0-\text{}_{\text{re}}E_0\right)\cos  \theta
\end{equation}
By combining Eq.\textbf{(}\ref{eq:45}\textbf{)} and Eq.\textbf{(}\ref{eq:47}\textbf{)}, we can get:
\begin{equation}
    \label{eq:48}
\, _{\text{tran}} E_0=\frac{2 (\, _{\text{in}}E_0) n_1 \mu _2 \cos  \theta }{n_2 \mu _1\cos  \theta'' +n_1 \mu _2\cos  \theta }
\end{equation}
\begin{equation}
    \label{eq:49}
\, _{\text{re}}E_0=\, _{\text{in}}E_0-\frac{2\text{   }n_2 \mu _1 \cos  \theta''}{n_2 \mu _1\cos  \theta''+n_1 \mu _2\cos  \theta }\left( \, _{\text{in}}E_0 \right)
\end{equation}
When an electromagnetic beam is incident from the free space perpendicular to the interface into a medium with refractive index $n$, there are $\theta=\theta''=0$, $\mu_1=\mu_0$, $\mu_2=\mu$ such that the above relation can be further simplified as:
\begin{equation}
    \label{eq:50}
\, _{\text{tran}} E_0=\frac{2  \mu  }{n  \mu _0 +  \mu }(\, _{\text{in}}E_0)
\end{equation}
\begin{equation}
    \label{eq:51}
\, _{\text{re}}E_0=\, _{\text{in}}E_0-\frac{2 n \mu _0 }{n  \mu _0 +  \mu  }\left( \, _{\text{in}}E _0\right)
\end{equation}
According to the wave vector relationship between the incident field and the reflected field, referring to Eq.\textbf{(}\ref{eq:38}\textbf{)}, and noting that the propagation direction of the incident wave and the reflected wave is opposite, we can get the electromagnetic field tensor of the incident field and the reflected field as follows:
\begin{equation}
    \label{eq:52}
\left(\, _{\text{in}}F^{\mu  \nu }\right)=\left( \begin{array}{cccc}  0 & 0 &  1 & 0 \\  0 & 0 &  1 & 0 \\  -1 & -1 & 0 & 0 \\  0 & 0 & 0 & 0 \end{array} \right)\frac{\, _{\text{in}}E_0}{c} f\left(-\frac{\omega  x^0}{c}+\frac{\omega  x^1}{c}\right)
\end{equation}
\begin{equation}
    \label{eq:53}
\left(\, _{\text{re}}F^{\mu  \nu }\right)=\left( \begin{array}{cccc}  0 & 0 &  1 & 0 \\  0 & 0 & -1 & 0 \\  -1 & 1 & 0 & 0 \\  0 & 0 & 0 & 0 \end{array} \right)\frac{\, _{\text{re}}E_0}{c} f\left(-\frac{\omega  x^0}{c}-\frac{\omega  x^1}{c}\right)
\end{equation}
In free space outside the interface, the electromagnetic field tensor is a superposition of the incident and reflected fields, namely $F^{\mu  \nu }=\, _{\text{in}}F^{\mu  \nu }+\, _{\text{re}}F^{\mu  \nu }$. By substituting $F^{\mu  \nu }$ into Eq.\textbf{(}\ref{eq:41}\textbf{)}, its corresponding energy-momentum tensor can be obtained and written as follows:
\begin{equation}
    \label{eq:54}
T^{\mu  \nu }=\left( \begin{array}{cccc}  \gamma ^2\left(\frac{p}{c^2}+\rho \right)c^2-p & \gamma ^2\left(\frac{p}{c^2}+\rho \right)c v & 0 & 0 \\  \gamma ^2\left(\frac{p}{c^2}+\rho \right)c v & \gamma ^2\left(\frac{p}{c^2}+\rho \right)v^2+p & 0 & 0 \\  0 & 0 & \pm p & 0 \\  0 & 0 & 0 & \mp p \end{array} \right)
\end{equation}
where:

\begin{equation}
    \label{eq:55}
\rho =\left|\frac{2(\, _{\text{in}}E_0)^2 \left(n \mu _0 -\mu\right)f\left(-\frac{\omega  x^0}{c}+\frac{\omega  x^1}{c}\right) f\left(-\frac{\omega  x^0}{c}-\frac{\omega  x^1}{c}\right)     }{c^4 \mu _0 \left(n \mu _0 +\mu\right)}\right|
\end{equation}

\begin{equation}
    \label{eq:56}
p=\rho c^2
\end{equation}

\begin{equation}
    \label{eq:57}
v=\left\{ \begin{array}{cccc}  c/k\text{     }\text{     }\text{     }(k\geq 1) \\  k c\text{     }\text{     }\text{     }\text{     }\text{     }(k<1) \end{array} \right.
\end{equation}

\begin{equation}
    \label{eq:58}
k=\frac{ \left(n \mu _0+\mu \right) f\left(-\frac{\omega  x^0}{c}+\frac{\omega  x^1}{c}\right) -\left(n \mu _0-\mu \right) f\left(-\frac{\omega  x^0}{c}-\frac{\omega  x^1}{c}\right) }{\left(n \mu _0+\mu \right)f\left(-\frac{\omega  x^0}{c}+\frac{\omega  x^1}{c}\right) +\left(n \mu _0-\mu \right) f\left(-\frac{\omega  x^0}{c}-\frac{\omega  x^1}{c}\right)}
\end{equation}

\begin{equation}
    \label{eq:59}
\gamma =\frac{1}{\sqrt{1-(v/c)^2}}
\end{equation}
It can be seen that the energy-momentum tensor of the electromagnetic wave is very similar to that of the fluid with density $\rho$, pressure $p$ and velocity $v$, except that $T^{22}$ and $T^{33}$ can be positive or negative, and their positive or negative values depend on the polarization of the electromagnetic wave. It is not surprising that a fluid can be seen as a collection of a large number of particles, such as molecules, and that electromagnetic waves can be seen as a collection of a large number of photons, but ordinary fluids do not have polarization, while electromagnetic waves do. In Eq.\textbf{(}\ref{eq:57}\textbf{)}, $v$ is the transmission rate of electromagnetic wave energy and information. In fact, there are two solutions for $\rho$ and $v$, respectively. The solution $\rho\geq0$ corresponds to $v\leq c$, while the solution $\rho<0$ corresponds to $v>c$. Since $\rho\geq0$ is always required, we can only observe $v\leq c$ and not the superluminal velocity. We can also see that the coordinate transformation relationship of the energy transmission speed satisfies the Lorentz transformation, that is, the von-Laue criterion\textcolor[rgb]{0.184313725,0.188235294117647,0.564705882}{\cite{vonLaue}}. We can also verify that $T^{\mu  \nu }$ makes $\partial _{\mu }T^{\mu  \nu }=0$ true, that is, it is conserved. At the interface between medium and free space, we can take $x^1=0$. In this case, we have:
\begin{equation}
    \label{eq:60}
\rho =\left|\frac{2 \left(n \mu _0-\mu \right)}{c^4 \mu _0 \left(n \mu _0+\mu \right)}(\, _{\text{in}}E)^2 \right|
\end{equation}
\begin{equation}
    \label{eq:61}
v=\left\{ \begin{array}{ll}  \frac{c \mu }{n \mu _0}\text{     }\text{     }\text{     }\text{     }\text{     }\text{     }( \frac{ \mu }{n \mu _0}\leq 1) \\  \frac{c n \mu_0}{\mu }\text{     }\text{     }\text{     }\text{     }\text{     }( \frac{ \mu }{n \mu _0}>1) \end{array} \right.
\end{equation}
where $\, _{\text{in}}E =\, _{\text{in}}E_0 f\left(-\frac{\omega  x^0}{c} \right)$.\par
In the medium, referring to Eq.\textbf{(}\ref{eq:38}\textbf{)}, we can get the electromagnetic field tensor of the transmitted field as:
\begin{equation}
    \label{eq:62}
\, _{\text{tran}}F^{\mu  \nu }=\left( \begin{array}{cccc}  0 & 0 & 1 & 0 \\  0 & 0 & n & 0 \\  -1 & -n & 0 & 0 \\  0 & 0 & 0 & 0 \end{array} \right)  \frac{\, _{\text{tran}}E}{c}
\end{equation}
where $\, _{\text{tran}}E =\, _{\text{tran}}E_0f \left(-\frac{\omega  x^0 }{c}+\frac{\omega  x^1}{c/n}\right)$. Substituting Eq.\textbf{(}\ref{eq:62}\textbf{)} into Eq.\textbf{(}\ref{eq:41}\textbf{)}, the energy-momentum tensor of the electromagnetic wave in the medium is:
\begin{equation}
    \label{eq:63}
\, _FT^{\mu  \nu }=\left( \begin{array}{cccc}  \frac{n^2+1}{2c^2 \mu _0} & \frac{n}{c^2\mu _0} & 0 & 0 \\  \frac{n}{c^2\mu _0} & \frac{n^2+1}{2 c^2\mu _0} & 0 & 0 \\  0 & 0 & \frac{n^2-1}{2c^2 \mu _0} & 0 \\  0 & 0 & 0 & -\frac{n^2-1}{2c^2 \mu _0} \end{array} \right)\left( \, _{\text{tran}}E\right) ^2
\end{equation}
If written in fluid form it is:
\begin{equation}
    \label{eq:64}
\, _FT^{\mu  \nu }=\left( \begin{array}{cccc}  \gamma ^2\left(\frac{p}{c^2}+\rho \right)c^2-p & \gamma ^2\left(\frac{p}{c^2}+\rho \right)c v & 0 & 0 \\  \gamma ^2\left(\frac{p}{c^2}+\rho \right)c v & \gamma ^2\left(\frac{p}{c^2}+\rho \right)v^2+p & 0 & 0 \\  0 & 0 & \pm p & 0 \\  0 & 0 & 0 & \mp p \end{array} \right)
\end{equation}
where:
\begin{equation}
    \label{eq:65}
 \rho =\left|\frac{n^2-1}{2 c^4  \mu_0}\left( \, _{\text{tran}}E\right) ^2 \right|
\end{equation}
\begin{equation}
    \label{eq:66}
p=\left|\frac{n^2-1}{2 c^2  \mu_0}\left( \, _{\text{tran}}E\right) ^2 \right|
\end{equation}
\begin{equation}
    \label{eq:67}
v=\left\{ \begin{array}{ll} u\text{     }\text{     }\text{     }\text{     }\text{     }\text{     }\text{     }\text{     }\text{     }\text{}(u\leq c) \\   c^2/u\text{     }\text{     }\text{     }\text{     }\text{     }(u>c) \end{array} \right.
\end{equation}
Where $u=c/n$ is the phase velocity of the electromagnetic wave. It can be seen that for monochromatic waves, when the phase velocity is less than or equal to the speed of light $c$ in vacuum, the rate of energy and information transfer is equal to the phase velocity. Conversely, if the phase velocity is larger than the speed of light in vacuum, $c$, then the speed of energy and information transfer is $c^2/u$. This relation also holds for free matter waves, where the phase velocity of the wave function of a free electron is $u>c$ and therefore the transport velocity of energy or information, that is, the kinematic velocity of the electron, is $c^2/u$. This relation can be verified quantum mechanically. Thus, not only electromagnetic waves can be considered as fluids, but all matter waves can be considered as fluids. This relationship of Eq.\textbf{(}\ref{eq:67}\textbf{)} is caused by the fact that the density $\rho$ must be greater than or equal to 0, which determines that it is impossible to transmit energy or information at the speed of superluminal light. Since the energy-momentum tensor of an electromagnetic wave in a medium is completely analogous to that of an ordinary fluid, except that the diagonal terms can be positive or negative. Thus the energy and momentum of an electromagnetic wave as a fluid composed of a large number of photons is not fundamentally different from that of a general fluid. Thus, as in ordinary fluids, the physical meaning of $cT^{0i}$ is the energy flow density $S^i$ and the physical meaning of $\frac{1}{c}T^{i0}$ is the momentum density $g^i$, so that Planck's principle $g^i = S^i/c^2$ holds. The Planck's principle is not valid in the Minkowski tensor\textcolor[rgb]{0.184313725,0.188235294117647,0.564705882}{\cite{Brevik1979}}.\par
In the free space outside the interface which is infinitely close to the interface, $x^1\rightarrow0$. In this case, $T^{\mu  \nu }$ can be written as:
\begin{equation}
    \label{eq:68}
 T^{\mu  \nu }=\left( \begin{array}{cccc} T^{00} & T^{01}& 0 & 0 \\ T^{10} & T^{11} & 0 & 0 \\  0 & 0 & T^{22} & 0 \\  0 & 0 & 0 & T^{33} \end{array} \right)
\end{equation}
where:\par
$T^{00}=T^{11}=\frac{2\left(n^2\mu _0^2+\mu ^2\right) }{c^2 \mu _0\left(n \mu _0+\mu \right){}^2}\left( \, _{\text{in}}E\right) ^2$\par
$T^{01}=T^{10}= \frac{4n \mu}{c^2\left(n \mu _0+\mu \right){}^2}\left( \, _{\text{in}}E\right) ^2$\par
$T^{22}=-T^{33}=  \frac{2 \left(n \mu _0-\mu \right) }{c^2 \mu _0\left(n \mu _0+\mu \right)}\left( \, _{\text{in}}E\right) ^2$\par
$\text{ }$\par\par
Similarly, in a medium infinitely close to the interface within the interface, $x^1\rightarrow0$. substituting Eq.\textbf{(}\ref{eq:50}\textbf{)} into Eq.\textbf{(}\ref{eq:64}\textbf{)}, $\, _FT^{\mu  \nu }$ can be written as:
In the free space outside the interface which is infinitely close to the interface, $x^1\rightarrow0$. In this case, $T^{\mu  \nu }$ can be written as:
\begin{equation}
    \label{eq:69}
 \, _FT^{\mu  \nu }=\left( \begin{array}{cccc} \, _FT^{00} & \, _FT^{01}& 0 & 0 \\ \, _FT^{10} & \, _FT^{11} & 0 & 0 \\  0 & 0 & \, _FT^{22} & 0 \\  0 & 0 & 0 & \, _FT^{33} \end{array} \right)
\end{equation}
where:\par
$\, _FT^{00}=\, _FT^{11}=\frac{2 \left(n^2+1\right) \mu ^2\text{  }}{c^2 \mu _0\left(n \mu _0+\mu \right){}^2} \left( \, _{\text{in}}E\right)^2$\par
$\, _FT^{01}=\, _FT^{10}= \frac{4 n \mu ^2\text{  }}{c^2 \mu _0\left(n \mu _0+\mu \right){}^2}\left( \, _{\text{in}}E\right) ^2$\par
$\, _FT^{22}=-\, _FT^{33}= \frac{2 \left(n^2-1\right) \mu ^2 }{c^2 \mu _0\left(n \mu _0+\mu \right)^2}\left( \, _{\text{in}}E\right) ^2$\par
$\text{ }$\par\par
We find that when considering $\mu\approx\mu_0$ for non-ferromagnetic linear media, we have:
\begin{equation}
    \label{eq:70}
 \, _FT^{00} / T^{00} =\frac{\left(1+n^2\right) \mu ^2}{\mu ^2+n^2 \mu _0^2}\approx1
\end{equation}
\begin{equation}
    \label{eq:71}
 \, _FT^{01} / T^{01} =\frac{\mu }{\mu _0}\approx1
\end{equation}
\begin{equation}
    \label{eq:72}
  \, _FT^{11} / T^{11} =\frac{\left(1+n^2\right) \mu ^2}{\mu ^2+n^2 \mu _0^2}\approx1
\end{equation}
\begin{equation}
    \label{eq:73}
  \, _FT^{22} / T^{22} = \, _FT^{33} / T^{33} =\frac{\left(n^2-1\right) \mu ^2}{n^2 \mu _0^2-\mu ^2}\approx1
\end{equation}
It can be seen that $\, _FT^{\mu  \nu }\approx T^{\mu  \nu }$ for non-ferromagnetic linear media. Because the definition of the energy-momentum tensor of electromagnetic wave in free space outside the medium is not controversial, and experiments show that the electromagnetic wave is almost lossless when propagating in a non-ferromagnetic linear medium, $\, _FT^{\mu  \nu }\approx T^{\mu  \nu }$ also confirms that it is reasonable for us to define $\, _FT^{\mu  \nu }$ as the energy-momentum tensor of electromagnetic wave in the medium. Due to conservation, we can regard $T^{\mu  \nu }$ as the total energy-momentum tensor flowing into the medium. Thus $ \, _FT^{01} / T^{01}\approx1$ indicates that the energy flow of the electromagnetic wave propagating in the medium is almost the energy flow input from the outside, and the energy flow carried and transmitted by the medium is almost negligible. However, although there is also $ \, _FT^{11} / T^{11} \approx$, since there is also pressure $p$, we cannot assume that the momentum of the electromagnetic wave propagating in the medium is approximately equal to the momentum input from the outside. As for the momentum of the electromagnetic wave in the medium, which we shall discuss below in Section. For $\, _FT^{\mu  \nu }$, we also find that the four-dimensional Lorentz force density $f^{\mu }=-\partial _{\nu } \left(T^{\mu  \nu }\right)$\textcolor[rgb]{0.184313725,0.188235294117647,0.564705882}{\cite{Brevik1979}} is:
\begin{equation}
    \label{eq:74}
 \begin{array}{ll}f^{\mu}=-\partial _{\nu} \left(\, _FT^{\mu  \nu }\right)=-(1,n,0,0)\frac{ n^2-1 }{c^3 \mu _0}\omega \left(\, _{\text{tran}}E_0\right)^2 \\ \text{  }\text{  }\text{  }\text{  }\text{  }\text{  }\text{  } \times f\left(-\frac{\omega  x^0}{c}+\frac{\omega  x^1}{c}\right) f'\left(-\frac{\omega  x^0}{c}+\frac{\omega  x^1}{c}\right)\ne0\end{array}
\end{equation}
That is, $\, _FT^{\mu  \nu }$ is not conserved. This is easily understood since the electromagnetic wave in the medium is not a closed system at this time, it exchanges energy and momentum with the medium. But we also found that when $ f\left(-\frac{\omega  x^0}{c}+\frac{\omega  x^1}{c}\right)=\sin\left(-\frac{\omega  x^0}{c}+\frac{\omega  x^1}{c}\right)$ is a sine wave, if we take the mean value of Eq.\textbf{(}\ref{eq:74}\textbf{)}  in a cycle, we will find that $\overline{\partial _{\mu } \left(\, _FT^{\mu  \nu }\right)}=0$. This indicates that in this case the exchange of energy and momentum between the electromagnetic wave and the medium is dynamically equilibrated. After the electromagnetic wave transfers energy and momentum to the medium, the medium in turn transfers energy and momentum back to the electromagnetic wave, and the energy and momentum are not lost in the medium. The same conclusion is reached  when the electromagnetic wave in the medium is a cosine wave.
\section{Mean energy and mean mechanical momentum of photons in the medium}\label{sec:s5}
The momentum of light in a medium is a point of contention between Minkowski and Abraham. This momentum of light in the medium is the point of contention between Minkowski and Abraham. This question has been debated for more than a century with no generally accepted result. The discussion in this paper will be useful to understand the momentum of photons in the medium. The discussion of the energy and momentum of a single photon in a medium is complicated by quantum effects, but the problem is much simpler if we discuss the mean energy and momentum of a large number of photons. In this paper we focus on the mean energy and momentum of a large number of photons. We find that if $\overline{\, _FT^{00}}$ is interpreted as the mean energy density of electromagnetic waves in the medium, and assuming that the energy of a single photon in the medium is still $\hbar  \omega$, then the mean photon number density $\overline{\, _{\gamma }n}=\overline{\, _FT^{00}}/(\hbar  \omega )$ in the medium is:
\begin{equation}
    \label{eq:75}
 \overline{\, _{\gamma }n}=\frac{n^2+1}{2c^2 \mu _0}\frac{1}{\hbar  \omega }\overline{\, _{\text{tran}}E^2}
\end{equation}
In this case, $\overline{\, _FT^{01}}/c$ should be interpreted as the mean-momentum density $\overline{g^1}=\frac{1}{c}\frac{n}{ c^2\mu _0}\overline{\, _{\text{tran}}E^2}$ of the electromagnetic wave in the medium, so we obtain that the mean-momentum $\overline{\, _{\gamma }P^1}=\overline{g^1}/\left(\overline{\, _{\gamma }n} \right)$ of the photon in the medium is:
\begin{equation}
    \label{eq:76}
 \overline{\, _{\gamma }P^1}=\frac{2 n\text{  }}{1+n^2}\hbar  \omega/c
\end{equation}
It is neither proportional to the refractive index $n$ nor inversely proportional to the refractive index $n$. However, from Eq.\textbf{(}\ref{eq:64}\textbf{)}, we can see that the energy-momentum tensor of electromagnetic waves in the medium contains pressure $p$, so like ordinary fluids with pressure, $\overline{\, _FT^{00}}$ should not be understood as the mean energy density of electromagnetic waves in the medium. Consider the propagation of a beam of light in a medium, the length of the beam is $d$. This beam flows through a section with cross-sectional area $S$ and into another space with volume $V=Sd$. In this case, the time required for the beam to flow through the cross section is $\Delta t=nd/c$. Since the physical meaning of $cT^{0i}$ is the energy flow density, the energy $E$ flowing into the volume $V$ is:
\begin{equation}
    \label{eq:77}
 E=c\overline{T^{01}}S\Delta t=\frac{n^2}{ c^2\mu _0}\overline{\, _{\text{tran}}E^2}Sd
\end{equation}
Due to the fact that during the process of this beam of light flowing into volume $V$, the medium acted on by pressure $p$ does not move with the beam at a macroscopic level but remains stationary, so pressure $p$ does not do any work on the medium at a macroscopic level. In this process, although the beam exchanges energy and momentum with the medium, the energy and momentum transmitted to the medium is always returned to the light, such that the energy transmitted to the medium by the light is zero in the macroscopic averaging effect. So, the energy flowing from the cross section into the volume $V$ is the energy of this beam in the volume $V$. Therefore, the mean energy density $\overline{\, _{\gamma }w}=E/V$ of this beam is:
\begin{equation}
    \label{eq:78}
 \overline{\, _{\gamma }w} =\frac{n^2}{ c^2\mu _0}\overline{\, _{\text{tran}}E^2}
\end{equation}
If one assumes that the energy of a single photon in the medium is still $\hbar  \omega$, then the mean photon number density $\overline{\, _{\gamma }n}=\overline{\, _{\gamma }w}/(\hbar  \omega )$ in the medium should be:
\begin{equation}
    \label{eq:79}
 \overline{\, _{\gamma }n}= \frac{n^2}{c^2 \mu _0}\frac{1}{\hbar  \omega }\overline{\, _{\text{tran}}E^2}
\end{equation}
Therefore, the mean momentum $\overline{\, _{\gamma }P^1}=\overline{g^1}/\left( \overline{\, _{\gamma }n} \right)$ of a single photon is:
\begin{equation}
    \label{eq:80}
 \overline{\, _{\gamma }P^1}=\frac{ \hbar  \omega }{c n}
\end{equation}
It follows that its value is the $1/n$ of its value in free space, in agreement with the Abraham momentum. However, while the mean energy of a photon corresponding to a free electromagnetic wave in free space is $\hbar  \omega$, there is no evidence that the mean energy of a photon corresponding to an electromagnetic wave interacting with a medium is still $\hbar  \omega$. The energy eigenvalue of the particle is $\hbar  \omega$ and the momentum eigenvalue is $2\pi\hbar  /\lambda$, which is based on quantum mechanics. We know that for the stationary state of the hydrogen atom with the principal quantum number $n$, we can obtain that its energy corresponding to the energy operator is $E_n=-\frac{m \alpha ^2}{2\hbar ^2 n^2}$ \textcolor[rgb]{0.184313725,0.188235294117647,0.564705882}{\cite{Landau}}, where $m$ is the mass of the electron, $\alpha =\frac{e^2}{ 4\pi  \varepsilon _0 }$. But this energy includes not only the kinetic energy of the electron, but also the potential energy due to the interaction. Similarly, in the presence of interactions, the mechanical quantities corresponding to momentum operators in quantum mechanics also include contributions from interactions. This momentum is called the canonical momentum of particles\textcolor[rgb]{0.184313725,0.188235294117647,0.564705882}{\cite{DaiYi}}. Relativistically, the momentum satisfying Planck's principle $g^i = S^i/c^2$ does not include contributions from interactions and is called mechanical momentum. Generally speaking, when a particle interacts with other fields and the interaction contributes to the canonical momentum of the particle, the canonical momentum of the particle and the mechanical momentum are no longer equal, and so is the relationship between the energy excluding the contribution of the interaction and the energy including the contribution of the interaction. Therefore, in the presence of interactions, we cannot use$\hbar  \omega$ directly when calculating the energy of a particle excluding the interaction contribution. Due to the interaction between in-medium and out-of-medium photons, we need to reconsider the energy of in-medium photons excluding the interaction contribution.\par
Substituting Eq.\textbf{(}\ref{eq:52}\textbf{)} into Eq.\textbf{(}\ref{eq:41}\textbf{)} and averaging the results, we can get the mean value of the energy-momentum tensor of the incident wave as:
\begin{equation}
    \label{eq:81}
 \left(\overline{\, _{\text{in}}T^{\mu  \nu }}\right)=\left( \begin{array}{cccc}  \frac{1}{c^2\mu _0} & \frac{1}{c^2\mu _0} & 0 & 0 \\  \frac{1}{c^2\mu _0} & \frac{1}{c^2\mu _0} & 0 & 0 \\  0 & 0 & 0 & 0 \\  0 & 0 & 0 & 0 \end{array} \right)\overline{\, _{\text{in}}E^2}
\end{equation}
Substituting Eq.\textbf{(}\ref{eq:53}\textbf{)} into Eq.\textbf{(}\ref{eq:41}\textbf{)} and considering Eq.\textbf{(}\ref{eq:51}\textbf{)}, averaging the results, we can get the mean value of the energy-momentum tensor of the reflected wave as:
\begin{equation}
    \label{eq:82}
 \left(\overline{\, _{\text{re}}T^{\mu  \nu }}\right)=\left( \begin{array}{cccc}  \frac{\left(\mu -n \mu _0\right){}^2 }{ c^2 \mu _0\left(\mu +n \mu _0\right){}^2} & -\frac{\left(\mu -n \mu _0\right){}^2}{ c^2 \mu _0 \left(\mu +n \mu _0\right){}^2} & 0 & 0 \\  -\frac{\text{  }\left(\mu -n \mu _0\right){}^2 }{c^2 \mu _0 \left(\mu +n \mu _0\right){}^2} & \frac{\text{  }\left(\mu -n \mu _0\right){}^2 }{ c^2 \text{$\mu $o} \left(\mu +n \mu _0\right){}^2} & 0 & 0 \\  0 & 0 & 0 & 0 \\  0 & 0 & 0 & 0 \end{array} \right)\overline{\, _{\text{in}}E^2}
\end{equation}
The incident and reflected waves are both in free space, so the mean photon number densities of the incident and reflected waves are $\overline{\, _{\text{$\gamma $in}}n}=\frac{1}{c^2\mu _0\hbar  \omega } \overline{\, _{\text{in}}E^2}$ and $\overline{\, _{\text{$\gamma $re}}n}=\frac{\left(\mu -n \mu _0\right){}^2 }{ c^2 \mu _0\left(\mu +n \mu _0\right){}^2\hbar  \omega } \overline{\, _{\text{in}}E^2}$, respectively. Because the propagation speed of the light beam in the medium is $1/n$ of that in the free space, the volume of the light beam in the medium is $1/n$ of that in the free space, so that the photon number density in the medium is $n$ times of that in the vacuum. Assuming that the total photon number is macroscopically conserved, the mean photon number density of the transmitted wave is $\overline{\, _{\text{$\gamma $tr}}n}=n(\overline{\, _{\text{$\gamma $in}}n}-\overline{\, _{\text{$\gamma $re}}n})$. Considering Eq.\textbf{(}\ref{eq:50}\textbf{)}, we get that the mean photon number density of the transmitted wave is:
\begin{equation}
    \label{eq:83}
 \overline{\, _{\text{$\gamma $tr}}n}=\frac{n^2}{c^2 \mu \text{  }} \frac{1}{\hbar  \omega }\overline{\, _{\text{tran}}E^2}
\end{equation}
Thus, the mean energy of a single photon in a non-ferromagnetic linear medium can be obtained as:
\begin{equation}
    \label{eq:84}
 \overline{\, _{\gamma }E}=\overline{\, _{\gamma }w}/\overline{\, _{\text{$\gamma $tr}}n}=\frac{\mu }{ \mu _0}\hbar  \omega\approx \hbar  \omega
\end{equation}
It can be seen that in a non-ferromagnetic linear medium, when the mean photon number is macroscopically conserved, the mean energy of a single photon is essentially the same as in free space. In this case, the mean mechanical momentum of a single photon in a non-ferromagnetic linear medium is:
\begin{equation}
    \label{eq:85}
 \overline{\, _{\gamma }P^1}=\overline{g^1}/\left( \overline{\, _{\text{$\gamma $tr}}n} \right)=\frac{\mu  }{ \mu _0}\frac{\hbar  \omega }{n c}\approx\frac{\hbar  \omega }{ n c}
\end{equation}
It can be seen that when the mean photon number is macroscopically conserved, the mean energy and the mean mechanical momentum of photons in the medium are $\frac{\mu }{ \mu _0}\hbar  \omega$ and $\frac{\mu  }{\mu _0}\frac{\hbar  \omega }{ n c}$, respectively. Since $\mu\approx\mu_0$ holds in non-ferromagnetic linear media, they can be approximated as $ \hbar  \omega$ and $ \frac{\hbar  \omega }{ n c}$, respectively. In this case, according to $\left(\overline{\, _{\gamma }E/c}\right){}^2-\left(\overline{\, _{\gamma }P}\right){}^2=\left(\overline{\, _{\gamma }m}\right){}^2 c^2$, the mean equivalent mass $\overline{\, _{\gamma }m}$ of photons is:
\begin{equation}
    \label{eq:86}
 \overline{\, _{\gamma }m}=\frac{\mu \sqrt{n^2-1} }{ n \mu _0}\frac{\hbar  \omega }{c^2}\approx\frac{\sqrt{n^2-1} }{  n}\frac{\hbar  \omega }{c^2}
\end{equation}
Thus, the mean energy propagation velocity of photons is:
\begin{equation}
    \label{eq:87}
 \overline{\, _{\gamma }v}=\frac{\overline{\, _{\gamma }P} }{\overline{\, _{\gamma }E/c^2}}= \frac{c}{n}=u
\end{equation}
This is consistent with our previous discussion. According to the above discussion, we can further write the energy-momentum tensor of the electromagnetic wave in the medium as:
\begin{equation}
    \label{eq:88}
 (\, _FT^{\mu  \nu })=\left( \begin{array}{cccc} w-p & \frac{wv}{c} & 0 & 0 \\  \frac{wv}{c} & \frac{wv^2}{c^2}+p & 0 & 0 \\  0 & 0 & \pm p & 0 \\  0 & 0 & 0 & \mp p \end{array} \right)
\end{equation}
where:
\begin{equation}
    \label{eq:89}
 w=\gamma ^2\left(\frac{p}{c^2}+\rho \right)c^2 =\frac{n^2}{c^2\mu _0} E^2
\end{equation}
\begin{equation}
    \label{eq:90}
\frac{wv}{c}=\frac{S^1}{c}=\frac{n }{c^2\mu _0} E^2
\end{equation}
\begin{equation}
    \label{eq:91}
p=w_0/2
\end{equation}
\begin{equation}
    \label{eq:92}
w_0=(1-v^2/c^2)w=\frac{n^2-1}{2 c^2\mu _0} E^2
\end{equation}
\begin{equation}
    \label{eq:93}
v=\left\{ \begin{array}{ll} u\text{     }\text{     }\text{     }\text{     }\text{     }\text{     }\text{     }\text{     }\text{     }\text{}(u\leq c) \\   c^2/u\text{     }\text{     }\text{     }\text{     }\text{     }(u>c) \end{array} \right.
\end{equation}
In the above formulas, $E$ is the electric field strength of the electromagnetic wave propagating in the medium, $w$ is the energy density of the electromagnetic wave, $p$ can be understood as pressure, $w_0 $ is the energy density of the electromagnetic wave in its center-of-momentum frame (that is rest frame), $v$ is the energy propagation velocity of the electromagnetic wave, which is equal to the phase velocity $u$ in the non dispersive medium with refractive index $n>1$. In the center-of-momentum frame of the electromagnetic wave, there are:
\begin{equation}
    \label{eq:94}
(\, _FT^{\mu  \nu })=\left( \begin{array}{cccc} w_0-p & 0 & 0 & 0 \\  0 & p & 0 & 0 \\  0 & 0 & \pm p & 0 \\  0 & 0 & 0 & \mp p \end{array} \right)
\end{equation}
From Eq.\textbf{(}\ref{eq:88}\textbf{)}, the physical meaning of each component of the energy-momentum tensor of the electromagnetic wave in the medium can be clearly seen. It can also be clearly seen that the momentum given by Eq.\textbf{(}\ref{eq:85}\textbf{)} is the mechanical momentum without the contribution of interaction. However, only the energy-momentum tensor of the electromagnetic wave can be written in the form of Eq.\textbf{(}\ref{eq:88}\textbf{)}, not the static electromagnetic field. Since, for an electromagnetic field that is stationary with respect to the medium, when it moves with velocity $v$ with respect to the observer, the medium also moves with the same velocity, there is a displacement of the medium in the pressure direction. Thus, in this case, the pressure $p$ does work on the medium, creating an energy flow that circulates in the medium. That is, in this case, $\gamma ^2\left(\frac{p}{c^2} \right)c^2$ in $T^{0i}$ belongs to the work done by the external medium, so it can no longer be considered as part of the energy of the electromagnetic field. The electromagnetic wave, on the other hand, is a collection of photons and can be viewed as a fluid composed of a large number of photons, while the static electromagnetic field is not a collection of photons and the two are essentially different. Therefore, the energy-momentum tensor of static electromagnetic field cannot be written in the form of Eq.\textbf{(}\ref{eq:88}\textbf{)}. In the presence of an static electromagnetic field, we should reconsider what form the energy-momentum tensor of the electromagnetic field can be written in, depending on the specific problem. For example, in an electrostatic field where the direction of the electric field is along the $x^1$ axis, its energy-momentum tensor can be written in the form of $(\, _FT^{\mu  \nu })=\left( \begin{array}{cccc}  w & 0 & 0 & 0 \\  0 & -p & 0 & 0 \\  0 & 0 & p & 0 \\  0 & 0 & 0 & p \end{array} \right)$, with $w=p=\frac{\left(E^1\right)^2}{2 c^2 \mu _0}$.\par
So far, according to the energy-momentum tensor Eq.\textbf{(}\ref{eq:88}\textbf{)} of electromagnetic waves in a medium different from the Abraham tensor, we have obtained that the mean energy and mean mechanical momentum of photons in the medium are $\frac{\mu }{ \mu _0}\hbar  \omega$ and $\frac{\mu  }{\mu _0}\frac{\hbar  \omega }{ n c}$, respectively. In non-ferromagnetic linear media, they can be approximated as $ \hbar  \omega$ and $ \frac{\hbar  \omega }{ n c}$, respectively. $ \frac{\hbar  \omega }{ n c}$ is the so-called Abraham momentum, which is the momentum that does not take into account the interaction contribution. In the following, we will see that we will obtain the Minkowski momentum when we consider the interaction contribution.

\section{The energy-momentum tensor with the contribution of the interaction term and mean canonical momentum of photons}\label{sec:s6}
As mentioned above, the canonical momentum related to quantum mechanics has the contribution of the interaction term. We find that we can also construct an energy-momentum tensor with the contribution of the interaction term, which is different from the Minkowski tensor. For a monochromatic plane wave electromagnetic wave, we can deduce from this energy-momentum tensor the canonical momentum consistent with the Minkowski viewpoint. The simplest idea for the energy-momentum tensor $\, _ {FI} T^{\mu  \nu }$ with the contribution of the interaction term is to make $\, _ {FI} T^{\mu  \nu }=\, _FT^{\mu  \nu }+\, _IT^{\mu  \nu }$, where $\, _FT^{\mu  \nu }$ is still given by Eq.\textbf{(}\ref{eq:41}\textbf{)}, and $\, _IT^{\mu  \nu }$ is given by $\, _IT^{\mu  \nu }=\frac{-2}{\sqrt{-g}}\frac{\delta ( \, _I\mathcal{L}\sqrt{-g})}{\delta  g_{\mu  \nu }}$. To make $\, _IT^{\mu  \nu }$ a symmetric tensor, we rewrite Eq.\textbf{(}\ref{eq:25}\textbf{)} as:
\begin{equation}
    \label{eq:95}
\, _I\mathcal{L}=\frac{1}{2}A^{\mu }j^{\nu }g_{\mu  \nu }+\frac{1}{2}A^{\nu }j^{\mu }g_{\mu  \nu }
\end{equation}
By substituting Eq.\textbf{(}\ref{eq:95}\textbf{)} into $\, _IT^{\mu  \nu }=\frac{-2}{\sqrt{-g}}\frac{\delta ( \, _I\mathcal{L}\sqrt{-g})}{\delta  g_{\mu  \nu }}$, we can obtain:
\begin{equation}
    \label{eq:96}
\\, _IT^{\mu  \nu }=-A^{\mu }j^{\nu }-A^{\nu }j^{\mu }+A_{\beta }j^{\beta }g^{\mu  \nu }
\end{equation}
so, we get:
\begin{equation}
    \label{eq:97}
\begin{array}{llll}
\, _ {FI} T^{\mu  \nu }=\frac{1}{\mu _0}\left(F^{\mu  \beta }F_{\text{   }\beta }^{\nu }-\frac{1}{4}F_{\alpha  \beta }F^{\alpha  \beta }g^{\mu  \nu }\right)\\
\text{ }\text{ }\text{ }\text{ }\text{ }\text{ }\text{ }\text{ }\text{ }\text{ }\text{ }\text{ }-A^{\mu }j^{\nu }-A^{\nu }j^{\mu }+A_{\beta }j^{\beta }g^{\mu  \nu }
\end{array}
\end{equation}
Because we have already considered the contribution of the interaction term, we hope that $\, _ {FI} T^{\mu  \nu }$ is conserved, that is, we hope that $\partial _{\nu }\left( \, _{F I}T^{\mu  \nu }\right)=0$ holds, but we find that $\, _ {FI} T^{\mu  \nu }$ given by Eq.\textbf{(}\ref{eq:97}\textbf{)} cannot make $\partial _{\nu }\left( \, _{F I}T^{\mu  \nu }\right)=0$ hold. We analyze the reason and find that this is because the $\, _IT^{\mu  \nu }$ given by the interaction term belongs to both sides of the interaction, so we should take only half of it. Thus, for a monochromatic plane wave, we obtain:
\begin{equation}
    \label{eq:98}
\, _ {FI} T^{\mu  \nu }=\, _FT^{\mu  \nu }+\frac{1}{2}(\, _IT^{\mu  \nu })
\end{equation}
So, we further obtain:
\begin{equation}
    \label{eq:99}
\begin{array}{llll}
\, _ {FI} T^{\mu  \nu }=\frac{1}{\mu _0}\left(F^{\mu  \beta }F_{\text{   }\beta }^{\nu }-\frac{1}{4}F_{\alpha  \beta }F^{\alpha  \beta }g^{\mu  \nu }\right)\\
\text{ }\text{ }\text{ }\text{ }\text{ }\text{ }\text{ }\text{ }\text{ }\text{ }\text{ }\text{ }+\frac{1}{2}(-A^{\mu }j^{\nu }-A^{\nu }j^{\mu }+A_{\beta }j^{\beta }g^{\mu  \nu })
\end{array}
\end{equation}
Assuming the electromagnetic four-dimensional potential as a monochromatic plane cosine wave propagating in the positive direction along the $x^1$ axis, Eq.\textbf{(}\ref{eq:37}\textbf{)} can be written as:
\begin{equation}
    \label{eq:100}
\left( \, _{\text{tran}}A^{\mu }\right)=\left(0,0, \, _{\text{tran}}A_0 \cos\left(-\frac{\omega  x^0}{c}+\frac{\omega n  x^1}{c}\right),0\right)
\end{equation}
In this case, the electromagnetic field tensor is:
\begin{equation}
    \label{eq:101}
F^{\mu  \nu }=\left( \begin{array}{cccc}  0 & 0 & 1 & 0 \\  0 & 0 & n & 0 \\  -1 & -n & 0 & 0 \\  0 & 0 & 0 & 0 \end{array} \right)\frac{ \, _{\text{tran}}E}{c}
\end{equation}
where $\, _{\text{tran}}E=\,_{\text{tran}}E_0 \sin\left(-\frac{\omega  x^0}{c}+\frac{\omega n  x^1}{c}\right)$ and $ \, _{\text{tran}}E_0=- (\, _{\text{tran}}A_0\omega)$. In this case, from Eq.\textbf{(}\ref{eq:32}\textbf{)} , we can obtain:
\begin{equation}
    \label{eq:102}
\left(j^{\mu }\right)=\left(0,0,\frac{\left(n^2-1\right) \omega ^2 }{c^2\mu _0} (\, _{\text{tran}}A_0) \cos\frac{\omega n  x^1-\omega  x^0}{c},0\right)
\end{equation}
Substituting Eq.\textbf{(}\ref{eq:100}\textbf{)}-Eq.\textbf{(}\ref{eq:102}\textbf{)} into Eq.\textbf{(}\ref{eq:99}\textbf{)}, we obtain:
\begin{equation}
    \label{eq:103}
     \begin{array}{llll}
\, _ {FI} T^{\mu  \nu }\\=\left( \begin{array}{cccc}  w-p-\, _Ip & \frac{w v}{c} & 0 & 0 \\  \frac{w v}{c} & \frac{w v^2}{c^2}+p+\, _Ip & 0 & 0 \\  0 & 0 & \pm p\mp \, _Ip & 0 \\  0 & 0 & 0 & \mp p\pm \, _Ip \end{array} \right)
\end{array}
\end{equation}
where:
\begin{equation}
    \label{eq:104}
 w= \frac{n^2}{c^2\mu _0}(\, _{\text{tran}}E_0)^2\sin ^2 \frac{\omega  \left(-x^0+n x^1\right)}{c}
\end{equation}
\begin{equation}
    \label{eq:105}
p=\left|\frac{n^2-1}{2 c^2  \mu_0}(\, _{\text{tran}}E_0)^2\sin ^2 \frac{\omega  \left(-x^0+n x^1\right)}{c} \right|
\end{equation}
\begin{equation}
    \label{eq:106}
\, _Ip=\left|\frac{n^2-1}{2 c^2  \mu_0}(\, _{\text{tran}}E_0)^2\cos ^2 \frac{\omega  \left(-x^0+n x^1\right)}{c} \right|
\end{equation}
\begin{equation}
    \label{eq:107}
v=\left\{ \begin{array}{ll} u\text{     }\text{     }\text{     }\text{     }\text{     }\text{     }\text{     }\text{     }\text{     }\text{}(u\leq c) \\   c^2/u\text{     }\text{     }\text{     }\text{     }\text{     }(u>c) \end{array} \right.
\end{equation}
Here $w$ is the energy density and $\, _ {FI} T^{00 }$ cannot be understood as the energy density; $\, _Ip$ is the contribution given by the interaction term; $u=c/n$ represents the phase velocity and $v$ represents the propagation velocity of energy and information. We also find that $\, _Ip$ and $p$ have the same amplitude, but different phases. After verification, it is indeed possible to obtain:
\begin{equation}
    \label{eq:108}
 \partial _{\nu } (\, _ {FI} T^{\mu  \nu })=0
\end{equation}
We take the mean of $\, _ {FI} T^{\mu  \nu }$ over a period and obtain:
\begin{equation}
    \label{eq:109}
 \overline{\, _ {FI} T^{\mu  \nu }}=\left( \begin{array}{cccc}  \frac{1}{2 \mu _0} & \frac{n}{2 \mu _0} & 0 & 0 \\  \frac{n}{2 \mu _0} & \frac{n^2}{2 \mu _0} & 0 & 0 \\  0 & 0 & 0 & 0 \\  0 & 0 & 0 & 0 \end{array} \right)\left(\frac{\, _{\text{tran}}E_0}{c}\right)^2
\end{equation}
According to Eq.\textbf{(}\ref{eq:83}\textbf{)}, the mean photon number density of a beam with a cross-sectional area of $S$ and a length of $d/n$ in the medium is:
\begin{equation}
    \label{eq:110}
 \overline{\, _{\text{$\gamma $tr}}n}=\frac{n^2}{2\mu \text{  }} \frac{1}{\hbar  \omega }\left(\frac{\, _{\text{tran}}E_0}{c}\right)^2
\end{equation}
Since the volume of this beam is $V=Sd/n$, the mean photon number $\overline{\, _{\text{$\gamma $tr}}N}$ is:
\begin{equation}
    \label{eq:111}
 \overline{\, _{\text{$\gamma $tr}}N}=\frac{n}{2\mu \text{  }} \frac{1}{\hbar  \omega }\left(\frac{\, _{\text{tran}}E_0}{c}\right)^2 Sd
\end{equation}
The canonical momentum includes the contribution from the interaction term, so $\overline{\, _ {FI} T^{11}}$ is the mean canonical momentum flow density and $\, _ {FI} T^{01}/c$ is not the canonical momentum density. Due to the time required for a beam of length $d/n$ to flow through the cross-section being $\Delta t=\frac{d}{n}/u=d/c$, the mean canonical momentum $\overline{\, _ {FI} P^{1}}$ in the volume $V$ can be obtained from $\overline{\, _ {FI} P^{1}}=\overline{\, _ {FI} T^{11}}S\Delta t$ as:
\begin{equation}
    \label{eq:112}
 \overline{\, _ {FI} P^{1}}=\frac{n^2}{2 c \mu _0}\left(\frac{\, _{\text{tran}}E_0}{c}\right)^2S d
\end{equation}
So, the mean canonical momentum of a single photon from $\overline{\, _ {\gamma FI} P^{1}}=\overline{\, _ {FI} P^{1}}/\overline{\, _{\text{$\gamma $tr}}N}$ is:
\begin{equation}
    \label{eq:113}
 \overline{\, _ {\gamma FI} P^{1}}=\frac{\mu  }{\mu _0} \frac{n \hbar  \omega }{c}
\end{equation}
In a non-ferromagnetic linear medium, $\mu\approx\mu_0$, so $\overline{\, _ {\gamma FI} P^{1}}\approx\frac{n \hbar  \omega }{c}$ is $n$ times its size in free space, which is slightly different from the strictly $\overline{\, _ {\gamma FI} P^{1}}=\frac{n \hbar  \omega }{c}$ case. In the following, we will see that this is related to the fact that we have not considered here the contribution of the polarization and magnetization energies of the medium. Although $\, _ {FI} T^{01} /c$ is not a canonical momentum density, $c(\, _ {FI} T^{01} )$ is still an energy flow density. So the mean energy of photons is still given by Eq.\textbf{(}\ref{eq:84}\textbf{)}. Unlike the Minkowski tensor, the energy-momentum tensor given by Eq.\textbf{(}\ref{eq:103}\textbf{)} is still symmetric. It can be clearly seen from Eq.\textbf{(}\ref{eq:103}\textbf{)} that the canonical momentum considering the contribution of interaction is directly proportional to the refractive index $n$ of the medium. For the mechanical momentum without considering the contribution of interaction, $\, _ {FI} T^{11 }-p-\, _Ip=\frac{w v^2}{c^2}$ is the momentum flow density. Therefore, according to Eq.\textbf{(}\ref{eq:104}\textbf{)}, Eq.\textbf{(}\ref{eq:107}\textbf{)}, and Eq.\textbf{(}\ref{eq:111}\textbf{)}, we can obtain:\par
\begin{equation}
    \label{eq:114}
\overline{\, _{\gamma }P^1}=\frac{w v^2 Sd}{c^2}/\left( \overline{\, _{\text{$\gamma $tr}}N} \right)=\frac{\mu  }{ \mu _0}\frac{\hbar  \omega }{n c}\approx\frac{\hbar  \omega }{ n c}
\end{equation}
The result is consistent with Eq.\textbf{(}\ref{eq:85}\textbf{)}. So, the energy-momentum tensor given by Eq.\textbf{(}\ref{eq:103}\textbf{)} can obtain both canonical momentum consistent with Minkowski momentum and mechanical momentum consistent with Abraham momentum.\par
So far we have shown that for a beam in a non-ferromagnetic linear medium, the mean mechanical and canonical momentum of the photons are $\overline{\, _{\gamma }P^1} \approx\frac{\hbar  \omega }{ n c}$ and $\overline{\, _ {\gamma FI} P^{1}}\approx\frac{n \hbar  \omega }{c}$, respectively. They differ only in whether the contribution of the interaction is included in the momentum. The mechanical momentum and canonical momentum of photons are two different definitions of momentum. When the photon interacts with other matter and contributes to the canonical momentum of the photon, the mechanical and canonical momentum of the photon are not equal. For experiments, when the measurement of momentum using experimental methods is less affected by the interaction of the medium, the experimental results will tend to support Eq.\textbf{(}\ref{eq:85}\textbf{)};  when the experimental method used has a significant impact on the measurement of momentum due to the interaction between the medium, the experimental results tend to support Eq.\textbf{(}\ref{eq:113}\textbf{)}. Stephen's 2010 paper\textcolor[rgb]{0.184313725,0.188235294117647,0.564705882}{\cite{Stephen}} also suggested that the mechanical momentum of the photon corresponds to the Abraham momentum and the canonical momentum to the Minkowski momentum, both of which make sense. This is consistent with the point of this paper, but our method of proof is not the same. We have also revealed in greater depth the connection and difference between these two types of momentum.\par
Whether we combine one half of $\, _IT^{\mu  \nu }=\frac{-2}{\sqrt{-g}}\frac{\delta ( \, _I\mathcal{L}\sqrt{-g})}{\delta  g_{\mu  \nu }}$ with $\, _FT^{\mu  \nu }=\frac{-2}{\sqrt{-g}}\frac{\delta ( \, _F\mathcal{L}\sqrt{-g})}{\delta  g_{\mu  \nu }}$ to form a new energy-momentum tensor $\, _ {FI} T^{\mu  \nu }$ and the other half with $\, _mT^{\mu  \nu }=\frac{-2}{\sqrt{-g}}\frac{\delta ( \, _m\mathcal{L}\sqrt{-g})}{\delta  g_{\mu  \nu }}$ to form other energy-momentum tensors, or split and combine in other forms, the total energy-momentum tensor $ T^{\mu  \nu }=\frac{-2}{\sqrt{-g}}\frac{\delta ( \mathcal{L}\sqrt{-g})}{\delta  g_{\mu  \nu }}$ will always be unique and determined, where $\mathcal{L}=\, _F\mathcal{L}+\, _I\mathcal{L}+\, _m\mathcal{L}$. The specific form of the Lagrangian density $\, _m\mathcal{L}$ for the medium can be quite complicated and is beyond the scope of this paper. This is consistent with the views of Brevik\textcolor[rgb]{0.184313725,0.188235294117647,0.564705882}{\cite{Brevik1979}}, Pfeifer\textcolor[rgb]{0.184313725,0.188235294117647,0.564705882}{\cite{Pfeifer}} and Stephen\textcolor[rgb]{0.184313725,0.188235294117647,0.564705882}{\cite{Stephen}}. The advantage of this method of segmentation and combination in this paper is that it satisfies the requirements of $\partial _{\nu  } (\, _ {FI} T^{\mu  \nu })=0$. That is, it is conserved and can profoundly reveal the relationship and differences between mechanical momentum and canonical momentum. We will see in Section 8 that $\partial _{\nu  } (  T^{\mu  \nu })=0$ does not hold for both Minkowski and Abraham tensors.

\section{Energy momentum-tensor containing the contribution of polarization energy and magnetization energy of the medium}\label{sec:s7}
As mentioned in the third Section.\ref{sec:s3}, the energy tensor of macroscopic electromagnetic waves in the medium given by Eq.\textbf{(}\ref{eq:64}\textbf{)} based on Eq.\textbf{(}\ref{eq:41}\textbf{)} does not include the contribution of polarization energy and magnetization energy stored in the medium. In the following, we will discuss the effect of the polarization energy and magnetization energy stored in a medium. From electrodynamics, it can be known that when the polarization energy and magnetization energy of the medium are included, for linear non-ferromagnetic media, its energy density can be expressed as\textcolor[rgb]{0.184313725,0.188235294117647,0.564705882}{\cite{Shuohong}}:
\begin{equation}
    \label{eq:115}
w=\frac{1}{2}\left(\boldsymbol{E}.\boldsymbol{D}+\boldsymbol{B}. \boldsymbol{H}\right)
\end{equation}
According to Brevik's paper published in 1979\textcolor[rgb]{0.184313725,0.188235294117647,0.564705882}{\cite{Brevik1979}}, when using the $(-,+,+,+)$ metric, the Minkowski tensor $\, _MT^{\mu  \nu }$ can be expressed as:
\begin{equation}
    \label{eq:116}
(\, _MT^{\mu  \nu })=\left( \begin{array}{llll}  \frac{1}{2}( \boldsymbol E. \boldsymbol D+ \boldsymbol H. \boldsymbol B) & \frac{1}{c} \boldsymbol E\times  \boldsymbol H \\  c  \boldsymbol D\times  \boldsymbol B & \, _MT^{i j} \end{array} \right)
\end{equation}
where $\, _MT^{i j}=-E^iD^j-H^iB^j+\frac{1}{2}( \boldsymbol E. \boldsymbol D+ \boldsymbol H. \boldsymbol B)g^{i j}$.\\
\\
And the Abraham tensor $\, _AT^{\mu  \nu }$ is:
\begin{equation}
    \label{eq:117}
(\, _AT^{\mu  \nu })=\left( \begin{array}{cc}  \frac{1}{2}( \boldsymbol E. \boldsymbol D+ \boldsymbol H. \boldsymbol B) & \frac{1}{c} \boldsymbol E\times  \boldsymbol H \\  \frac{1}{c} \boldsymbol E\times  \boldsymbol H & \, _AT^{i j} \end{array} \right)
\end{equation}
where:\\\\
$\begin{array}{llll}\, _AT^{i j}=-\frac{1}{2}\left(E^iD^j+E^jD^i\right)-\frac{1}{2}\left(H^iB^j+H^jB^i\right)\\
\text{  }\text{  }\text{  }\text{  }\text{  }\text{  }\text{  }\text{  }\text{  }\text{  }\text{  }+\frac{1}{2}( \boldsymbol E. \boldsymbol D+ \boldsymbol H. \boldsymbol B)g^{i j}\end{array}$
\\\\
It can be seen that both the Minkowski tensor and the Abraham tensor take into account contributions from the polarization and magnetization energies of the medium. We found that by changing $\frac{1}{\mu_0}F_{\text{   }\beta }^{\nu }$ to $K_{\text{   }\beta }^{\nu }$ in Eq.\textbf{(}\ref{eq:41}\textbf{)}, we immediately obtain the Minkowski tensor as:
\begin{equation}
    \label{eq:118}
\, _MT^{\mu  \nu } =F^{\mu  \beta }K_{\text{   }\beta }^{\nu }-\frac{1}{4}F_{\alpha  \beta }K^{\alpha  \beta }g^{\mu  \nu }
\end{equation}
where:
\begin{equation}
    \label{eq:119}
K^{\mu  \nu }=\left( \begin{array}{cccc}  0 & c D^1 & c D^2 & c D^3 \\  -c D^1  & 0 & H^3 & -H^2 \\  -c D^2  & -H^3 & 0 & H^1 \\  -c D^3 & H^2 & -H^1 & 0 \end{array} \right)
\end{equation}
Therefore, the Minkowski tensor can be written in the concise form of Eq.\textbf{(}\ref{eq:118}\textbf{)}. The Minkowski tensor is not symmetric. The Abraham tensor is symmetric with respect to the Minkowski tensor. However, symmetrizing Eq.\textbf{(}\ref{eq:118}\textbf{)} to $\frac{1}{2}F^{\mu  \beta }K_{\text{   }\beta }^{\nu }+\frac{1}{2}K_{\text{   }\beta }^{\mu }F^{\nu  \beta }-\frac{1}{4}F_{\alpha  \beta }K^{\alpha  \beta }g^{\mu  \nu }$ does not yield the Abraham tensor. However, the difference is only in the energy flow density. Simply replacing the energy flow density by $\frac{1}{c} \boldsymbol E\times  \boldsymbol H$ yields the Abraham tensor. There is:
\begin{equation}
    \label{eq:120}
\left\{ \begin{array}{ll}   \, _AT^{\mu 0 }  =\, _AT^{0 \mu }= \, _MT^{0 \mu } \\  \, _AT^{i j}=  \frac{1}{2}(\, _MT^{i  j } +\, _MT^{ji } ) \end{array} \right.
\end{equation}
Since neither the Minkowski tensor nor the Abraham tensor can be written in the form of Eq.\textbf{(}\ref{eq:16}\textbf{)}, we hope to construct an energy-momentum tensor that can be written in this form. It not only corresponds to Eq.\textbf{(}\ref{eq:64}\textbf{)}, but also has an energy density that corresponds to Eq.\textbf{(}\ref{eq:115}\textbf{)}. We found that by substituting the electric field intensity vector $\boldsymbol E$ and magnetic induction intensity vector $\boldsymbol B $ given by Eq.\textbf{(}\ref{eq:38}\textbf{)}, as well as the corresponding electric displacement vector $\boldsymbol D$ and magnetic field intensity vector $\boldsymbol H$, into Eq.\textbf{(}\ref{eq:115}\textbf{)}, we can obtain $w=\frac{n^2E^2 }{c^2 \mu }$.  Therefore, for $ \,_FT^{\mu\nu}$, simply rewrite $\mu_0$ in Eq.\textbf{(}\ref{eq:89}\textbf{)} to $\mu$, and we can obtain an energy density consistent with Eq.\textbf{(}\ref{eq:115}\textbf{)}. We also found that by substituting the electric field intensity vector $\boldsymbol E$ and magnetic induction intensity vector $\boldsymbol B $ given by Eq.\textbf{(}\ref{eq:38}\textbf{)}, as well as the corresponding electric displacement vector $\boldsymbol D$ and magnetic field intensity vector $\boldsymbol H$, into $\boldsymbol S=\frac{1}{c}\boldsymbol E\times \boldsymbol H$, we can obtain $S^1/c=\frac{n E^2}{c^2 \mu }$. We can see again that by simply replacing $\mu_0$ in $ \,_FT^{\mu\nu}$ with $\mu$ we can make $T^{01}=\frac{1}{c}\boldsymbol E\times \boldsymbol H$ hold. We can also see this relationship from Eq.\textbf{(}\ref{eq:68}\textbf{)} to Eq.\textbf{(}\ref{eq:73}\textbf{)}. Due to conservation, the total energy density and total energy flow density input into the medium should be continuous with Eq.\textbf{(}\ref{eq:68}\textbf{)} at the interface of the medium. Therefore, for electromagnetic waves propagating in the medium, simply transform $\mu_0$ in $ \,_FT^{\mu\nu}$ to $\mu $, and we can immediately obtain the total energy density and total energy flow density at the interface that are continuous with Eq.\textbf{(}\ref{eq:68}\textbf{)}. They include the contributions of polarization energy and magnetization energy. Finally, we construct the energy-momentum tensor satisfying these conditions as:
\begin{equation}
    \label{eq:121}
(\, _KT^{\mu  \nu })=\frac{\mu_0}{\mu  }(\, _FT^{\mu  \nu })=\left( \begin{array}{cccc}  w-p & \frac{w v}{c} & 0 & 0 \\  \frac{w v}{c} & w\frac{v^2}{c^2}+p & 0 & 0 \\  0 & 0 & \pm p & 0 \\  0 & 0 & 0 & \mp p \end{array} \right)
\end{equation}
where:
\begin{equation}
    \label{eq:122}
w=\frac{1}{2}\left(\boldsymbol{E}.\boldsymbol{D}+\boldsymbol{B}. \boldsymbol{H}\right)
\end{equation}
\begin{equation}
    \label{eq:123}
p= \frac{\left(c^2-v^2\right) w}{2 c^2}
\end{equation}
Eq.\textbf{(}\ref{eq:121}\textbf{)} where $\, _FT^{\mu  \nu }$ is the energy-momentum tensor of the macro electromagnetic wave excluding the polarization energy and magnetization energy, and $\, _KT^{\mu  \nu }$ is the energy-momentum tensor of the electromagnetic wave in the motionless coordinate system of the medium including the contribution of the polarization energy and magnetization energy. Similarly, there is also:
\begin{equation}
    \label{eq:124}
 \begin{array}{ll}\, _{KI}T^{\mu  \nu } =\frac{\mu_0}{\mu  }(\, _{FI}T^{\mu  \nu })\\=\left( \begin{array}{cccc}  w-p-\, _Ip & \frac{w v}{c} & 0 & 0 \\  \frac{w v}{c} & \frac{w v^2}{c^2}+p+\, _Ip & 0 & 0 \\  0 & 0 & \pm p\mp \, _Ip & 0 \\  0 & 0 & 0 & \mp p\pm \, _Ip \end{array} \right)\end{array}
\end{equation}
where:\par
$w= \frac{1}{2}\left(\boldsymbol{E}.\boldsymbol{D}+\boldsymbol{B}. \boldsymbol{H}\right) $\par
$p=\left|\frac{n^2-1}{2 c^2  \mu }(\, _{\text{tran}}E_0)^2\sin ^2 \frac{\omega  \left(-x^0+n x^1\right)}{c} \right| $\par
$_Ip=\left|\frac{n^2-1}{2 c^2  \mu }(\, _{\text{tran}}E_0)^2\cos ^2 \frac{\omega  \left(-x^0+n x^1\right)}{c} \right|$\\\\
Therefore, according to Eq.\textbf{(}\ref{eq:113}\textbf{)} and Eq.\textbf{(}\ref{eq:114}\textbf{)}, it can be seen when the contributions from polarization and magnetization are taken into account, the mean mechanical momentum of the electromagnetic wave is $\overline{\, _{\gamma }P^1} =\frac{\hbar  \omega }{ n c}$ and the mean canonical momentum is $\overline{\, _ {\gamma FI} P^{1}}=\frac{n \hbar  \omega }{c}$, which is strictly true.  For linear non-ferromagnetic media, $\mu\approx\mu_0$, so there are:
\begin{equation}
    \label{eq:125}
\left\{ \begin{array}{ll}  (\, _FT^{\mu  \nu })\approx (\, _KT^{\mu  \nu })\\  (\, _{FI}T^{\mu  \nu })\approx (\, _{KI}T^{\mu  \nu })  \end{array} \right.
\end{equation}
It can be seen that for electromagnetic waves propagating in such a medium, the effects of polarization and magnetization of the medium are negligible.\par
However, as mentioned at the end of the fifth Section.\ref{sec:s5}, Eq.\textbf{(}\ref{eq:121}\textbf{)} is not applicable to static electromagnetic fields in media. If Eq.\textbf{(}\ref{eq:121}\textbf{)} is applied to an electrostatic field:
\begin{equation}
    \label{eq:126}
\left(F^{\mu  \nu }\right)=\left( \begin{array}{cccc}  0 & \left.E^1\right/c & 0 & 0 \\  \left.-E^1\right/c & 0 & 0 & 0 \\  0 & 0 & 0 & 0 \\  0 & 0 & 0 & 0 \end{array} \right)
\end{equation}
We will obtain:
\begin{equation}
    \label{eq:127}
(\, _KT^{\mu  \nu })=\left( \begin{array}{cccc}  \frac{\left(E^1\right)^2}{2 c^2 \mu } & 0 & 0 & 0 \\  0 & -\frac{\left(E^1\right)^2}{2 c^2 \mu } & 0 & 0 \\  0 & 0 & \frac{\left(E^1\right)^2}{2 c^2 \mu } & 0 \\  0 & 0 & 0 & \frac{\left(E^1\right)^2}{2 c^2 \mu } \end{array} \right)
\end{equation}
We can see that in this case the polarization energy in the medium is lost. We found that on the basis of Eq.\textbf{(}\ref{eq:127}\textbf{)}, the following term can be added to compensate for the loss of polarization energy:
\begin{equation}
    \label{eq:128}
(\, _ET^{\mu  \nu })=\frac{n^2-1}{\mu  }\left(E^{\mu  \beta }E_{\text{   }\beta }^{\nu }-\frac{1}{4}E_{\alpha  \beta }E^{\alpha  \beta }g^{\mu  \nu }\right)
\end{equation}
where:
\begin{equation}
    \label{eq:129}
E^{\mu  \nu }=\left( \begin{array}{cccc}  0 & \left.E^1\right/c & \left.E^2\right/c & \left.E^3\right/c \\  \left.-E^1\right/c & 0 & 0 & 0 \\  \left.-E^2\right/c & 0 & 0 & 0 \\  \left.-E^3\right/c & 0 & 0 & 0 \end{array} \right)
\end{equation}
In this way, for the static electromagnetic field, we obtain:
\begin{equation}
    \label{eq:130}
 \, _KT^{\mu  \nu } =\frac{\mu_0}{\mu  }(\, _FT^{\mu  \nu })+ \, _ET^{\mu  \nu }
\end{equation}
We find that in isotropic linear non-ferromagnetic media, what we get from Eq.\textbf{(}\ref{eq:130}\textbf{)} is exactly the Abraham tensor. Therefore, for the analysis of electrostatic magnetic fields, the Abraham tensor is preferred. Based on the above discussion, we believe that Eq.\textbf{(}\ref{eq:64}\textbf{)} given on the basis of Eq.\textbf{(}\ref{eq:16}\textbf{)} is reasonable when only considering the macro electromagnetic field, because only it conforms to the form of Einstein field equations. When considered together with the polarization and magnetization of the medium, it cannot be forced to conform to the symmetric form of the Einstein field equations since it is not a single field energy-momentum tensor but a composite energy-momentum tensor. In this case, either Minkowski tensor or Abraham tensor, or Einstein-Laub tensor and other forms of tensors\textcolor[rgb]{0.184313725,0.188235294117647,0.564705882}{\cite{Brevik1979}} can be used to analyze some experimental phenomena. However, one should recognize that for an electromagnetic wave in a medium, since it interacts with the medium, its momentum has two different forms, one is the mechanical momentum which does not take into account the interaction contribution, and the other is the canonical momentum which takes into account the interaction contribution. The so-called Abrahams momentum is the counterpart of the mechanical momentum. The Minkowski momentum is the counterpart of the canonical momentum. We wish to find such an energy-momentum tensor that is symmetric and reflects both the mechanical momentum and canonical momentum of electromagnetic waves in the medium. In this paper, we find the energy-momentum tensor that meet these requirements, which are Eq.\textbf{(}\ref{eq:130}\textbf{)} and Eq.\textbf{(}\ref{eq:124}\textbf{)}. The energy-momentum tensors $ (\, _{FI}T^{\mu  \nu })$ and $ (\, _{KI}T^{\mu  \nu })$ given by Eq.\textbf{(}\ref{eq:103}\textbf{)} and Eq.\textbf{(}\ref{eq:124}\textbf{)} can derive these two different forms of momentum at the same time, and clarify their physical meaning. This is an important finding of this paper. We also find that for non-ferromagnetic media, the influence of the polarization and magnetization of the media on the energy-momentum tensor of the electromagnetic waves in the media is negligible.
\section{The relation between the pressure of a beam on its side and the polarization of the beam}\label{sec:s8}
For electromagnetic waves propagating in a medium, when considering polarization energy and magnetization energy, substituting the electromagnetic field tensor represented by Eq.\textbf{(}\ref{eq:62}\textbf{)} into Eq.\textbf{(}\ref{eq:116}\textbf{)} yields the Minkowski tensor as:
\begin{equation}
    \label{eq:131}
 (\, _MT^{\mu  \nu }) =\left( \begin{array}{cccc}  \frac{n^2}{c^2 \mu } & \frac{n}{c^2 \mu } & 0 & 0 \\  \frac{n^3}{c^2 \mu } & \frac{n^2}{c^2 \mu } & 0 & 0 \\  0 & 0 & 0 & 0 \\  0 & 0 & 0 & 0 \end{array} \right)\left(\, _{\text{tran}}E\right)^2
\end{equation}
For a monochromatic plane wave, The four-dimensional Lorentz force density is $\, _Mf^\mu=-\partial _{\nu } \left(\, _MT^{\mu  \nu }\right)=\left(\frac{(n-1) n }{c^3 \mu },\frac{(n-1) n^2\text{  }}{c^3 \mu },0,0\right)\left(\, _{\text{tran}}E_0\right){}^2 \omega  \sin  \frac{2 \left(x^1-x^0\right) \omega }{c}\ne0$.\par
Similarly, according to Eq.\textbf{(}\ref{eq:117}\textbf{)}, we can obtain the Abraham tensor as:
\begin{equation}
    \label{eq:132}
 (\, _AT^{\mu  \nu } )=\left( \begin{array}{cccc}  \frac{n^2}{c^2 \mu } & \frac{n}{c^2 \mu } & 0 & 0 \\   \frac{n}{c^2 \mu } & \frac{n^2}{c^2 \mu } & 0 & 0 \\  0 & 0 & 0 & 0 \\  0 & 0 & 0 & 0 \end{array} \right)\left(\, _{\text{tran}}E\right)^2
\end{equation}
For a monochromatic plane wave, The four-dimensional Lorentz force density is $\, _Af^\mu=-\partial _{\nu} \left(\, _AT^{\mu  \nu }\right)=\left(\frac{(n-1) n }{c^3 \mu },\frac{(1-n) n }{c^3 \mu },0,0\right)\left(\, _{\text{tran}}E_0\right){}^2 \omega  \sin  \frac{2 \left(x^1-x^0\right) \omega }{c}\ne0$.\par
We find that neither the Minkowski nor the Abraham tensor can make $\partial _{\nu} (T^{\mu  \nu })=0$ hold, that is, they are both non-conserved. We also found that $\, _{\text{Abr}}f^i=\, _Af^i-\, _Mf^i=\frac{n^2-1}{c^2}\frac{\partial S^i}{\partial t}$ is exactly the so-called Abraham force\textcolor[rgb]{0.184313725,0.188235294117647,0.564705882}{\cite{Pfeifer}}. We can see that both the Minkowski tensor and the Abraham tensor have $T^{22}=T^{33}=0$. According to the Minkowski tensor and the Abraham tensor, a beam propagating in a medium has no pressure on its side. However, whether based on Eq.\textbf{(}\ref{eq:88}\textbf{)} or Eq.\textbf{(}\ref{eq:121}\textbf{)}, the energy-momentum tensor obtained in this paper can always be written as:
\begin{equation}
    \label{eq:133}
(T^{\mu  \nu })=\left( \begin{array}{cccc}  w-p & \frac{w v}{c} & 0 & 0 \\  \frac{w v}{c} & w\frac{v^2}{c^2}+p & 0 & 0 \\  0 & 0 & \pm p & 0 \\  0 & 0 & 0 & \mp p \end{array} \right)
\end{equation}
Whether considering the contribution of polarization and magnetization or not, the beam propagating in the medium has a pressure effect on its side. This is one of the important differences between the energy-momentum tensor we obtain and the Minkowski tensor and the Abraham tensor. For media with refractive index $n>1$, without considering the contributions of polarization and magnetization, there are:
\begin{equation}
    \label{eq:134}
w=\frac{n^2}{c^2 \mu_0 }\left(\, _{\text{tran}}E\right){}^2
\end{equation}
\begin{equation}
    \label{eq:135}
p=\frac{n^2-1}{2 c^2 \mu_0 }\left(\, _{\text{tran}}E\right){}^2
\end{equation}
We also find that the pressure of the beam on its side is not isotropic and is related to the polarization of the light. For electromagnetic waves propagating along the x1 axis in a medium, the electromagnetic field tensor can be written as:
\begin{equation}
    \label{eq:136}
(\, _{\text{tran}}F^{\mu  \nu })=\left( \begin{array}{cccc}  0 & 0 & \left.E^2\right/c & \left.E^3\right/c \\  0 & 0 & B^3 & -B^2 \\  \left.-E^2\right/c & -B^3 & 0 & 0 \\  \left.-E^3\right/c & B^2 & 0 & 0 \end{array} \right)
\end{equation}
Substituting it into Eq.\textbf{(}\ref{eq:41}\textbf{)}, we can obtain:
\begin{equation}
    \label{eq:137}
(\, _FT^{\mu  \nu })=\left( \begin{array}{cccc}  w-p & \frac{w v}{c} & 0 & 0 \\  \frac{w v}{c} & w\frac{v^2}{c^2}+p & 0 & 0 \\  0 & 0 & p^{22} & p^{23}  \\  0 & 0 & p^{32}  & p^{33} \end{array} \right)
\end{equation}
where:
\begin{equation}
    \label{eq:138}
 p=\sqrt{(p^{22})^2+(p^{23})^2}
\end{equation}
\begin{equation}
    \label{eq:139}
p^{22}=-p^{33}=\frac{\left(B^3\right)^2c^2-\left(B^2\right)^2c^2+\left(E^3\right)^2-\left(E^2\right)^2}{2 c^2 \mu_0 }
\end{equation}
\begin{equation}
    \label{eq:140}
p^{23}=p^{32}=-\frac{\left(B^2\right) \left(B^3\right) c^2+\left(E^2\right)\left(E^3\right)}{ c^2\mu_0 }
\end{equation}
We can see that the contribution of the electromagnetic field component to the pressure in the parallel direction of the component is negative for the terms on the main diagonal;  In contrast, the contribution in the vertical direction is positive. For electromagnetic waves propagating in a medium, the electric field intensity vector and the magnetic induction intensity vector are not independent of each other. Let the electromagnetic wave propagate along the $x^1$ axis, the angle between the electric field strength and the $x^2$ axis be $\theta$, and the coordinate system be the right-handed system. The electromagnetic field tensor can actually be written as:
\begin{equation}
    \label{eq:141}
(\, _{\text{tran}}F^{\mu  \nu })=\left( \begin{array}{cccc}  0 & 0 & \cos  \theta  & -\sin  \theta  \\  0 & 0 & n \cos  \theta  & -n \sin  \theta  \\  -\cos  \theta  & -n \cos  \theta  & 0 & 0 \\  \sin  \theta  & n \sin  \theta  & 0 & 0 \end{array} \right)\frac{\, _{\text{tran}}E}{c}
\end{equation}
Substituting it into Eq.\textbf{(}\ref{eq:41}\textbf{)}, we can obtain:
\begin{equation}
    \label{eq:142}
(\, _FT^{\mu  \nu })=\left( \begin{array}{cccc}  w-p & \frac{w v}{c} & 0 & 0 \\  \frac{w v}{c} & w\frac{v^2}{c^2}+p & 0 & 0 \\  0 & 0 & p^{22} & p^{23}  \\  0 & 0 & p^{32}  & p^{33} \end{array} \right)
\end{equation}
where:
\begin{equation}
    \label{eq:143}
p=\frac{n^2-1}{2 c^2 \mu_0 }\left(\, _{\text{tran}}E\right){}^2
\end{equation}
\begin{equation}
    \label{eq:144}
 p^{22}=-p^{33}=p \cos (2 \theta )
\end{equation}
\begin{equation}
    \label{eq:145}
 p^{23}=p^{32}=-p \sin (2 \theta )
\end{equation}
We can see that the pressure of the beam on its side is not isotropic but depends on the polarization of the light. In the direction parallel to the electric field intensity vector, the barotropic pressure is highest and its magnitude is $p=\frac{n^2-1}{2 c^2 \mu_0 }\left(\, _{\text{tran}}E\right){}^2$. In the direction at an angle of $\theta$ with the electric field intensity vector, the magnitude of the positive pressure is $p \cos (2 \theta )$. The presence of transverse pressure on the beam is one of the important differences between the energy-momentum tensor of light in the medium that we have obtained, and the Minkowski tensor and the Abraham tensor. We also see that for Eq.\textbf{(}\ref{eq:103}\textbf{)} considering the contribution of the interaction term, the mean value of the lateral pressure will be $0$, but the instantaneous value is
\begin{equation}
    \label{eq:146}
 \, _{\text{side}}p=\frac{ n^2-1 }{2 c^2 \mu _0}\left(\, _{\text{tran}}E_0\right){}^2\cos \frac{2 \omega \left(n x^1-x^0\right)}{c}
\end{equation}
It is not 0. We hope that experimental workers can design experiments to verify the existence of these lateral pressures.

\section{Bernoulli effect of light beams}\label{sec:s9}
Since a beam of light can be viewed as a fluid with pressure, we can also discuss the Bernoulli effect of the beam. We write the energy-momentum tensor in the following form:
\begin{equation}
    \label{eq:147}
 \left(\,_KT^{\mu \nu }\right)=\left( \begin{array}{cccc} \, _{\gamma }\rho c^2-p &\, _{\gamma }\rho c v & 0 & 0 \\ \, _{\gamma }\rho ^2c v & \, _{\gamma }\rho v^2+p & 0 & 0 \\  0 & 0 & \pm p & 0 \\  0 & 0 & 0 & \mp p\end{array} \right)
\end{equation}
where $\, _{\gamma }\rho=\gamma ^2\left(\frac{p}{c^2}+\rho \right)$. At the beam streamline tube $a$, the energy flow density is:
\begin{equation}
    \label{eq:148}
\, _aS^1 =c( \,_a\ T^{01})=\left(\frac{\, _ap}{c^2}+\, _a\rho \right)\left(\, _a\gamma \right){}^2c^2 \, \left.(_av\right)
\end{equation}
At the beam streamline tube $b$, the energy flow density is:
\begin{equation}
    \label{eq:149}
\, _bS^1=c( \,_b\ T^{01})=\left(\frac{\, _bp}{c^2}+\, _b\rho \right)\left(\, _b\gamma \right){}^2c^2 \, \left.(_bv\right)
\end{equation}
Based on the conservation of energy, we obtain:
\begin{equation}
    \label{eq:150}
\, _aS \left( \, _aS^1 \right)=\, _bS \left( \, _bS^1 \right)+\, _{ab}P
\end{equation}
Here, $\, _aS $ is the cross-sectional area of the streamline tube perpendicular to the energy flow density vector at $a$, and $\, _bS $ is the cross-sectional area of the streamline tube perpendicular to the energy flow density vector at $b$. $\, _{ab}P$ is the power of the external work done by the beam of light propagating from $a$ to $b$ in the same streamline tube. It represents the loss of energy, which may be caused by reflection, etc. From Eq.\textbf{(}\ref{eq:148}\textbf{)} - Eq.\textbf{(}\ref{eq:150}\textbf{)}, we can solve to obtain:
\begin{equation}
    \label{eq:151}
\, _bp=\frac{\, _a\gamma ^2}{ \, _b\gamma ^2}\frac{\, _av}{\, _bv}\frac{\, _aS}{\, _bS}\left(\, _ap+\, _a\rho  c^2\right)-\, _b\rho  c^2-\frac{1}{\text{  }\, _b\gamma ^2}\frac{1}{ \, _bv}\frac{\, _{\text{ab}}P}{ \, _bS}
\end{equation}
This is the Bernoulli effect equation for the beam. For a continuous flow that is incompressible and neglects energy loss, there are $\frac{\, _a\gamma  }{ \, _b\gamma  }\frac{\, _av}{\, _bv}\frac{\, _aS}{\, _bS}=1$, $ \,_a\rho =\,_b\rho$, $\, _{\text{ab}}P=0$, so Eq.\textbf{(}\ref{eq:151}\textbf{)} can be transformed as:
\begin{equation}
    \label{eq:152}
\, _bp=\frac{\, _a\gamma }{ \, _b\gamma } \left(\, _ap+ \rho  c^2\right)- \rho  c^2
\end{equation}
At the Newton mechanics limit, we obtain:
\begin{equation}
    \label{eq:153}
\, _bp=\, _ap+\left(\frac{\, _av^2}{2 }-\frac{\, _bv^2}{2 } \right)\rho
\end{equation}
This is the usual Bernoulli effect equation seen for ordinary fluids under Newtonian mechanics. But the beam is not exactly the same as an ordinary fluid. On the one hand, the energy propagation rate of light in a medium can be compared to the speed of light in vacuum, and therefore cannot be approximated by Newtonian mechanics. On the other hand, a beam of light is not an incompressible fluid and its density $\rho$ is not a constant. The variation of the speed of light is generally caused by its incidence from one medium to another, where the loss of energy by reflection is almost inevitable. According to Eq.\textbf{(}\ref{eq:65}\textbf{)} and Eq.\textbf{(}\ref{eq:66}\textbf{)}, for a beam of light in a medium, there is $p=\rho c^2$, so Eq.\textbf{(}\ref{eq:151}\textbf{)} can be further written as:
\begin{equation}
    \label{eq:154}
\, _bp=\frac{\, _a\gamma ^2}{ \, _b\gamma ^2}\frac{\, _av}{\, _bv}\frac{\, _aS}{\, _bS}\left(\, _ap \right) -\frac{1}{2}\frac{1}{\text{  }\, _b\gamma ^2}\frac{1}{ \, _bv}\frac{\, _{\text{ab}}P}{ \, _bS}
\end{equation}
Let us now analyze the simple case in detail. Assuming that the beam is vertically incident from a medium with refractive index $\, _1n$ to a medium with refractive index $\, _2n$, the reflectivity is $x3$, so that:
\begin{equation}
    \label{eq:155}
\frac{\, _{\text{ab}}P}{ \, _bS}=R\left(\, _1S^1\right)=2\left(\frac{\, _1n-\, _2n}{\, _1n+\, _2n}\right) ^2\left(\, _1\gamma ^2\, \right)\left.(_1p\right)\left( \, _1v\right)
\end{equation}
Due to the vertical incidence, the cross-sectional area is equal at both ends of the streamline tube. Therefore, after substituting Eq.\textbf{(}\ref{eq:155}\textbf{)} into Eq.\textbf{(}\ref{eq:154}\textbf{)}, we obtain:
\begin{equation}
    \label{eq:156}
\, _2p=\frac{\, _1\gamma ^2}{ \, _2\gamma ^2}\frac{\, _1v}{\, _2v} \left[1-\left(\frac{\, _1n-\, _2n}{\, _1n+\, _2n}\right){}^2\right]\left(\, _1p \right)
\end{equation}
After simplification and organization, we obtain:
\begin{equation}
    \label{eq:157}
\, _2p=\frac{4 \left(\, _1n\right){}^2 }{ \left(\, _1n+\, _2n\right){}^2}\frac{\left(\, _2n^2-1\right)}{\left(\, _1n^2-1\right) }\left(\, _1p\right)
\end{equation}
Due to:
\begin{equation}
    \label{eq:158}
\frac{d \left(\frac{4 \left(\, _1n\right){}^2 }{ \left(\, _1n+\, _2n\right){}^2}\frac{\left(\, _2n^2-1\right)}{\left(\, _1n^2-1\right)\text{  }}\right)}{d\left(\, _2n\right)}=\frac{8 \left(\, _1n\right){}^2 \left(1+\, _1n \, _2n \right)}{\left(\, _1n-1\right) \left(\, _1n+\, _2n\right){}^3}>0
\end{equation}
And when $\, _2n=\, _1n$, $\frac{4 \left(\, _1n\right){}^2 }{ \left(\, _1n+\, _2n\right){}^2}\frac{\left(\, _2n^2-1\right)}{\left(\, _1n^2-1\right) }=1$. Thus, the pressure increases in a medium with a higher refractive index. Since the refractive index is higher, the energy transfer rate of the beam is lower. The smaller the velocity, the larger the pressure, which is characteristic of the Bernoulli effect. It should be noted that the pressure here is different from the radiation pressure due to the collision of the beam with the reflecting surface. The beam does not have pressure in free space, so when the beam is incident vertically from vacuum, considering that the cross-sectional area of the streamline tube is equal, Eq.\textbf{(}\ref{eq:151}\textbf{)} can be rewritten as:
\begin{equation}
    \label{eq:159}
\, _bp=\frac{1}{2}\frac{1}{ \, _b\gamma ^2}\frac{1}{\, _bv} \left(\, _aS^1-\frac{\, _{\text{ab}}P}{ \, _bS}\right)
\end{equation}
Here $\, _aS^1=\, _{in}S^1$ is the energy flow density of the incident beam. Since $R=\frac{(n-1)^2}{(1+n)^2}$ for reflectivity, $\frac{\, _{\text{ab}}P}{\, _bS}=\frac{(n-1)^2}{(1+n)^2}\left(\, _{in}S^1\right)$. Substituting it into Eq.\textbf{(}\ref{eq:159}\textbf{)} and noting $\, _bv=c/n$ and $ \, _b\gamma =1/\sqrt{1-1/n^2}$, the pressure of the beam in the medium can be obtained as:
\begin{equation}
    \label{eq:160}
p=\frac{2(n-1)}{n+1}\left(\frac{\, _{in}S^1}{c}\right)
\end{equation}
In 1973, Ashkin performed an experiment in which an incident laser beam caused the surface of a liquid medium to deform\textcolor[rgb]{0.184313725,0.188235294117647,0.564705882}{\cite{1973Radiation}}. He focused a strong laser beam on the water surface and found that the surface protruded outward under the irradiation of the laser beam. Eq.\textbf{(}\ref{eq:160}\textbf{)} can be used to explain Ashkin's experimental results. We hope that experimental workers can design experiments to further verify the Bernoulli effect of the beam.
\section{Energy-momentum tensor of electromagnetic waves in moving media}\label{sec:s10}
In the laboratory reference frame, we imagine that a beam of light with angular frequency $\omega_0$ propagating in the positive direction along the $x^1$ axis is incident on a medium moving in the positive direction along the $x^1$ axis with velocity $V$. Assuming that in the laboratory reference frame, the electromagnetic field tensor of the incident beam is:
\begin{equation}
    \label{eq:161}
(\, _{\text{oin}}F^{\mu  \nu })=\left( \begin{array}{cccc}  0 & 0 & 1 & 0 \\  0 & 0 & 1 & 0 \\  -1 & -1 & 0 & 0 \\  0 & 0 & 0 & 0 \end{array} \right)\frac{\, _{\text{oin}}E }{c}
\end{equation}
We use the Lorentz transformation matrix:
\begin{equation}
    \label{eq:162}
\left(\Lambda _{\text{   }\nu }^{\mu }\right)=\left( \begin{array}{cccc}  \gamma  & -\gamma \frac{V}{c} & 0 & 0 \\  -\gamma \frac{V}{c} & \gamma  & 0 & 0 \\  0 & 0 & 1 & 0 \\  0 & 0 & 0 & 1 \end{array} \right)
\end{equation}
transform Eq.\textbf{(}\ref{eq:161}\textbf{)} to the rest reference frame of the medium. Assuming that the incident electromagnetic field tensor is $(\, _{\text{in}}F^{\mu  \nu })$ in the rest reference frame of the medium, we have:
\begin{equation}
    \label{eq:163}
(\, _{\text{in}}F^{\mu  \nu })=\left( \begin{array}{cccc}  0 & 0 & 1 & 0 \\  0 & 0 & 1 & 0 \\  -1 & -1 & 0 & 0 \\  0 & 0 & 0 & 0 \end{array} \right)\frac{\, _{\text{in}}E }{c}
\end{equation}
where:
\begin{equation}
    \label{eq:164}
\, _{\text{in}}E=\gamma  (1-V/c)(\, _{\text{oin}}E)
\end{equation}
By using the boundary condition Eq.\textbf{(}\ref{eq:50}\textbf{)}, it can be obtained that the electromagnetic field tensor of electromagnetic waves propagating in the medium is:
\begin{equation}
    \label{eq:165}
(\, _{\text{tran}}F^{\mu  \nu })=\left( \begin{array}{cccc}  0 & 0 & 1 & 0 \\  0 & 0 & n & 0 \\  -1 & -n & 0 & 0 \\  0 & 0 & 0 & 0 \end{array} \right)\frac{\, _{\text{tran}}E }{c}
\end{equation}
where $\, _{\text{tran}} E=\frac{2  \mu  }{n  \mu _0 +  \mu }(\, _{\text{in}}E)$. Substituting Eq.\textbf{(}\ref{eq:165}\textbf{)} into Eq.\textbf{(}\ref{eq:41}\textbf{)}, we obtain that the energy-momentum tensor of electromagnetic waves propagating in the medium in a rest reference frame is:
\begin{equation}
    \label{eq:166}
 (\, _FT^{\mu  \nu })=\left( \begin{array}{cccc} \, _FT^{00} & \, _FT^{01}& 0 & 0 \\ \, _FT^{10} & \, _FT^{11} & 0 & 0 \\  0 & 0 & \, _FT^{22} & 0 \\  0 & 0 & 0 & \, _FT^{33} \end{array} \right)
\end{equation}
where:\par
$\, _FT^{00}=\, _FT^{11}=\frac{2 \left(n^2+1\right) \mu ^2\text{  }}{c^2 \mu _0\left(n \mu _0+\mu \right){}^2} \left( \, _{\text{in}}E\right)^2$\par
$\, _FT^{01}=\, _FT^{10}= \frac{4 n \mu ^2\text{  }}{c^2 \mu _0\left(n \mu _0+\mu \right){}^2}\left( \, _{\text{in}}E\right) ^2$\par
$\, _FT^{22}=-\, _FT^{33}= \frac{2 \left(n^2-1\right) \mu ^2 }{c^2 \mu _0\left(n \mu _0+\mu \right)^2}\left( \, _{\text{in}}E\right) ^2$\par
$\text{ }$\par\par
We transform the four-dimensional wave vector $\left(K^{\mu }\right)=\left(\frac{\omega_0 }{c},\frac{\omega_0 }{c},0,0\right)$ of the incident wave in the laboratory reference frame using the transformation matrix Eq.\textbf{(}\ref{eq:162}\textbf{)} to the rest reference frame of the medium, and obtain the transformation relationship between angular frequency and wavelength as follows:
\begin{equation}
    \label{eq:167}
\omega =\gamma  (1-V/c)\omega _0
\end{equation}
\begin{equation}
    \label{eq:168}
\lambda =\frac{2\pi  c}{\gamma  (1-V/c)\omega _0}
\end{equation}
According to Eq.\textbf{(}\ref{eq:166}\textbf{)} and $wv/c=\, _FT^{01}$, the energy density of electromagnetic waves propagating in the medium is:
\begin{equation}
    \label{eq:169}
w=\frac{4 n^2\mu ^2}{c^2\mu _0\left(n \mu _0+\mu \right){}^2}\left( \, _{\text{in}}E\right) ^2
\end{equation}
If the medium is at rest in the laboratory reference frame, the energy density is $w_0=\frac{4 n^2\mu ^2}{c^2\mu _0\left(n \mu _0+\mu \right){}^2}\left( \, _{\text{oin}}E\right) ^2$. In the rest reference frame of the medium, the wavelength of the incident beam is given by Eq.\textbf{(}\ref{eq:168}\textbf{)}, so the wavelength in the medium is $\lambda /n$. If the medium is at rest in the laboratory reference frame, the wavelength of the electromagnetic wave propagating in the medium is $2\pi c/\omega_0$. Thus, when the volume of the beam in the moving medium is $\Omega= S\lambda/n=\frac{2\pi  c}{n\gamma  (1-V/c)\omega _0}S$, the volume in the stationary medium will be $\Omega_0= \frac{2\pi c}{n\omega_0}S $. Since the energy $E=\overline{w} \Omega$ is proportional to the angular frequency, the relation between angular frequencies is:
\begin{equation}
    \label{eq:170}
\omega/\omega_0=\frac{\overline{w} \Omega}{\overline{w_0 }\Omega_0}=\gamma  (1-V/c)
\end{equation}
The results are consistent with Eq.\textbf{(}\ref{eq:167}\textbf{)}. This indicates that the transformation relationship between energy and angular frequency is self-consistent for the energy-momentum tensor obtained in this paper. In the rest reference frame of the medium, the frequency of the electromagnetic wave does not change after it is incident on the medium. Due to the phase velocity $u=c/n1$ in the medium, the four-dimensional wave vector of the electromagnetic wave in the rest reference frame of the medium is:
\begin{equation}
    \label{eq:171}
\left(K^{\mu }\right)=\left(\frac{1 }{c},\frac{n}{c},0,0\right)\gamma  (1-V/c)\omega _0
\end{equation}
By inverting it into the laboratory reference frame, we obtain that the angular frequency and wavelength of an electromagnetic wave in a moving medium in the laboratory reference frame are:
\begin{equation}
    \label{eq:172}
\omega=\frac{(c-V) (c+n V) \gamma ^2  \omega_0}{c^2}
\end{equation}
\begin{equation}
    \label{eq:173}
\lambda =\frac{2\pi  c}{ \gamma ^2(1-V/c) ( n+V/c) \omega _0}
\end{equation}
The above is the frequency and wavelength conversion relation for an electromagnetic wave incident vertically from vacuum into a medium. It can be seen that when an electromagnetic wave is incident into a moving medium, the angular frequency $\omega$ of the electromagnetic wave in the medium is no longer equal to the angular frequency $\omega_0$ of the incident wave. Only in the rest reference frame of the medium, the angular frequency $\omega$ of the electromagnetic wave in the medium is equal to the angular frequency $\omega_0$ of the incident wave. From $u'=\frac{\omega  \lambda }{2\pi }$, we can conclude that the phase velocity of an electromagnetic wave in a moving medium is $u'= \frac{ u+ V }{1+\frac{V u}{c^2}}$, which is consistent with the relativistic velocity transformation relationship. By inverting Eq.\textbf{(}\ref{eq:166}\textbf{)} into the laboratory system and writing it in the form of Eq.\textbf{(}\ref{eq:88}\textbf{)}, we obtain:
\begin{equation}
    \label{eq:174}
(\, _FT^{\mu  \nu })=\left( \begin{array}{cccc} w-p & \frac{wv}{c} & 0 & 0 \\  \frac{wv}{c} & \frac{wv^2}{c^2}+p & 0 & 0 \\  0 & 0 & \pm p & 0 \\  0 & 0 & 0 & \mp p \end{array} \right)
\end{equation}
where:
\begin{equation}
    \label{eq:175}
 w=\frac{ (c+V/n)^2\text{  }}{\text{  }(c+V)^2\text{  }}\frac{4 n^2\mu ^2}{c^2\mu _0\left(n \mu _0+\mu \right){}^2}(\, _{\text{oin}}E)^2
\end{equation}
\begin{equation}
    \label{eq:176}
p=\gamma ^2(1-V/c)^2 \frac{2 \left(n^2-1\right) \mu ^2}{c^2\mu _0\left(n \mu _0+\mu \right){}^2}(\, _{\text{oin}}E)^2
\end{equation}
\begin{equation}
    \label{eq:177}
v=\frac{ u+ V }{ 1+\frac{V u}{c^2}}
\end{equation}
It can be seen that the speed of energy propagation of electromagnetic waves in a moving medium also obeys the relativistic velocity transformation relation. In the stationary reference frame of the medium, based on $E_0=E/\gamma$ and $\gamma=1/\sqrt{1-1/n^2}$, the rest energy corresponding to an electromagnetic beam with a length of one wavelength can be obtained from Eq.\textbf{(}\ref{eq:166}\textbf{)} and Eq.\textbf{(}\ref{eq:168}\textbf{)} as follows:
\begin{equation}
    \label{eq:178}
E_0=\sqrt{1-\frac{1}{n^2}}\frac{4 n^2(c-V)^2 \gamma ^2 \mu ^2}{c^4 \mu _0 \left(\mu +n \mu _0\right){}^2}\overline{(\, _{\text{oin}}E)^2}\frac{2\pi  c}{n \gamma  (1-V/c)\omega _0}S
\end{equation}
Here $x1$ is the cross-sectional area of the electromagnetic beam. In the laboratory reference frame, the energy $E$ corresponding to an electromagnetic wave with a length of one wavelength can be obtained from Eq.\textbf{(}\ref{eq:175}\textbf{)}, Eq.\textbf{(}\ref{eq:172}\textbf{)}, and $E=\overline{\omega}\Omega=\overline{\omega}\lambda S$ as follows:
\begin{equation}
    \label{eq:179}
\begin{array}{ll}E=\frac{ (c+V/n)^2\text{  }}{\text{  }(c+V)^2\text{  }}\frac{4 n^2\mu ^2}{c^2\mu _0\left(n \mu _0+\mu \right){}^2}\overline{(\, _{\text{oin}}E)^2}\\
\text{ }\text{ }\text{ }\text{ }\text{ }\times \frac{2\pi  c}{ \gamma ^2(1-V/c) ( n+V/c) \omega _0}S
\end{array}
\end{equation}
From Eq.\textbf{(}\ref{eq:178}\textbf{)} and Eq.\textbf{(}\ref{eq:179}\textbf{)}, we obtain:
\begin{equation}
    \label{eq:180}
E=\frac{1}{\sqrt{1- \frac{v^2}{c^c}}}E_0
\end{equation}
where $v=\frac{ u+ V }{ 1+\frac{V u}{c^2}}$. The energy of a visible beam satisfies the relativistic transformation relation $E=\gamma E_0$. In the rest reference frame of the medium, according to Eq.\textbf{(}\ref{eq:165}\textbf{)} and boundary condition Eq.\textbf{(}\ref{eq:48}\textbf{)}, the electromagnetic field tensor of the reflected wave is:
\begin{equation}
    \label{eq:181}
(\, _{\text{out}}F^{\mu  \nu })=\left( \begin{array}{cccc}  0 & 0 & 1 & 0 \\  0 & 0 & 1 & 0 \\  -1 & -1 & 0 & 0 \\  0 & 0 & 0 & 0 \end{array} \right)\frac{\, _{\text{out}}E }{c}
\end{equation}
where $\, _{\text{out}}E=\frac{2 n \mu _0 }{\mu  +n \mu _0}(\, _{\text{tran}}E)$.
Substituting Eq.\textbf{(}\ref{eq:181}\textbf{)} into Eq.\textbf{(}\ref{eq:41}\textbf{)}, we obtain the energy-momentum tensor of electromagnetic waves emitted into vacuum in a stationary reference frame of the medium as follows:
\begin{equation}
    \label{eq:182}
\left(\, _{\text{out}}T^{\mu  \nu }\right)=\left( \begin{array}{cccc}  \frac{1}{c^2 \mu _0} & \frac{1}{c^2 \mu _0} & 0 & 0 \\  \frac{1}{c^2 \mu _0} & \frac{1}{c^2 \mu _0} & 0 & 0 \\  0 & 0 & 0 & 0 \\  0 & 0 & 0 & 0 \end{array} \right)\left(\, _{\text{out}}E\right)^2
\end{equation}
Inverted into the laboratory reference frame, we obtain:
\begin{equation}
    \label{eq:183}
\left(\, _{\text{out}}T^{\mu  \nu }\right)=\left( \begin{array}{cccc}  \frac{\gamma ^2 (1+V/c)^2 }{c^2 \mu _0} & \frac{\gamma ^2 (1+V/c)^2 }{c^2 \mu _0} & 0 & 0 \\  \frac{\gamma ^2 (1+V/c)^2 }{c^2 \mu _0} & \frac{\gamma ^2 (1+V/c)^2 }{c^2 \mu _0} & 0 & 0 \\  0 & 0 & 0 & 0 \\  0 & 0 & 0 & 0 \end{array} \right)\left(\, _{\text{out}}E\right)^2
\end{equation}
Using $\, _{\text{tran}} E$ to represent equation Eq.\textbf{(}\ref{eq:174}\textbf{)} and inverting it into a laboratory reference frame, we obtain:
\begin{equation}
    \label{eq:184}
(\, _FT^{\mu  \nu })=\left( \begin{array}{cccc} w-p & \frac{wv}{c} & 0 & 0 \\  \frac{wv}{c} & \frac{wv^2}{c^2}+p & 0 & 0 \\  0 & 0 & \pm p & 0 \\  0 & 0 & 0 & \mp p \end{array} \right)
\end{equation}
where:
\begin{equation}
    \label{eq:185}
 w=\frac{ (c+V/n)^2}{c^2-V^2 }\frac{4 n^2 }{c^2\mu _0 }(\, _{\text{tran}}E)^2
\end{equation}
\begin{equation}
    \label{eq:186}
p=\frac{n^2-1 }{2 c^2 \mu _0}(\, _{\text{tran}}E)^2
\end{equation}
\begin{equation}
    \label{eq:187}
v=\frac{ u+ V }{ 1+\frac{V u}{c^2}}
\end{equation}
Assuming that the angular frequency of an electromagnetic wave propagating in a medium is $\,_{tran}\omega$ and the wavelength is $\,_{tran}\lambda=\frac{2\pi c/n }{\,_{tran}\omega}$ in a rest reference frame, and then converting to a laboratory reference frame, we obtain:
\begin{equation}
    \label{eq:188}
\omega =\gamma (1+n V/c)\left( \, _{\text{tran}}\omega \right)
\end{equation}
\begin{equation}
    \label{eq:189}
\lambda=\frac{n}{ \gamma (n+V/c)\text{  }}\left( \, _{\text{tran}}\lambda \right)
\end{equation}
In the laboratory reference frame, the time required for a beam of wavelength $\lambda$ to fully cross the interface is $ \Delta t=\frac{\lambda }{v-V}$, due to the medium moving with velocity $V$ in the positive direction of the $x^1$ axis. Thus, in the laboratory reference frame, the wavelength of the outgoing beam is $\, _{\text{out}}\lambda =\frac{ c-V}{v-V}\lambda$. Substituting Eq.\textbf{(}\ref{eq:187}\textbf{)} and Eq.\textbf{(}\ref{eq:189}\textbf{)} into $\frac{ c-V}{v-V}\lambda$, we obtain:
\begin{equation}
    \label{eq:190}
\, _{\text{out}}\lambda =\frac{n c}{\gamma (c +V)}\, \left.(_{\text{tran}}\lambda \right)
\end{equation}
Thus we obtain the volume is $\Omega=\frac{n c}{\gamma (c +V)}\, \left.(_{\text{tran}}\lambda \right)S$ for a beam with an output length of one wavelength in the laboratory reference frame, while the corresponding volume for the beam in the rest frame of the medium is $\Omega_0=n \left.(_{\text{tran}}\lambda \right)S$. According to Eq.\textbf{(}\ref{eq:183}\textbf{)}, obtain $ \overline{\, _{\text{out}}w }=\frac{\gamma ^2 (1+V/c)^2 }{c^2 \mu _0}\overline{\left(\, _{\text{out}}E\right){}^2}$ and $ \overline{w_0 }=\frac{ \overline{\left(\, _{\text{out}}E\right){}^2}}{c^2 \mu _0}$. Thus, based on $\, _{\text{out}}\omega =\frac{\overline{\, _{\text{out}}w} \Omega}{\overline{w_0 }\Omega_0} (\, _{\text{tran}}\omega)$, we obtain:
\begin{equation}
    \label{eq:191}
\, _{\text{out}}\omega =\frac{ \gamma (c+V)}{c}\, \left.(_{\text{tran}}\omega \right)
\end{equation}
From Eq.\textbf{(}\ref{eq:188}\textbf{)} and Eq.\textbf{(}\ref{eq:191}\textbf{)}, we can obtain that when the frequency of electromagnetic waves in a moving medium is $\omega$, the frequency of the exit wave is:
\begin{equation}
    \label{eq:192}
\, _{\text{out}}\omega =\frac{ c+V }{c+n V}\omega
\end{equation}
It can be verified that the above Eq.\textbf{(}\ref{eq:190}\textbf{)} and Eq.\textbf{(}\ref{eq:191}\textbf{)} are completely consistent with the results obtained using four-dimensional wave vector transformation. This is a sufficient indication that the energy-momentum tensor for electromagnetic waves in the medium obtained in this paper is self-consistent and supports the relativistic transformation relation for both velocity and energy. From Eq.\textbf{(}\ref{eq:192}\textbf{)}, it can be seen that when an electromagnetic wave exits from a moving medium, the frequency of the electromagnetic wave before and after exit is not equal, and the frequency of the electromagnetic wave before and after exit is equal only when the medium is stationary.
\section{Theoretical analysis of several important experiments}\label{sec:s11}
\subsection{Jones' experiment of light pressure in a medium in 1951}\label{sec:s11-1}
The first experiment to measure the radiation pressure of light in a medium was conducted by Jones (1951) at the University of Aberdeen\textcolor[rgb]{0.184313725,0.188235294117647,0.564705882}{\cite{JONES}}. Jones attempted to test this prediction by showing that when a mirror is immersed in a medium, the radiation pressure exerted on the mirror will be proportional to the refractive index $n$ of the medium. Later, Jones and Richards improved this experiment and published a paper confirming this effect, with an accuracy of $\sigma =\pm 1.2\%$\textcolor[rgb]{0.184313725,0.188235294117647,0.564705882}{\cite{JONES1954}}. Brevik discussed this experiment theoretically in 1979\textcolor[rgb]{0.184313725,0.188235294117647,0.564705882}{\cite{Brevik1979}}, but his discussion of the experiment was not comprehensive enough.\par
Due to the fact that the medium used in the experiment has $\mu\approx\mu_0$, the effects of polarization and magnetization are relatively small. Therefore, we can use the energy-momentum tensor of electromagnetic waves defined based on Eq.\textbf{(}\ref{eq:16}\textbf{)} and given by Eq.\textbf{(}\ref{eq:41}\textbf{)} to analyze Jones' photopressure experiment. For the Jones experiment, we assume that the amplitude of the electric field strength of the beam propagating in a liquid medium is $\, _{\text{tran}}E_0$, and we assume that the reflectivity of the rhodium-plated reflector is $R$. We take the incident light to be incident along the $x^1$ axis in the horizontal direction, that is, the wave vector direction of the incident wave is $(\, _ke)=(1,0,0)$. When the light hits the reflector, the deflection angle of the reflector is $\theta$ and the normal direction of the reflector becomes $(\, _re)=(-cos\theta,0,-sin\theta)$. Therefore, the wave vector direction of the reflected beam is $(\, _{rek}e) =(-cos(2\theta),0,-sin(2\theta))$. The relationship between incident and reflected waves should satisfy the boundary conditions of the metal interface. Therefore, according to Brevik's paper\textcolor[rgb]{0.184313725,0.188235294117647,0.564705882}{\cite{Brevik1979}}, we can obtain that the electromagnetic field tensor in the liquid at the interface $x^1=0$ and $x^3=0$ infinitely close to the rhodium plated reflector in the experiment is:
\begin{equation}
    \label{eq:193}
(\, _{\text{in}}F^{\mu  \nu })=\left( \begin{array}{cccc}  0 & 0 & 1 & 0 \\  0 & 0 & n & 0 \\  -1 & -n & 0 & 0 \\  0 & 0 & 0 & 0 \end{array} \right)\frac{\, _{\text{tran}}E_0}{c} \sin (- \frac{\omega x^0}{c})
\end{equation}
\begin{equation}
    \label{eq:194}
\begin{array}{ll}(\, _{\text{re}}F^{\mu  \nu })=\left( \begin{array}{cccc}  0 & 0 & -1 & 0 \\  0 & 0 & n \cos(2\theta) & 0 \\  1 & -n\cos(2\theta) & 0 & n \sin(2\theta) \\  0 & 0 & -n \sin(2\theta) & 0 \end{array} \right)\\
\text{ }\text{ }\text{ }\text{ }\text{ }\text{ }\text{ }\text{ }\text{ }\text{ }\text{ }\text{ }\text{ }\times \frac{\, _{\text{tran}}E_0}{c} \sqrt{R}\sin (- \frac{\omega x^0}{c}+\delta )\end{array}
\end{equation}
Among them, $n$ is the refractive index of the liquid medium used in the experiment, $\, _{\text{in}}F^{\mu  \nu }$ is the electromagnetic field tensor of the incident wave, $\, _{\text{re}}F^{\mu  \nu }$ is the electromagnetic field tensor of the reflected wave, $\, _{\text{tran}}E_0$ is the amplitude of the electric field intensity of the electromagnetic wave in the experimental liquid, $R$ is the reflection coefficient at the interface, and for rhodium metal, there are $\text{tan} \delta\approx-0.1$ and $R\approx70\%$. At infinitely close to the interface, $x^1=0$ and $x^3=0$, and the total electromagnetic field tensor is $F^{\mu  \nu }=\, _{\text{in}}F^{\mu  \nu }+\, _{\text{re}}F^{\mu  \nu }$. By substituting $F^{\mu  \nu }$ into Eq.\textbf{(}\ref{eq:41}\textbf{)}, we obtain the energy-momentum tensor $(\,_FT^{\mu  \nu })$ of electromagnetic waves at the interface in a liquid. The mean value of its $\,_FT^{11}$ component over a period is:
\begin{equation}
    \label{eq:195}
\begin{array}{ll}\overline{\,_FT^{11}}=\frac{1+n^2+R+2 \sqrt{R} \cos (\delta )\left[-1+n^2 \cos (2 \theta )\right]+n^2 R \cos (4 \theta )}{4 c^2  \mu_0 }\\
\text{ }\text{ }\text{ }\text{ }\text{ }\text{ }\text{ }\text{ }\times (\, _{\text{tran}}E_0)^2\end{array}
\end{equation}
Due to the high reflectivity of rhodium, there is virtually no momentum flow across the interface at this time, so $\overline{\,_FT^{11}}$ can be approximated as the mean pressure $\overline{p ^1}$ acting on the interface . When $\theta=0$ , $\delta=0$ , and $R=1$, the mean pressure is:
\begin{equation}
    \label{eq:196}
\overline{p ^1}=\frac{n^2}{c^2 \mu _0}\left(\, _{\text{tran}}E_0\right)^2=2n\overline{\, _{\text{tran}}S^1}
\end{equation}
This is consistent with the results obtained by Brevik\textcolor[rgb]{0.184313725,0.188235294117647,0.564705882}{\cite{Brevik1979}}, but there are slight differences in other cases. Note that light is incident vertically from the air into the glass, and then vertically into the experimental liquid from the glass. When light enters the experimental liquid from the glass, there is $\, _{\text{tran}}E_0=\chi \left( \, _{\text{in}}E_0\right)$, where $ \, _{\text{in}}E_0$ is the amplitude of the electric field intensity of the beam in the glass, which remains constant throughout the entire experimental process. By according to the boundary conditions, $\chi$ is given by:
\begin{equation}
    \label{eq:197}
\chi =\frac{2 (\,_{\text{gla}}n)}{n+ (\,_{\text{gla}}n)}
\end{equation}
where $  \,_{\text{gla}}n$ is the refractive index of the glass used in the experiment, and we take $ \,_{\text{gla}}n \approx1.5$. We assume that the deflection angle $\theta$ of the reflector is proportional to the mean pressure $\overline{p ^1}$, so we have $\overline{p ^1}=\kappa \theta$. Expand $cos (2 \theta )$ and $cos (4 \theta )$ to the second order and make $\frac{4\kappa c^2 \mu _0}{(\, _{\text{in}}E_0)^2}=k$. After substituting them into Eq.\textbf{(}\ref{eq:195}\textbf{)} and organizing them, we obtain:
\begin{equation}
    \label{eq:198}
\chi ^2 \left(b-a \theta ^2\right)\approx k \theta
\end{equation}
where:\par
$a=4 n^2\left( \sqrt{R}\cos  \delta +2R\right)$\par
$b=1+R-2 \sqrt{R} \cos  \delta +n^2\left(1+R+2 \sqrt{R} \cos  \delta \right)$\par
$ $\\
Solve Eq.\textbf{(}\ref{eq:198}\textbf{)} and take a positive solution to obtain:
\begin{equation}
    \label{eq:199}
\theta= \frac{-k+\sqrt{k^2+4 a b \chi ^4}}{2 a \chi ^2}
\end{equation}
Clearly, the experimental results depend on the value of $k=\frac{4\kappa c^2 \mu _0}{(\, _{\text{in}}E_0)^2}$. The power of the light source used in the experiment is $\, _{\text{in}}P=30W$\textcolor[rgb]{0.184313725,0.188235294117647,0.564705882}{\cite{Brevik1979}}, so we can obtain $ \, _{\text{in}}E_0 =\frac{2}{ {  \,_{\text{gla}}n }+1}\sqrt{c \mu _0\left( \left.\, _{\text{in}}P\right/S\right)}$ based on the boundary conditions of air and glass, where $S$ is the cross-sectional area of the light beam emitted by the light source. The reflectivity of rhodium metal is approximately $R\approx0.7$. According to the data provided by Brevik, the phase factor $\delta$ can be estimated using $\delta =\text{arctan}(-0.1)$\textcolor[rgb]{0.184313725,0.188235294117647,0.564705882}{\cite{Brevik1979}}.\par
\begin{figure}[h]
\centering
\includegraphics[scale=0.95]{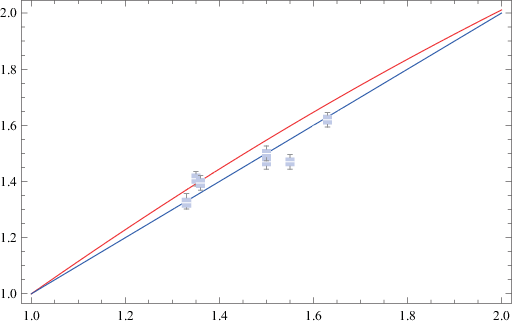}
\caption{Change curve of $\, _n\theta/\, _{\text{air}}\theta$ with $n$. The red line is $\, _n\theta/\, _{\text{air}}\theta$, and the blue line is $n$. Where $\kappa =1Pa/rad$ and $S=0.01m^2$. Box-plot is Jones' experimental data.  }
\label{fig:1}
\end{figure}
\begin{figure}[h]
\centering
\includegraphics[scale=0.95]{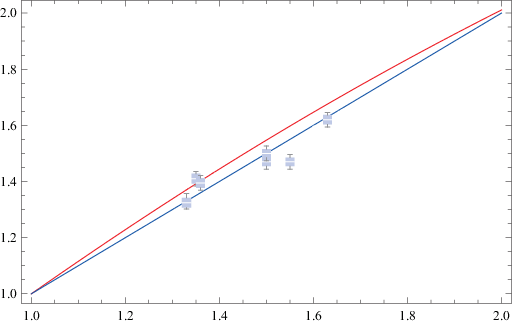}
\caption{Change curve of $\, _n\theta/\, _{\text{air}}\theta$ with $n$. The red line is $\, _n\theta/\, _{\text{air}}\theta$, and the blue line is $n$. Where $\kappa =300Pa/rad$ and $S=0.002m^2$. Box-plot is Jones' experimental data. }
\label{fig:2}
\end{figure}
\begin{figure}[h]
\centering
\includegraphics[scale=0.95]{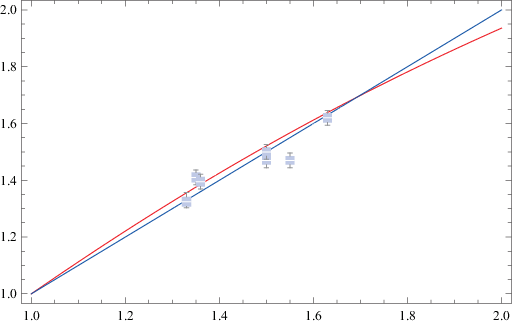}
\caption{Change curve of $\, _n\theta/\, _{\text{air}}\theta$ with $n$. The red line is $\, _n\theta/\, _{\text{air}}\theta$, and the blue line is $n$. Where $\kappa =0.01Pa/rad$ and $S=0.0001m^2$. Box-plot is Jones' experimental data. }
\label{fig:3}
\end{figure}
\begin{figure}[h]
\centering
\includegraphics[scale=0.95]{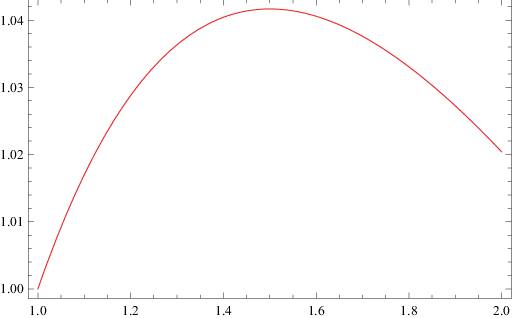}
\caption{The curve of $\overline{\, _{\text{tran}}S^1}/\overline{\, _{\text{air}}S^1}$ changing with refractive index $n$ }
\label{fig:4}
\end{figure}

At present, the parameters we have not yet determined include the cross-sectional area of the beam $S$, which is related to the light source used in the experiment, and the value $\kappa$, which is related to the elasticity of the fine wire used in the experimental apparatus. We can find that, within certain limits, the experimental results do not depend heavily on these two parameters. We estimateand take $\kappa =1Pa/rad$ and $S=0.01m^2$. Assuming that the deflection angle of the reflector is $\, _n\theta$ when the liquid with refractive index $n$ is used in the experiment and $\, _{\text{air}}\theta$ when the reflector is in air. The refractive index of air is taken as $n\approx1$. The curve of $\, _n\theta/\, _{\text{air}}\theta$ as a function of refractive index $n$ is shown in (\hyperref[fig:1]{fig.1}).
The red line represents the theoretical value curve obtained in this paper, the blue line represents a curve strictly proportional to $n$, and the Box-plot represents the experimental data obtained by Jones. In this case, the maximum relative deviation $ (\frac{\, _n\theta}{\, _{\text{air}}\theta}-n)/n$ between $\, _n\theta/\, _{\text{air}}\theta$ and refractive index $n$ is about $ 3.6\%$, indicating some discrepancy between theoretical and experimental values. This could be related to experimental errors or to the values of $\kappa$ and $S$. However, within a certain range, the theoretical values do not depend on the values of $\kappa$ and $S$. For example, when $\kappa =300Pa/rad$ and $S=0.002m^2$ are taken, there is still the $\, _n\theta/\, _{\text{air}}\theta$ versus refractive index $n$ curve shown in (\hyperref[fig:2]{fig.2}).
It can be seen that, within certain limits, the experimental results are approximately independent of the values of the cross-sectional areas $S$ and $\kappa$. However, the values of $\kappa$ and $S$ that we provide here are artificially estimated, and actual experimental parameters may deviate significantly from our estimates. For example, if we take $\kappa =350Pa/rad$ and $S=0.02m^2$, the curve of $\, _n\theta/\, _{\text{air}}\theta$ as a function of refractive index $n$ is shown in (\hyperref[fig:3]{fig.3}).
It can be seen that in this case the theoretical values are in good agreement with the experimental ones in the range of refractive indices used in the experiment. If Jones's experimental data is accurate enough, it is likely due to the small elastic modulus of the filaments he used. Theoretically, the pressure on the reflector is not strictly proportional to the refractive index $n$. From Eq.\textbf{(}\ref{eq:196}\textbf{)}, it can be seen that the pressure received on the reflector is not only related to the refractive index $n$, but also to the energy flow density of the beam in the liquid. In experiments with different liquids, the energy flow density of the beam in the liquid is not constant, but varies with the refractive index of the liquid used. Therefore, from Eq.\textbf{(}\ref{eq:196}\textbf{)}, it can be seen that the pressure acting on the reflector is not strictly proportional to the refractive index $n$. For electromagnetic waves propagating in experimental liquid media, based on boundary conditions, the mean energy flow density $\overline{\, _{\text{tran}}S^1}=\overline{c(\, _FT^{01})}$ is:
\begin{equation}
    \label{eq:200}
\overline{\, _{\text{tran}}S^1}=\frac{n}{2 c \mu _0}\left(\frac{2\, _{\text{gla}}n}{n+\, _{\text{gla}}n}\right)^2\left(\, _{\text{in}}E_0\right)^2
\end{equation}
So, the curve of $\overline{\, _{\text{tran}}S^1}/\overline{\, _{\text{air}}S^1}$ changing with refractive index $n$ is shown in (\hyperref[fig:4]{fig.4}).\par
\par
\begin{figure}[h]
\centering
\includegraphics[scale=0.95]{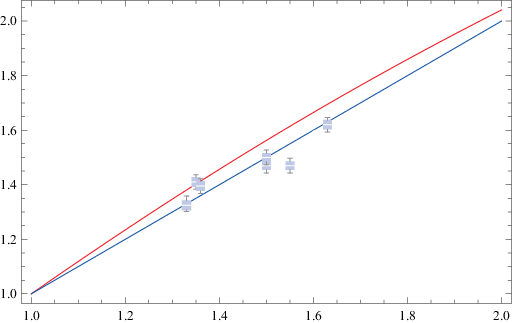}
\caption{The curve of $\, _n\theta/\, _{\text{air}}\theta$ changing with n directly calculated from Eq.\textbf{(}\ref{eq:196}\textbf{)}}
\label{fig:5}
\end{figure}

It follows that $\overline{\, _{\text{tran}}S^1}/\overline{\, _{\text{air}}S^1}$ is approximately 1, but it is not really equal to 1. Substituting Eq.\textbf{(}\ref{eq:200}\textbf{)} into Eq.\textbf{(}\ref{eq:196}\textbf{)} yields $\, _n\theta/\, _{\text{air}}\theta=\overline{\, _np ^1}/\overline{\, _{\text{air}}p ^1}=\frac{n^2 \left(1+\, _{\text{gla}}n\right){}^2}{\left(n+\, _{\text{gla}}n\right){}^2}$. The curve of its variation with refractive index x3 is shown in (\hyperref[fig:5]{fig.5}).
\par
The results do not differ significantly from (\hyperref[fig:1]{fig.1}). According to Brevik's analysis, Eq.\textbf{(}\ref{eq:196}\textbf{)} can be obtained from the Minkowski tensor\textcolor[rgb]{0.184313725,0.188235294117647,0.564705882}{\cite{Brevik1979}}. Now we can see that even if the calculation is performed in terms of the Minkowski tensor, the result obtained is only as shown in (\hyperref[fig:5]{fig.5}), and the difference between it and the experimental value is almost the same as in (\hyperref[fig:1]{fig.1}). Thus, the difference is not caused by using our own energy-momentum tensor instead of the Minkowski tensor. Since it has been recognized that the Minkowski tensor can explain this experiment, the energy-momentum tensor obtained in this paper is equally effective in explaining this experiment.\par
\begin{figure}[h]
\centering
\includegraphics[scale=0.95]{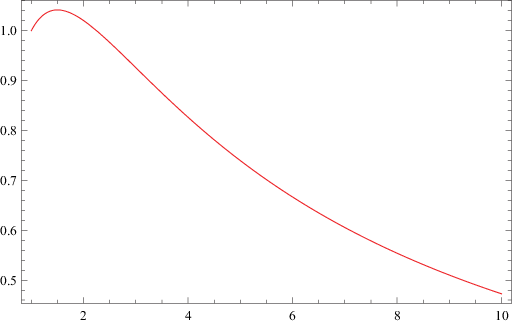}
\caption{$\overline{\, _{\text{tran}}S^1}/\overline{\, _{\text{air}}S^1}$ varies over a wider range with $n$}
\label{fig:6}
\end{figure}

For the Jones experiment, the pressure on the reflector is approximately proportional to the refractive index $n$ only when $\overline{\, _{\text{tran}}S^1}/\overline{\, _{\text{air}}S^1}\approx1$ holds. Brevik did not further analyze the relationship and differences between Eq.\textbf{(}\ref{eq:196}\textbf{)} and experimental results after obtaining Eq.\textbf{(}\ref{eq:196}\textbf{)}. Therefore, he did not reveal that the pressure on the reflector is approximately proportional to the refractive index is closely related to $\overline{\, _{\text{tran}}S^1}/\overline{\, _{\text{air}}S^1}\approx1$. Therefore, Brevik's analysis is not comprehensive enough. $\overline{\, _{\text{tran}}S^1}/\overline{\, _{\text{air}}S^1}\approx1$ is only approximately true in the range of refractive indices used in the experiment, but it does not hold for a wider range of refractive indices, as shown in (\hyperref[fig:6]{fig.6}).
\par
\begin{figure}[h]
\centering
\includegraphics[scale=0.95]{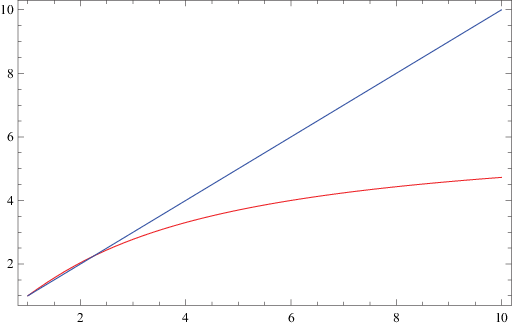}
\caption{$\, _n\theta/\, _{\text{air}}\theta$ varies over a wider range with $n$}
\label{fig:7}
\end{figure}

The $\, _n\theta/\, _{\text{air}}\theta$ also deviates significantly from linear growth over a wider range of refractive indices. See (\hyperref[fig:7]{fig.7}) for an illustration.
\par
We hope that experimental workers can design experiments to verify this conclusion. So, the linear growth result of the Jones experiment is only approximately valid in a certain range of refractive indices. From Eq.\textbf{(}\ref{eq:64}\textbf{)}, it can be seen that electromagnetic waves in a medium can be regarded as a fluid composed of a large number of particles. When this fluid encounters a reflector and is rebounded, not only are the particles of which the fluid is composed rebounded and pressure is applied to the reflector, but the fluid flowing in the medium is naturally accompanied by a pure pressure. The pressure exerted by the particle being rebounded is combined with this "pure" pressure to apply the total pressure to the reflector. Due to the presence of this "pure" pressure, although the mean mechanical momentum of photons in the medium decreases to $1/n$ in vacuum, it works together with the "pure" pressure, resulting in the results observed in Jones' experiment. The "pure" pressure $p$ is caused by the interaction between the beam and the medium, so for this experiment, the influence of the medium's interaction is significant. Therefore, the obtained results tend to support the canonical momentum given by Eq.\textbf{(}\ref{eq:113}\textbf{)}. Within the refractive index range of the medium used in the Jones experiment, the pressure on the reflector is approximately proportional to the refractive index $n$ of the medium.
\subsection{Ashkin's free liquid surface deformation experiment in 1973}\label{sec:s11-2}
Ashkin conducted an experiment in 1973 in which an incident laser beam deformed the surface of a liquid medium\textcolor[rgb]{0.184313725,0.188235294117647,0.564705882}{\cite{1973Radiation}}. He focused a strong laser beam on the water surface and found that the water surface protruded outward under the irradiation of the laser beam. Some people believe that Ashkin's experiment confirms that the momentum of light in a medium is greater than it is in a vacuum. But that may not be the case. On the one hand, as pointed out in Leonhardt's paper published in 2006\textcolor[rgb]{0.184313725,0.188235294117647,0.564705882}{\cite{Leonhardt}}, the beam of light shining on the surface of a liquid is uneven, resulting in a transverse force. This transverse force affects the experimental results of Ashkin, so it cannot be concluded that the momentum of light in the medium is larger than its momentum in vacuum. On the other hand, in Section.\ref{sec:s9} we analyze the Bernoulli effect of the beam and deduce that the pressure increases when the beam enters a high-index medium from a low-index medium. The light in the Ashkin experiment is actually incident from air into water, which has a smaller refractive index than water, so that the pressure of the beam in water is greater than in air. We can generally approximate air as a vacuum. The beam does not have a pressure in vacuum, but from Eq.\textbf{(}\ref{eq:64}\textbf{)} and the analysis of the Bernoulli effect of the beam, it can be seen that when the wavefront of the beam enters the medium, the beam will generate a pressure $p$. This pressure causes the beam to attach a momentum flow in the positive direction of the $x^1$ axis. Due to momentum conservation, it is inevitable that a momentum flow of equal size and opposite direction will be attached to the medium. At this point, the interface of the medium will be subjected to a pressure directed towards the outer part of the medium, causing the interface to protrude outward. This pressure can be given by Eq.\textbf{(}\ref{eq:66}\textbf{)}:
\begin{equation}
    \label{eq:201}
p= \frac{n^2-1}{2 c^2 \mu_0}\left(\, _{\text{tran}}E \right) ^2
\end{equation}
Brevik carried out a detailed analysis of Ashkin's experiment in 1979. In Brevik's analysis, he argued that the so-called Abraham-Minkowski surface force density of light acting on the interface is\textcolor[rgb]{0.184313725,0.188235294117647,0.564705882}{\cite{Brevik1979}}:
\begin{equation}
    \label{eq:202}
p= \frac{n^2-1}{2 c^2 \mu_0}   E  ^2
\end{equation}
Here $E$ is actually the electric field strength $ \, _{\text{tran}}E  $ of the transmitted electromagnetic wave in the medium. As can be seen, our results are in perfect agreement with Brevik's analysis. Due to:
\begin{equation}
    \label{eq:203}
\, _{\text{tran}} E=\frac{2  \mu  }{n  \mu _0 +  \mu }(\, _{\text{in}}E)\approx\frac{2   }{n   + 1 }(\, _{\text{in}}E)
\end{equation}
Substituting Eq.\textbf{(}\ref{eq:203}\textbf{)} into Eq.\textbf{(}\ref{eq:201}\textbf{)} yields:
\begin{equation}
    \label{eq:204}
p=\frac{2(n-1)}{n+1}\frac{\text{  }\, _{\text{in}}E^2}{c^2\mu_0}
\end{equation}
Here $\frac{\text{  }\, _{\text{in}}E^2}{c^2\mu_0}$ is actually $\, _{in}S^1/c$. So this is completely consistent with Eq.\textbf{(}\ref{eq:160}\textbf{)} obtained when we analyzed the Bernoulli effect of the beam in Section.\ref{sec:s9}. So, Ashkin's experimental result can be explained by the Bernoulli effect of the beam. From this it can be seen that Ashkin's experimental result is not due to the fact that the mechanical momentum of light in the medium is larger than in vacuum, but rather to the interaction of light with the medium when it enters, which creates a pressure inside the light. This pressure reacts on the medium, causing the medium to also generate a pressure towards the outside of the medium, causing the interface to protrude outward. Conversely, when a beam of light is emitted from the medium into the free space outside the medium, there is this pressure while the wavefront of the beam is still inside the medium, and this pressure vanishes when the wavefront passes through the medium. This indicates that the momentum flow of the pressure is transferred to the interface of the medium, thus also causing the interface to protrude outward. So, Ashkin's experimental results do not contradict the fact that the mechanical momentum of light in a medium is $1/n$ of that in free space. Of course, we cannot exclude that the transverse forces pointed out by Leonhardt may also have an effect on the experimental results of Ashkin. From the analysis of the Bernoulli effect of a beam, it can be seen that when a beam is incident from a medium with a low refractive index into a medium with a high refractive index, the surface of the medium with a high refractive index will be subjected to a force that causes it to protrude outward. The pressure difference between the two media at the interface is:
\begin{equation}
    \label{eq:205}
\Delta p=[\frac{4 \left(\, _1n\right){}^2 }{ \left(\, _1n+\, _2n\right){}^2}\frac{\left(\, _2n^2-1\right)}{\left(\, _1n^2-1\right) }-1]\left(\, _1p\right)
\end{equation}
It's a relatively complicated relationship. It is hoped that more precise experiments can be carried out to verify whether this relation holds. Since the influence of pressure is the main factor in this experiment, it supports the canonical momentum given by Eq.\textbf{(}\ref{eq:113}\textbf{)}.
\subsection{Weilong's optical fiber deformation experiment in 2008}\label{sec:s11-3}
Weilong conducted an experiment in 2008\textcolor[rgb]{0.184313725,0.188235294117647,0.564705882}{\cite{Weilong2008}} in which he fixed one end of a fiber while the other end was free and then caused a laser beam to be incident from the fixed end and emitted from the free end. Experimentally, the fiber is found to be curved in a bow-shaped shape, indicating that the free end of the fiber is subjected to a force in the opposite direction of light propagation. To analyze this experiment, we will calculate the change of the mechanical momentum of the transmitted beam as it passes through the interface of the medium.\par
If the light transmitted into the medium is a beam of length $d$ and cross-sectional area $S$ in free space before transmission, then from Eq.\textbf{(}\ref{eq:54}\textbf{)} to Eq.\textbf{(}\ref{eq:59}\textbf{)} and after calculation, it can be obtained that in a free space slightly outside the interface that is infinitely close to the interface $x^1=0$, the mean mechanical momentum density $\overline{g^1} =\overline{T^{01}/c}$ of the transmitted beam is:
\begin{equation}
    \label{eq:206}
\overline{g^1}=\frac{4n \mu  }{c^3 \left(n \mu _0+\mu \right){}^2}\overline{\left(\, _{\text{in}}E_0 f\left(-\frac{\omega  x^0}{c}\right)\right) ^2}
\end{equation}
So its mechanical momentum is:
\begin{equation}
    \label{eq:207}
P^1=\frac{4n \mu  }{c^3 \left(n \mu _0+\mu \right){}^2}S d\overline{\left(\, _{\text{in}}E_0 f\left(-\frac{\omega  x^0}{c}\right)\right) ^2}
\end{equation}
According to Eq.\textbf{(}\ref{eq:63}\textbf{)} and Eq.\textbf{(}\ref{eq:50}\textbf{)}, when this beam of light is transmitted into the medium, the mean mechanical momentum density $\overline{\, _Fg^1}=\overline{\, _FT^{01}/c}$ of the transmitted beam is:
\begin{equation}
    \label{eq:208}
\overline{\, _Fg^1}=\frac{4 n \mu ^2}{c^3 \mu _0 \left(n \mu _0+\mu \right){}^2}\overline{\left(\, _{\text{in}}E_0 f\left(-\frac{\omega  x^0}{c}\right)\right) ^2}
\end{equation}
For $\mu\approx\mu_0$, the mechanical momentum density of the beam before crossing the interface is approximately equal to the mechanical momentum density after crossing the interface. However, a beam of length $d$ before crossing the interface becomes shorter after crossing the interface due to the slower wave speed. After passing through the interface, the length of the beam becomes $d/n$. So, the mechanical momentum of this beam after crossing the interface becomes:
\begin{equation}
    \label{eq:209}
\, _FP^1 =\frac{4 n \mu ^2}{c^3 \mu _0 \left(n \mu _0+\mu \right){}^2}\frac{Sd}{n}\overline{\left(\, _{\text{in}}E_0 f\left(-\frac{\omega  x^0}{c}\right)\right) ^2}\approx P^1/n
\end{equation}
In this way, the momentum difference between the beam before and after crossing the interface is $\Delta  P^1\approx(1-1/n)P^1$. This will inevitably result in the medium gaining momentum $\Delta  P^1$ due to momentum conservation. However, the medium is generally kept at rest during the experiment, so the external environment must exert a force on the medium with an mean size of:
\begin{equation}
    \label{eq:210}
\overline{f^1}=\frac{\Delta  P^1}{\Delta  t}\approx\frac{4(n-1)}{c^2 \mu _0(1+n)^2 }S\overline{\left(\, _{\text{in}}E_0 f\left(-\frac{\omega  x^0}{c}\right)\right) ^2}
\end{equation}
The direction of this force is opposite to the direction of propagation of the beam. $\Delta t=d/c$ is applied here. Similarly, when a beam of light is emitted into free space from a medium, it can be calculated that within the medium, infinitely close to the interface, there are:
\begin{equation}
    \label{eq:211}
\bar{\rho }=\frac{n^2\left(\mu _0^2-\mu ^2\right)}{c^4 \mu _0 \left(\mu +n \mu _0\right){}^2}\left(\, _{\text{tran}}E\right){}^2\approx 0
\end{equation}
\begin{equation}
    \label{eq:212}
\bar{p}=\frac{n^2\left(\mu _0^2-\mu ^2\right)}{c^2 \mu _0 \left(\mu +n \mu _0\right){}^2}\left(\, _{\text{tran}}E\right){}^2\approx 0
\end{equation}
\begin{equation}
    \label{eq:213}
v=\frac{\mu }{\mu _0}c\approx c
\end{equation}
Where $\, _{\text{tran}}E=\, _{\text{tran}}E_0 f\left(-\frac{\omega  x^0}{c}\right)$. It can be seen that in a medium that is infinitely close to the interface, the mechanical quantities of light begin to approximately recover to be consistent with those in free space. After the calculation, it can be concluded that the mean mechanical momentum density of the beam about to cross the interface in the medium is:
\begin{equation}
    \label{eq:214}
\overline{\, _Fg^1}=\frac{4 n^2\mu \text{  }}{c^3 \left(n \mu _0+\mu \right){}^2}\overline{\left(\, _{\text{tran}}E\right)^2}
\end{equation}
According to calculations, the mean momentum density of light that has already emitted from the interface into free space is:
\begin{equation}
    \label{eq:215}
\overline{g^1}=\frac{4n^2 \mu _0}{c^3 \left(n \mu _0 +\mu\right){}^2}\overline{\left(\, _{\text{tran}}E\right) ^2}
\end{equation}
It follows that when $\mu\approx\mu_0$, there is also $\overline{\, _Fg^1}\approx\overline{g^1}$. Since the propagation speed of light in a medium is $1/n$ of that in free space, when this beam has a length of $d$ in free space, its length in the medium is $d/n$. Thus, the mechanical momentum is:
\begin{equation}
    \label{eq:216}
\, _FP^1=\frac{4n \mu }{c^3 \left(\mu +n \mu _0\right)^2}Sd\overline{\left(\, _{\text{tran}}E\right)^2}
\end{equation}
When light is emitted into free space, its mechanical momentum becomes:
\begin{equation}
    \label{eq:217}
P^1=\frac{4n^2 \mu _0 }{c^3 \left(\mu +n \mu _0\right){}^2}\overline{\left(\, _{\text{tran}}E\right)^2}Sd\approx n\left(\, _FP^1\right)
\end{equation}
It can be seen that when a beam of light is emitted from the medium into free space, the mechanical momentum of the beam becomes approximately $n$ times larger than its mechanical momentum in the medium. The momentum difference due to the beam emitted from the medium into free space is $\Delta P^1=P^1-\, _FP^1=(n-1)\, _FP^1$, so in this case the medium must be subjected to a force. Using $\Delta t=d/c$, the mean magnitude of this force is:
\begin{equation}
    \label{eq:218}
\overline{f^1}=\frac{\Delta  P^1}{\Delta  t}\approx\frac{4n(n-1) }{c^2 \mu _0(1+n)^2}S\overline{\left(\, _{\text{tran}}E\right)^2}
\end{equation}
Here, we have not calculated the force exerted on the fiber due to the reflection of light. According to Eq.\textbf{(}\ref{eq:63}\textbf{)}, the mean total mechanical momentum density $\overline{\,_{aF}g^1}$ of a beam propagating in a medium is:
\begin{equation}
    \label{eq:219}
\overline{\,_{aF}g^1}=\frac{ n}{c^3 \mu _0}\overline{\left(\, _{\text{tran}}E\right) ^2}
\end{equation}
Thus the mean mechanical momentum $\,_{re}P^1=(\overline{\,_{aF}g^1}-\overline{\, _Fg^1})Sd/n$ of the reflected light at the interface of the medium is:
\begin{equation}
    \label{eq:220}
\,_{re}P^1\approx\frac{(n-1)^2 }{c^3 \mu _0(n+1)^2 }Sd\overline{\left(\, _{\text{tran}}E\right) ^2}
\end{equation}
According to $\overline{\,_{re}f^1}=\frac{2(\,_{re}P^1)}{\Delta t}$, the force exerted on the free end of the fiber due to reflection is:
\begin{equation}
    \label{eq:221}
\overline{\,_{re}f^1}\approx\frac{2(n-1)^2 }{c^2 \mu _0(n+1)^2 }S\overline{\left(\, _{\text{tran}}E\right) ^2}
\end{equation}
Due to the fact that $\overline{\,_{re}f^1}$ is oriented in the same direction as the propagation direction of the light, while $\overline{f^1}$ is oriented in the opposite direction, the total force $\overline{\,_{all}f^1}=\overline{ f^1}-\overline{\,_{re}f^1}$ exerted on the fiber as the beam exits the fiber into free space is:
\begin{equation}
    \label{eq:222}
\overline{\,_{all}f^1}\approx\frac{2 (n-1)}{c^2 \mu _0(n+1) }S\overline{\left(\, _{\text{tran}}E\right) ^2}
\end{equation}
In the fiber, the mean power of light is $\overline{ \, _{\text{tran}}P }=\frac{S\overline{\left(\, _{\text{tran}}E\right) ^2}}{c^2 \mu _0} c n $, so that:
\begin{equation}
    \label{eq:223}
\overline{\,_{all}f^1}\approx\frac{2 (n-1)}{ c n(n+1)  } \overline{ \, _{\text{tran}}P }
\end{equation}
Thus the total force acting on the fiber is opposite to the direction of light propagation. Our results are consistent with those of Weilong, but the energy-momentum tensor we use is not the Abraham tensor. From the experiments of Weilong, it is seen that when the power of light exceeds a certain value, the force acting on the free end of the fiber causes the fiber to bend in a bow-like shape\textcolor[rgb]{0.184313725,0.188235294117647,0.564705882}{\cite{Weilong2008}}. On the contrary, if the mechanical momentum of light emitted from the medium decreases in free space, the force acting on the medium will be in the same direction as the propagation of the beam, and the effect of the force is to make the fiber more straight rather than bent. Thus, by looking at the deformation of the fiber as the beam is emitted from the fiber into free space, it is possible to determine whether the mechanical momentum of the beam is larger in the medium or in free space.We believe that Weilong's experiment confirmed that the mechanical momentum of a beam in a medium is smaller than that in free space. Due to the fact that mixing with light propagates mechanical momentum rather than canonical momentum, this experiment is not affected by the "pure" pressure of the beam. Therefore, the experimental results support the mechanical momentum given by Eq.\textbf{(}\ref{eq:114}\textbf{)}.
\subsection{Frequency shift measurement experiment}\label{sec:s11-4}
\begin{figure}[h]
\centering
\includegraphics[scale=0.6]{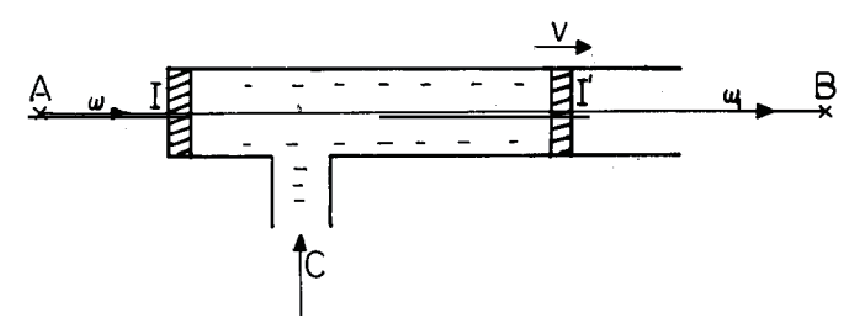}
\caption{The experimental setup for frequency shift measurement experiment}
\label{fig:8}
\end{figure}

The experimental setup for frequency shift measurement experiment is shown in (\hyperref[fig:8]{fig.8}).
\par
The experimental beam enters at the fixed end $I$ and exits at the mobile end $I'$. Between $IC$, it can be seen as a stationary medium with refractive index $n$; between $CI'$, it can be seen as a medium with refractive index $n$ moving at rate $V$. The result of the experiment is that the angular frequency received by the receiver is $\omega _1\approx[1-(n -1)\beta ] \omega<\omega$, where $\beta=V/c$. This experiment can be explained by the coordinate transformation of four-dimensional wave vectors. Brevik analyzed this experiment\textcolor[rgb]{0.184313725,0.188235294117647,0.564705882}{\cite{Brevik1979}} and believes that it can also be explained by energy-momentum tensors, and believes that it confirms the applicability of the Minkowski tensor. Brevik also argued that it would be complicated to interpret this experiment in terms of other energy-momentum tensors. Brevik argued that the four-dimensional momentum of a beam in a laboratory medium can be expressed as:
\begin{equation}
    \label{eq:224}
\left(P^{\mu }\right)=\left(\frac{N \hbar  \omega }{c},\frac{N n \hbar  \omega }{c},0,0\right)
\end{equation}
Here N is the mean number of photons in the beam. After transforming to the moving reference frame, the time component of the four-dimensional momentum is given by:
\begin{equation}
    \label{eq:225}
P'^0=\frac{N (c-n V) \gamma  \hbar  \omega  }{c^2}
\end{equation}
Brevik argued that in the moving reference frame, when the beam is emitted from the medium and returns to free space, the time component $P'^0$ of this four-dimensional momentum will remain the same, but the momentum will become $P'^1=P'^0$. As a result, the four-dimensional momentum of the beam emitted into free space becomes:
\begin{equation}
    \label{eq:226}
\left(P'^{\mu }\right)=\left(\frac{N (c-nV) \gamma  \hbar  \omega  }{c^2},\frac{N (c-nV) \gamma  \hbar  \omega }{c^2},0,0\right)
\end{equation}
Translating it back into the laboratory reference frame, we obtain:
\begin{equation}
    \label{eq:227}
\begin{array}{ll}\left(P^{\mu }\right)=\left(\frac{ N(1-n \beta )(1+\beta ) \gamma ^2\hbar  \omega }{c},\frac{ N(1-n \beta )(1+\beta ) \gamma ^2\hbar  \omega }{c},0,0\right)\end{array}
\end{equation}
So, the angular frequency received by the receiver is:
\begin{equation}
    \label{eq:228}
 \omega _1=(1-n \beta )(1+\beta ) \gamma ^2 \omega\approx[1-(n-1)\beta ] \omega
\end{equation}
Thus, Brevik argues that this experiment confirms that the Minkowski tensor is more applicable in this respect than other forms of energy-momentum tensors. This, however, is not the case. In Section.\ref{sec:s10}, we discuss the energy-momentum tensor of a beam in a moving medium and analyze the frequency conversion relation of the beam in the moving medium in terms of the energy-momentum tensor. The conclusion of this paper is that the energy-momentum tensor of electromagnetic waves in a medium is self-consistent, which supports the relativistic transformation relation of velocity and energy, the resulting frequency conversion relations are consistent with those obtained using four-dimensional wave vectors. We will now apply the results obtained in Section.\ref{sec:s10} to analyze the frequency shift experiment here. For segment $IC$, the medium is stationary in the laboratory reference frame, so the angular frequency of the beam in this segment is the same as the angular frequency of the incident wave, both of which are $\omega_0$. For segment $CI'$, it is in motion relative to the laboratory reference frame. Therefore, we first need to perform a coordinate transformation to transform it to the rest reference frame of the medium, at which point the $IC$ segment becomes moving with velocity $-V$. Therefore, according to Eq.\textbf{(}\ref{eq:188}\textbf{)}, the angular frequency is:
\begin{equation}
    \label{eq:229}
 \omega _0'=\gamma (1-n V/c) \omega _0
\end{equation}
Due to the fact that the $CI'$ segment is stationary in this case, the angular frequency of light incident from the $IC$ segment to the $CI'$ segment remains constant. So, in the rest reference frame of the medium, the angular frequency of the light in the $CI'$ segment is $ \, _{\text{tran}}\omega=\gamma (1-n V/c) \omega _0$. In this case, the $CI'$ segment moves at the speed $V$ in the laboratory reference frame. According to Eq.\textbf{(}\ref{eq:188}\textbf{)} again, in the laboratory reference system, the angular frequency of light in the $CI'$ segment is:
\begin{equation}
    \label{eq:230}
  \omega =\gamma^2 (1+n V/c) (1-n V/c) \omega _0
\end{equation}
Substituting Eq.\textbf{(}\ref{eq:230}\textbf{)} into Eq.\textbf{(}\ref{eq:192}\textbf{)} yields the angular frequency received by the receiver as:
\begin{equation}
    \label{eq:231}
\,_{out}\omega =(1-n V/c)(1+ V/c)  \gamma^2 \omega _0
\end{equation}
It is completely consistent with Eq.\textbf{(}\ref{eq:228}\textbf{)}. As can be seen, the energy-momentum tensor for electromagnetic waves in media obtained in this paper is fully applicable to the analysis of such frequency shift experiments. In addition, the four-dimensional momentum represented by Eq.\textbf{(}\ref{eq:224}\textbf{)} does not satisfy the relativistic relationship of $E=\gamma E_0$ when its corresponding energy undergoes Lorentz transformation in different inertial frames. This is also understandable because the momentum given by Eq.\textbf{(}\ref{eq:224}\textbf{)} is actually canonical momentum, which includes the contribution of the interaction term, so it no longer satisfies the $E=\gamma E_0$ relationship, but it is consistent with the transformation relationship of the four-dimensional wave vector. The energy corresponding to the mechanical momentum satisfies the $E=\gamma E_0$ relation under the Lorentz transformation, but it differs from the transformation relation for the four-dimensional wave vector. This experiment and the analysis presented in this paper will help to further understand the connection and differences between mechanical momentum and canonical momentum.
\subsection{Other experiments}\label{sec:s11-5}
In the literature\textcolor[rgb]{0.184313725,0.188235294117647,0.564705882}{\cite{Brevik1979}}, some experiments on static electromagnetic fields such as "Condenser partially immersed in a dielectric liquid" were also analyzed. Since in Section.\ref{sec:s7} we argued that it is appropriate to use the Abraham tensor for the analysis of static electromagnetic fields, we will not repeat the analysis of these experiments here. Leonhardt's 2006 papere\textcolor[rgb]{0.184313725,0.188235294117647,0.564705882}{\cite{Leonhardt}} also introduced Campbell et al.'s experiment to measure the momentum transfer of light using the precise tools of atomic optics. Experimental results support Minkowski momentum. Leonhardt argues that further research is needed on when the Minkowski momentum and the Abraham momentum apply, and light will continue to surprise people\textcolor[rgb]{0.184313725,0.188235294117647,0.564705882}{\cite{Leonhardt}}. Leonhardt argued that the fundamental pillar behind the Minkowski momentum is quantum mechanics, while the pillar behind the Abraham tensor is relativity. Based on the previous discussion in this paper, our point of view is that the momentum based on quantum mechanics is the canonical momentum and the momentum based on relativity is the mechanical momentum. They are two types of momenta with different properties. Canonical momenta include contributions from interactions with the medium, while mechanical momenta do not. So, we believe that when the momentum detected in the experiment includes the contribution of the interaction with the medium, that is, the pressure $p$ significantly affects the experimental results, the experimental results will tend to show support for Minkowski momentum, that is, canonical momentum.  On the contrary, when the momentum detected in the experiment does not include the contribution of the interaction with the medium, that is, the influence of pressure $p$ on the experimental results can be ignored, the experimental results will tend to show support for the Abraham momentum, that is, mechanical momentum. Of course, some experimental phenomena that belong purely to the medium, such as electrostriction or magnetostriction, do not fall within the description of the energy-momentum tensor given in this paper.
\section{Conclusion}\label{sec:s12}
Although the form of the energy-momentum tensor may not be unique, we believe that a unified definition of the energy-momentum tensor of a material field using Eq.\textbf{(}\ref{eq:16}\textbf{)} is a better choice, as only this form of definition is consistent with the right side of the Einstein field equations, and the physical meaning on the right side of the Einstein field equations must be the energy-momentum tensor of the material field. The energy-momentum tensor of macroscopic pure electromagnetic fields in a medium defined based on Eq.\textbf{(}\ref{eq:16}\textbf{)} is given by Eq.\textbf{(}\ref{eq:41}\textbf{)}. According to the boundary conditions, from the perspective of the energy and momentum flow densities input from the outside into the medium, for non-ferromagnetic media, the energy and momentum flow densities of macroscopic pure electromagnetic waves given by Eq.\textbf{(}\ref{eq:41}\textbf{)} account for the vast majority of the total energy and momentum flow densities of the input. This makes it possible to reasonably explain the relevant experiments in general using only the energy-momentum tensor of macroscopic pure electromagnetic waves in the medium given by Eq.\textbf{(}\ref{eq:41}\textbf{)}. This is our new discovery. From Eq.\textbf{(}\ref{eq:64}\textbf{)}, it can be seen that electromagnetic waves in a medium can be regarded as a fluid composed of a large number of photons. In addition to the momentum of the photons that make up the fluid, it also has a "pure" pressure related to the polarization of the electromagnetic waves, and this pressure also has the Bernoulli effect. At the same time, this pressure is also applied to the side of the beam. Linking the energy-momentum tensor of electromagnetic waves to that of ordinary fluids is our important innovation. From the fluid point of view, we find that the Abraham momentum is the mechanical momentum, which does not take into account the contribution of the interaction term between the macroscopic pure electromagnetic wave and the medium. When considering the contribution of the interaction term between macroscopic pure electromagnetic waves and media, we obtain a new energy-momentum tensor Eq.\textbf{(}\ref{eq:103}\textbf{)} that is different from the Minkowski tensor. According to the energy-momentum tensor given by Eq.\textbf{(}\ref{eq:103}\textbf{)}, we find that Minkowski momentum is a canonical momentum that considers the contribution of the interaction term between macroscopic pure electromagnetic waves and the medium. We also found that according to Eq.\textbf{(}\ref{eq:103}\textbf{)}, we can simultaneously obtain canonical momentum consistent with Minkowski momentum and mechanical momentum consistent with Abraham momentum. This is also our important new discovery. We have discussed the energy-momentum tensor of a non-pure macroscopic electromagnetic field when the effects of polarization energy and magnetization of the medium are taken into account, and provided connections and differences between it and the energy-momentum tensor of a macroscopic pure electromagnetic field. We also discuss the energy-momentum tensor of electromagnetic waves in moving media and find that the energy-momentum tensor obtained in this paper is highly self-consistent with relativity, further clarifying the connection and distinction between mechanical momentum and canonical momentum. Based on the energy-momentum tensor we have obtained in this paper, it can explain both experiments supporting the Minkowski momentum and those supporting the Abraham momentum. We found that when the "pressure" given in this article significantly affects the experimental results, the experiment will support Minkowski momentum, namely canonical momentum. When the influence of the "pressure" given in this article on the experimental results is negligible, the experiment will support the Abraham momentum, namely mechanical momentum. According to Section.\ref{sec:s8}, there is also a pressure on the side of the beam, which is related to the polarization of the light.This is one of the important differences between the energy-momentum tensor we obtain and the Minkowski and Abraham tensors. We hope that experimental workers will be able to design experiments to verify the presence of beam pressure on its side. The findings of this paper may shed new light on the application of light.

\bibliographystyle{apsrev4-1}
\bibliography{reference}

\begin{thebibliography}{43}%
\makeatletter
\providecommand \@ifxundefined [1]{%
 \@ifx{#1\undefined}
}%
\providecommand \@ifnum [1]{%
 \ifnum #1\expandafter \@firstoftwo
 \else \expandafter \@secondoftwo
 \fi
}%
\providecommand \@ifx [1]{%
 \ifx #1\expandafter \@firstoftwo
 \else \expandafter \@secondoftwo
 \fi
}%
\providecommand \natexlab [1]{#1}%
\providecommand \enquote  [1]{``#1''}%
\providecommand \bibnamefont  [1]{#1}%
\providecommand \bibfnamefont [1]{#1}%
\providecommand \citenamefont [1]{#1}%
\providecommand \href@noop [0]{\@secondoftwo}%
\providecommand \href [0]{\begingroup \@sanitize@url \@href}%
\providecommand \@href[1]{\@@startlink{#1}\@@href}%
\providecommand \@@href[1]{\endgroup#1\@@endlink}%
\providecommand \@sanitize@url [0]{\catcode `\\12\catcode `\$12\catcode
  `\&12\catcode `\#12\catcode `\^12\catcode `\_12\catcode `\%12\relax}%
\providecommand \@@startlink[1]{}%
\providecommand \@@endlink[0]{}%
\providecommand \url  [0]{\begingroup\@sanitize@url \@url }%
\providecommand \@url [1]{\endgroup\@href {#1}{\urlprefix }}%
\providecommand \urlprefix  [0]{URL }%
\providecommand \Eprint [0]{\href }%
\providecommand \doibase [0]{http://dx.doi.org/}%
\providecommand \selectlanguage [0]{\@gobble}%
\providecommand \bibinfo  [0]{\@secondoftwo}%
\providecommand \bibfield  [0]{\@secondoftwo}%
\providecommand \translation [1]{[#1]}%
\providecommand \BibitemOpen [0]{}%
\providecommand \bibitemStop [0]{}%
\providecommand \bibitemNoStop [0]{.\EOS\space}%
\providecommand \EOS [0]{\spacefactor3000\relax}%
\providecommand \BibitemShut  [1]{\csname bibitem#1\endcsname}%
\let\auto@bib@innerbib\@empty
\bibitem [{\citenamefont {Brevik}(1979)}]{Brevik1979}%
  \BibitemOpen
  \bibfield  {author} {\bibinfo {author} {\bibfnamefont {I.}~\bibnamefont
  {Brevik}},\ }\href {\doibase 10.1016/0370-1573(79)90074-7} {\bibfield
  {journal} {\bibinfo  {journal} {Physics Reports}\ }\textbf {\bibinfo {volume}
  {52}},\ \bibinfo {pages} {133} (\bibinfo {year} {1979})}\BibitemShut
  {NoStop}%
\bibitem [{\citenamefont {Pfeifer}\ \emph {et~al.}(2007)\citenamefont
  {Pfeifer}, \citenamefont {Nieminen}, \citenamefont {Heckenberg},\ and\
  \citenamefont {Rubinsztein-Dunlop}}]{Pfeifer}%
  \BibitemOpen
  \bibfield  {author} {\bibinfo {author} {\bibfnamefont {R.~N.~C.}\
  \bibnamefont {Pfeifer}}, \bibinfo {author} {\bibfnamefont {T.~A.}\
  \bibnamefont {Nieminen}}, \bibinfo {author} {\bibfnamefont {N.~R.}\
  \bibnamefont {Heckenberg}}, \ and\ \bibinfo {author} {\bibfnamefont
  {H.}~\bibnamefont {Rubinsztein-Dunlop}},\ }\href {\doibase
  10.1103/RevModPhys.79.1197} {\bibfield  {journal} {\bibinfo  {journal}
  {Reviews of Modern Physics}\ }\textbf {\bibinfo {volume} {79}},\ \bibinfo
  {pages} {1197} (\bibinfo {year} {2007})}\BibitemShut {NoStop}%
\bibitem [{\citenamefont {U.Leonhardt}(2006)}]{Leonhardt}%
  \BibitemOpen
  \bibfield  {author} {\bibinfo {author} {\bibnamefont {U.Leonhardt}},\ }\href
  {\doibase 10.1038/444823a} {\bibfield  {journal} {\bibinfo  {journal}
  {Nature}\ }\textbf {\bibinfo {volume} {444}},\ \bibinfo {pages} {823}
  (\bibinfo {year} {2006})}\BibitemShut {NoStop}%
\bibitem [{\citenamefont {Obukhov}(2022)}]{Obukhov}%
  \BibitemOpen
  \bibfield  {author} {\bibinfo {author} {\bibfnamefont {Y.~N.}\ \bibnamefont
  {Obukhov}},\ }\href {\doibase 10.48550/arXiv.2208.10951} {\  (\bibinfo {year}
  {2022}),\ 10.48550/arXiv.2208.10951}\BibitemShut {NoStop}%
\bibitem [{\citenamefont {Crenshaw}(2023)}]{Michael}%
  \BibitemOpen
  \bibfield  {author} {\bibinfo {author} {\bibfnamefont {M.~E.}\ \bibnamefont
  {Crenshaw}},\ }\href {\doibase 10.48550/arXiv.2211.09871} {\  (\bibinfo
  {year} {2023}),\ 10.48550/arXiv.2211.09871}\BibitemShut {NoStop}%
\bibitem [{\citenamefont {Yaghjian}(2023)}]{Arthur}%
  \BibitemOpen
  \bibfield  {author} {\bibinfo {author} {\bibfnamefont {A.~D.}\ \bibnamefont
  {Yaghjian}},\ }\href {\doibase 10.48550/arXiv.2210.12267} {\  (\bibinfo
  {year} {2023}),\ 10.48550/arXiv.2210.12267}\BibitemShut {NoStop}%
\bibitem [{\citenamefont {Ortega-Gomez}\ \emph {et~al.}(2023)\citenamefont
  {Ortega-Gomez}, \citenamefont {M.Lobet}, \citenamefont {J.E.V.Lozano},\ and\
  \citenamefont {I.Liberal}}]{Angel}%
  \BibitemOpen
  \bibfield  {author} {\bibinfo {author} {\bibfnamefont {A.}~\bibnamefont
  {Ortega-Gomez}}, \bibinfo {author} {\bibnamefont {M.Lobet}}, \bibinfo
  {author} {\bibnamefont {J.E.V.Lozano}}, \ and\ \bibinfo {author}
  {\bibnamefont {I.Liberal}},\ }\href {\doibase 10.48550/arXiv.2301.03333} {\
  (\bibinfo {year} {2023}),\ 10.48550/arXiv.2301.03333}\BibitemShut {NoStop}%
\bibitem [{\citenamefont {Koivurova}\ \emph {et~al.}(2023)\citenamefont
  {Koivurova}, \citenamefont {Robson},\ and\ \citenamefont
  {Ornigotti}}]{Matias}%
  \BibitemOpen
  \bibfield  {author} {\bibinfo {author} {\bibfnamefont {M.}~\bibnamefont
  {Koivurova}}, \bibinfo {author} {\bibfnamefont {C.~W.}\ \bibnamefont
  {Robson}}, \ and\ \bibinfo {author} {\bibfnamefont {M.}~\bibnamefont
  {Ornigotti}},\ }\href {\doibase 10.48550/arXiv.2304.08108} {\  (\bibinfo
  {year} {2023}),\ 10.48550/arXiv.2304.08108}\BibitemShut {NoStop}%
\bibitem [{\citenamefont {Minkowski}(1908)}]{Minkowski1}%
  \BibitemOpen
  \bibfield  {author} {\bibinfo {author} {\bibfnamefont {H.}~\bibnamefont
  {Minkowski}},\ }\href@noop {} {\bibfield  {journal} {\bibinfo  {journal}
  {Math Phys}\ ,\ \bibinfo {pages} {53}} (\bibinfo {year} {1908})}\BibitemShut
  {NoStop}%
\bibitem [{\citenamefont {Minkowski}(1910)}]{Minkowski2}%
  \BibitemOpen
  \bibfield  {author} {\bibinfo {author} {\bibfnamefont {H.}~\bibnamefont
  {Minkowski}},\ }\href {\doibase 10.1007/BF01455871} {\bibfield  {journal}
  {\bibinfo  {journal} {Mathematische Annalen}\ }\textbf {\bibinfo {volume}
  {68}},\ \bibinfo {pages} {472} (\bibinfo {year} {1910})}\BibitemShut
  {NoStop}%
\bibitem [{\citenamefont {M.Abraham}(1909)}]{Abraham1}%
  \BibitemOpen
  \bibfield  {author} {\bibinfo {author} {\bibnamefont {M.Abraham}},\ }\href
  {\doibase 10.1007/bf03018208} {\bibfield  {journal} {\bibinfo  {journal}
  {Rendiconti Del Circolo Matematico Di Palermo}\ }\textbf {\bibinfo {volume}
  {28}},\ \bibinfo {pages} {1} (\bibinfo {year} {1909})}\BibitemShut {NoStop}%
\bibitem [{\citenamefont {M.Abraham}(1910)}]{Abraham2}%
  \BibitemOpen
  \bibfield  {author} {\bibinfo {author} {\bibnamefont {M.Abraham}},\
  }\href@noop {} {\bibfield  {journal} {\bibinfo  {journal} {Rend Circ Mat
  Palermo}\ ,\ \bibinfo {pages} {30}} (\bibinfo {year} {1910})}\BibitemShut
  {NoStop}%
\bibitem [{\citenamefont {Livens}(1918)}]{LinZonghan}%
  \BibitemOpen
  \bibfield  {author} {\bibinfo {author} {\bibfnamefont {G.~H.}\ \bibnamefont
  {Livens}},\ }\href@noop {} {\emph {\bibinfo {title} {The Theory of
  Electricity}}}\ (\bibinfo  {publisher} {Cambridge University Press,
  Cambridge},\ \bibinfo {year} {1918})\BibitemShut {NoStop}%
\bibitem [{\citenamefont {Marx}\ and\ \citenamefont {Gyorgyi}(1955)}]{Marx}%
  \BibitemOpen
  \bibfield  {author} {\bibinfo {author} {\bibfnamefont {G.}~\bibnamefont
  {Marx}}\ and\ \bibinfo {author} {\bibfnamefont {G.}~\bibnamefont {Gyorgyi}},\
  }\href@noop {} {\bibfield  {journal} {\bibinfo  {journal} {Ann Phys}\ ,\
  \bibinfo {pages} {16}} (\bibinfo {year} {1955})}\BibitemShut {NoStop}%
\bibitem [{\citenamefont {Grot}\ and\ \citenamefont {Eringen}(1966)}]{Grot}%
  \BibitemOpen
  \bibfield  {author} {\bibinfo {author} {\bibfnamefont {R.~A.}\ \bibnamefont
  {Grot}}\ and\ \bibinfo {author} {\bibfnamefont {A.~C.}\ \bibnamefont
  {Eringen}},\ }\href@noop {} {\bibfield  {journal} {\bibinfo  {journal} {Int J
  Eng Sci}\ ,\ \bibinfo {pages} {4}} (\bibinfo {year} {1966})}\BibitemShut
  {NoStop}%
\bibitem [{\citenamefont {de~Groot}\ and\ \citenamefont
  {Suttorp}(1967)}]{Groot}%
  \BibitemOpen
  \bibfield  {author} {\bibinfo {author} {\bibfnamefont {S.~R.}\ \bibnamefont
  {de~Groot}}\ and\ \bibinfo {author} {\bibfnamefont {L.~G.}\ \bibnamefont
  {Suttorp}},\ }\href@noop {} {\emph {\bibinfo {title} {Physica}}}\ (\bibinfo
  {publisher} {Amsterdam},\ \bibinfo {year} {1967})\BibitemShut {NoStop}%
\bibitem [{\citenamefont {de~Groot}\ and\ \citenamefont
  {Suttorp}(1968)}]{Groot2}%
  \BibitemOpen
  \bibfield  {author} {\bibinfo {author} {\bibfnamefont {S.~R.}\ \bibnamefont
  {de~Groot}}\ and\ \bibinfo {author} {\bibfnamefont {L.~G.}\ \bibnamefont
  {Suttorp}},\ }\href@noop {} {\emph {\bibinfo {title} {Physica}}}\ (\bibinfo
  {publisher} {Amsterdam},\ \bibinfo {year} {1968})\BibitemShut {NoStop}%
\bibitem [{\citenamefont {de~Groot}\ and\ \citenamefont
  {Suttorp}(1972)}]{Groot3}%
  \BibitemOpen
  \bibfield  {author} {\bibinfo {author} {\bibfnamefont {S.~R.}\ \bibnamefont
  {de~Groot}}\ and\ \bibinfo {author} {\bibfnamefont {L.~G.}\ \bibnamefont
  {Suttorp}},\ }\href@noop {} {\emph {\bibinfo {title} {Foundations of
  Electrodynamics}}}\ (\bibinfo  {publisher} {North-Holland, Amsterdam},\
  \bibinfo {year} {1972})\BibitemShut {NoStop}%
\bibitem [{\citenamefont {Penfield}\ and\ \citenamefont
  {Haus}(1967)}]{Penfield}%
  \BibitemOpen
  \bibfield  {author} {\bibinfo {author} {\bibfnamefont {J.}~\bibnamefont
  {Penfield}, \bibfnamefont {P.}}\ and\ \bibinfo {author} {\bibfnamefont
  {H.~A.}\ \bibnamefont {Haus}},\ }\href@noop {} {\emph {\bibinfo {title}
  {Electrodynamics of Moving Media}}}\ (\bibinfo  {publisher} {MIT, Cambridge,
  MA},\ \bibinfo {year} {1967})\BibitemShut {NoStop}%
\bibitem [{\citenamefont {Grot}\ and\ \citenamefont {Eringen}(1976)}]{Peierls}%
  \BibitemOpen
  \bibfield  {author} {\bibinfo {author} {\bibfnamefont {R.~A.}\ \bibnamefont
  {Grot}}\ and\ \bibinfo {author} {\bibfnamefont {A.~C.}\ \bibnamefont
  {Eringen}},\ }\href@noop {} {\bibfield  {journal} {\bibinfo  {journal} {Proc.
  R. Soc. London, Ser.A}\ ,\ \bibinfo {pages} {347}} (\bibinfo {year}
  {1976})}\BibitemShut {NoStop}%
\bibitem [{\citenamefont {Barnett}(2010)}]{Stephen}%
  \BibitemOpen
  \bibfield  {author} {\bibinfo {author} {\bibfnamefont {S.~M.}\ \bibnamefont
  {Barnett}},\ }\href {\doibase 10.1103/physrevlett.104.070401} {\bibfield
  {journal} {\bibinfo  {journal} {Physical Review Letters}\ }\textbf {\bibinfo
  {volume} {104}} (\bibinfo {year} {2010}),\
  10.1103/physrevlett.104.070401}\BibitemShut {NoStop}%
\bibitem [{\citenamefont {Gordon}\ and\ \citenamefont
  {James}(1973)}]{Gordon1973Radiation}%
  \BibitemOpen
  \bibfield  {author} {\bibinfo {author} {\bibnamefont {Gordon}}\ and\ \bibinfo
  {author} {\bibfnamefont {P.}~\bibnamefont {James}},\ }\href {\doibase
  10.1103/PhysRevA.8.14} {\bibfield  {journal} {\bibinfo  {journal} {Physical
  Review A}\ }\textbf {\bibinfo {volume} {8}},\ \bibinfo {pages} {14} (\bibinfo
  {year} {1973})}\BibitemShut {NoStop}%
\bibitem [{\citenamefont {Mlkura}(1976)}]{1976Variational}%
  \BibitemOpen
  \bibfield  {author} {\bibinfo {author} {\bibfnamefont {Z.}~\bibnamefont
  {Mlkura}},\ }\href {\doibase 10.1103/PhysRevA.13.2265} {\bibfield  {journal}
  {\bibinfo  {journal} {Physical Review A}\ }\textbf {\bibinfo {volume} {13}},\
  \bibinfo {pages} {2265} (\bibinfo {year} {1976})}\BibitemShut {NoStop}%
\bibitem [{\citenamefont {Miroslav}(1979)}]{1979About}%
  \BibitemOpen
  \bibfield  {author} {\bibinfo {author} {\bibfnamefont {K.}~\bibnamefont
  {Miroslav}},\ }\href {\doibase 10.1139/p79-140} {\bibfield  {journal}
  {\bibinfo  {journal} {Can. J. Phys}\ }\textbf {\bibinfo {volume} {57}},\
  \bibinfo {pages} {1022} (\bibinfo {year} {1979})}\BibitemShut {NoStop}%
\bibitem [{\citenamefont {Maugin}(1980)}]{Maugin1980Further}%
  \BibitemOpen
  \bibfield  {author} {\bibinfo {author} {\bibfnamefont {G.~A.}\ \bibnamefont
  {Maugin}},\ }\href {\doibase 10.1139/p80-155} {\bibfield  {journal} {\bibinfo
   {journal} {Canadian Journal of Physics}\ }\textbf {\bibinfo {volume} {58}},\
  \bibinfo {pages} {1163} (\bibinfo {year} {1980})}\BibitemShut {NoStop}%
\bibitem [{\citenamefont {Dereli}\ \emph {et~al.}(2007)\citenamefont {Dereli},
  \citenamefont {Gratus},\ and\ \citenamefont {Tucker}}]{2007New}%
  \BibitemOpen
  \bibfield  {author} {\bibinfo {author} {\bibfnamefont {T.}~\bibnamefont
  {Dereli}}, \bibinfo {author} {\bibfnamefont {J.}~\bibnamefont {Gratus}}, \
  and\ \bibinfo {author} {\bibfnamefont {R.~W.}\ \bibnamefont {Tucker}},\
  }\href {\doibase 10.1088/1751-8113/40/21/016} {\bibfield  {journal} {\bibinfo
   {journal} {Journal of Physics A Mathematical and Theoretical}\ }\textbf
  {\bibinfo {volume} {40}},\ \bibinfo {pages} {5695} (\bibinfo {year}
  {2007})}\BibitemShut {NoStop}%
\bibitem [{\citenamefont {Mahdy}\ \emph {et~al.}(2020)\citenamefont {Mahdy},
  \citenamefont {Rivy}, \citenamefont {Jony},\ and\ \citenamefont
  {et~al}}]{2020Dielectric}%
  \BibitemOpen
  \bibfield  {author} {\bibinfo {author} {\bibfnamefont {M.}~\bibnamefont
  {Mahdy}}, \bibinfo {author} {\bibfnamefont {H.~M.}\ \bibnamefont {Rivy}},
  \bibinfo {author} {\bibfnamefont {Z.~R.}\ \bibnamefont {Jony}}, \ and\
  \bibinfo {author} {\bibnamefont {et~al}},\ }\href {\doibase
  10.1088/1751-8113/40/21/016} {\bibfield  {journal} {\bibinfo  {journal}
  {Chinese Physics B}\ }\textbf {\bibinfo {volume} {29}},\ \bibinfo {pages}
  {014211 (13pp)} (\bibinfo {year} {2020})}\BibitemShut {NoStop}%
\bibitem [{\citenamefont {Jones}(1951)}]{JONES}%
  \BibitemOpen
  \bibfield  {author} {\bibinfo {author} {\bibfnamefont {R.~V.}\ \bibnamefont
  {Jones}},\ }\href {\doibase 10.1038/167439a0} {\bibfield  {journal} {\bibinfo
   {journal} {Radiation Pressure in a Refracting Medium}\ }\textbf {\bibinfo
  {volume} {167}},\ \bibinfo {pages} {4246} (\bibinfo {year}
  {1951})}\BibitemShut {NoStop}%
\bibitem [{\citenamefont {Jones}\ and\ \citenamefont
  {Richards}(1954)}]{JONES1954}%
  \BibitemOpen
  \bibfield  {author} {\bibinfo {author} {\bibfnamefont {R.~V.}\ \bibnamefont
  {Jones}}\ and\ \bibinfo {author} {\bibfnamefont {J.~C.~S.}\ \bibnamefont
  {Richards}},\ }\href@noop {} {\bibfield  {journal} {\bibinfo  {journal}
  {Proc. R. Soc. London,Ser.}\ }\textbf {\bibinfo {volume} {A}},\ \bibinfo
  {pages} {480} (\bibinfo {year} {1954})}\BibitemShut {NoStop}%
\bibitem [{\citenamefont {Ashkin}\ and\ \citenamefont
  {Dziedzic}(1973)}]{1973Radiation}%
  \BibitemOpen
  \bibfield  {author} {\bibinfo {author} {\bibfnamefont {A.}~\bibnamefont
  {Ashkin}}\ and\ \bibinfo {author} {\bibfnamefont {J.~M.}\ \bibnamefont
  {Dziedzic}},\ }\href {\doibase 10.1103/PhysRevLett.30.139} {\bibfield
  {journal} {\bibinfo  {journal} {Physical Review Letters}\ }\textbf {\bibinfo
  {volume} {30}},\ \bibinfo {pages} {139} (\bibinfo {year} {1973})}\BibitemShut
  {NoStop}%
\bibitem [{\citenamefont {Walker}\ \emph {et~al.}(1975)\citenamefont {Walker},
  \citenamefont {Lahoz},\ and\ \citenamefont {Walker}}]{G1975Measurement}%
  \BibitemOpen
  \bibfield  {author} {\bibinfo {author} {\bibfnamefont {G.~B.}\ \bibnamefont
  {Walker}}, \bibinfo {author} {\bibfnamefont {D.~G.}\ \bibnamefont {Lahoz}}, \
  and\ \bibinfo {author} {\bibfnamefont {G.}~\bibnamefont {Walker}},\ }\href
  {\doibase 10.1139/p75-313} {\bibfield  {journal} {\bibinfo  {journal}
  {Canadian Journal of Physics}\ }\textbf {\bibinfo {volume} {53}},\ \bibinfo
  {pages} {139} (\bibinfo {year} {1975})}\BibitemShut {NoStop}%
\bibitem [{\citenamefont {A.F.Gibson}\ \emph {et~al.}(1980)\citenamefont
  {A.F.Gibson}, \citenamefont {M.F.Kimmitt}, \citenamefont {A.O.Koohian},\ and\
  \citenamefont {et~al}}]{A1980Walker}%
  \BibitemOpen
  \bibfield  {author} {\bibinfo {author} {\bibnamefont {A.F.Gibson}}, \bibinfo
  {author} {\bibnamefont {M.F.Kimmitt}}, \bibinfo {author} {\bibnamefont
  {A.O.Koohian}}, \ and\ \bibinfo {author} {\bibnamefont {et~al}},\ }\href
  {\doibase 10.1098/rspa.1980.0035} {\bibfield  {journal} {\bibinfo  {journal}
  {Proceedings of the Royal Society A Mathematical Physical and Engineering
  Sciences}\ }\textbf {\bibinfo {volume} {370}},\ \bibinfo {pages} {303}
  (\bibinfo {year} {1980})}\BibitemShut {NoStop}%
\bibitem [{\citenamefont {Campbell}\ \emph {et~al.}(2005)\citenamefont
  {Campbell}, \citenamefont {Leanhardt}, \citenamefont {Mun},\ and\
  \citenamefont {et~al}}]{Campbell2005}%
  \BibitemOpen
  \bibfield  {author} {\bibinfo {author} {\bibfnamefont {G.~K.}\ \bibnamefont
  {Campbell}}, \bibinfo {author} {\bibfnamefont {A.~E.}\ \bibnamefont
  {Leanhardt}}, \bibinfo {author} {\bibfnamefont {J.}~\bibnamefont {Mun}}, \
  and\ \bibinfo {author} {\bibnamefont {et~al}},\ }\href {\doibase
  10.1103/physrevlett.94.170403} {\bibfield  {journal} {\bibinfo  {journal}
  {Physical Review Letters}\ }\textbf {\bibinfo {volume} {94}} (\bibinfo {year}
  {2005}),\ 10.1103/physrevlett.94.170403}\BibitemShut {NoStop}%
\bibitem [{\citenamefont {Weilong}\ \emph {et~al.}(2008)\citenamefont
  {Weilong}, \citenamefont {Jianhui},\ and\ \citenamefont
  {Feng}}]{Weilong2008}%
  \BibitemOpen
  \bibfield  {author} {\bibinfo {author} {\bibfnamefont {S.}~\bibnamefont
  {Weilong}}, \bibinfo {author} {\bibfnamefont {Y.}~\bibnamefont {Jianhui}}, \
  and\ \bibinfo {author} {\bibfnamefont {R.}~\bibnamefont {Feng}},\ }\href
  {\doibase 10.1103/physrevlett.101.243601} {\bibfield  {journal} {\bibinfo
  {journal} {Physical Review Letters}\ }\textbf {\bibinfo {volume} {101}}
  (\bibinfo {year} {2008}),\ 10.1103/physrevlett.101.243601}\BibitemShut
  {NoStop}%
\bibitem [{\citenamefont {Bellantoni}(2010)}]{2010Noether}%
  \BibitemOpen
  \bibfield  {author} {\bibinfo {author} {\bibfnamefont {L.}~\bibnamefont
  {Bellantoni}},\ }\href@noop {} {\emph {\bibinfo {title} {Noether's
  Theorem}}}\ (\bibinfo  {publisher} {Modern Physics},\ \bibinfo {year}
  {2010})\BibitemShut {NoStop}%
\bibitem [{\citenamefont {E.Noether}(1971)}]{Noether}%
  \BibitemOpen
  \bibfield  {author} {\bibinfo {author} {\bibnamefont {E.Noether}},\ }\href
  {\doibase doi:10.1080/00411457108231446} {\bibfield  {journal} {\bibinfo
  {journal} {Transport Theory and Statistical Physics}\ }\textbf {\bibinfo
  {volume} {1}},\ \bibinfo {pages} {186} (\bibinfo {year} {1971})}\BibitemShut
  {NoStop}%
\bibitem [{\citenamefont {Canbin}\ and\ \citenamefont
  {Bin}(2006)}]{LiangCanbin2006}%
  \BibitemOpen
  \bibfield  {author} {\bibinfo {author} {\bibfnamefont {L.}~\bibnamefont
  {Canbin}}\ and\ \bibinfo {author} {\bibfnamefont {Z.}~\bibnamefont {Bin}},\
  }\href@noop {} {\emph {\bibinfo {title} {Introduction to Differential
  geometry and General Relativity}}}\ (\bibinfo  {publisher} {Science Press},\
  \bibinfo {year} {2006})\BibitemShut {NoStop}%
\bibitem [{\citenamefont {Dirac}(1979)}]{Dirac}%
  \BibitemOpen
  \bibfield  {author} {\bibinfo {author} {\bibfnamefont {P.~A.~M.}\
  \bibnamefont {Dirac}},\ }\href@noop {} {\emph {\bibinfo {title} {General
  Theory of RelAtivity}}}\ (\bibinfo  {publisher} {Science Press},\ \bibinfo
  {year} {1979})\BibitemShut {NoStop}%
\bibitem [{\citenamefont {C.Misner}\ \emph {et~al.}(1997)\citenamefont
  {C.Misner}, \citenamefont {K.Thorne},\ and\ \citenamefont
  {J.Wheeler}}]{Wheeler}%
  \BibitemOpen
  \bibfield  {author} {\bibinfo {author} {\bibnamefont {C.Misner}}, \bibinfo
  {author} {\bibnamefont {K.Thorne}}, \ and\ \bibinfo {author} {\bibnamefont
  {J.Wheeler}},\ }\href@noop {} {\emph {\bibinfo {title} {Gravitation}}}\
  (\bibinfo  {publisher} {Zhengzhong Bookstore Co., Ltd},\ \bibinfo {year}
  {1997})\BibitemShut {NoStop}%
\bibitem [{\citenamefont {Shuohong}(2008)}]{Shuohong}%
  \BibitemOpen
  \bibfield  {author} {\bibinfo {author} {\bibfnamefont {G.}~\bibnamefont
  {Shuohong}},\ }\href@noop {} {\emph {\bibinfo {title} {Electrodynamics}}}\
  (\bibinfo  {publisher} {Higher Education Press},\ \bibinfo {year}
  {2008})\BibitemShut {NoStop}%
\bibitem [{\citenamefont {von Laue}(1950)}]{vonLaue}%
  \BibitemOpen
  \bibfield  {author} {\bibinfo {author} {\bibnamefont {von Laue}},\
  }\href@noop {} {\bibfield  {journal} {\bibinfo  {journal} {Z. Phys}\ }\textbf
  {\bibinfo {volume} {128}} (\bibinfo {year} {1950})}\BibitemShut {NoStop}%
\bibitem [{\citenamefont {Landau}\ and\ \citenamefont {E.M.}(1980)}]{Landau}%
  \BibitemOpen
  \bibfield  {author} {\bibinfo {author} {\bibnamefont {Landau}}\ and\ \bibinfo
  {author} {\bibnamefont {E.M.}},\ }\href@noop {} {\emph {\bibinfo {title}
  {Quantum mechanics}}}\ (\bibinfo  {publisher} {Higher Education Press},\
  \bibinfo {year} {1980})\BibitemShut {NoStop}%
\bibitem [{\citenamefont {Yi}()}]{DaiYi}%
  \BibitemOpen
  \bibfield  {author} {\bibinfo {author} {\bibfnamefont {D.}~\bibnamefont
  {Yi}},\ }\emph {\bibinfo {title} {Research on Abraham Minkowski momentum in
  media}},\ \href@noop {} {Ph.D. thesis},\ \bibinfo  {school} {Wuhan University
  of Science and Technology}\BibitemShut {NoStop}%
\end{thebibliography}%

\end{document}